\documentclass[aps,prb,twocolumn,superscriptaddress,longbibliography]{revtex4-2}
\usepackage{amsmath,amssymb,amsthm}
\usepackage{graphicx}
\usepackage[colorlinks=true,urlcolor=blue,linkcolor=blue,citecolor=blue]{hyperref}
\usepackage{physics}
\usepackage[dvipsnames]{xcolor}
\usepackage{bm}
\usepackage{soul}
\usepackage{bbold}
\usepackage{pifont}
\usepackage{mathtools}
\usepackage{tikz}
\usetikzlibrary{tikzmark,arrows.meta,positioning,shapes.geometric,calc}
\usepackage{array}
\usepackage[ruled,noline]{algorithm2e}
\usepackage[normalem]{ulem}
\usepackage{mdframed}
\usepackage{enumitem}
\usepackage{float}
\usepackage{makecell}
\usepackage{comment}

\DeclareMathOperator{\Sgn}{Sgn}

\newcommand{\ii}{\boldsymbol{\mathrm{i}}}
\newcommand{\mc}[1]{\mathcal{#1}}
\newcommand{\mr}[1]{\mathrm{#1}}
\newcommand{\mf}[1]{\mathfrak{#1}}
\newcommand{\mb}[1]{\mathbb{#1}}
\newcommand{\mbf}[1]{\mathbf{#1}}
\newcommand{\bs}[1]{\boldsymbol{#1}}
\newcommand{\h}[1]{\hat{#1}}
\newcommand{\Id}{\mb{1}}

\newcommand{\cT}{\mc{T}}
\newcommand{\cC}{\mc{C}}

\newcommand{\cU}{\mc{U}}
\newcommand{\hcT}{\h{\mc{T}}}

\newcommand{\hcN}{\h{\mc{N}}}

\newcommand{\bfr}{{\bf r}}
\newcommand{\TS}{|{\mr{TS}} \rangle\!\rangle}
\newcommand{\CI}{|{\mr{CI}} \rangle\!\rangle}
\newcommand{\vac}{|\mr{vac}\rangle\!\rangle}

\newcommand{\AZK}[1]{\mc{K}_{\mr{#1}}}

\newcommand{\cmark}{\textcolor{green}{\ding{51}}}
\newcommand{\xmark}{\textcolor{red}{\ding{55}}}

\theoremstyle{plain}
\newtheorem{theorem}{Theorem}
\theoremstyle{definition}
\newtheorem{definition}[theorem]{Definition}
\theoremstyle{remark}
\newtheorem{remark}[theorem]{Remark}

\setstcolor{magenta}

\begin{document}

\title{Free-Fermion Dynamics with Measurements: Topological Classification and Adaptive Preparation of Topological States}

\author{Asadullah Bhuiyan}
\affiliation{Department of Physics, Cornell University, Ithaca, New York 14853, USA}

\author{Haining Pan}
\affiliation{Department of Physics and Astronomy, Center for Materials Theory, Rutgers University, Piscataway, New Jersey 08854 USA}

\author{Chao-Ming Jian}
\affiliation{Department of Physics, Cornell University, Ithaca, New York 14853, USA}

\begin{abstract}
We develop a general framework for classifying fermionic dynamical systems with measurements using symmetry and topology. We introduce two complementary classification schemes based on the Altland–Zirnbauer tenfold way: (1) the many-body evolution operator (mEO) symmetry class, which classifies fermionic dynamics at the many-body level and naturally extends to interacting dynamics, and (2) the single-particle transfer matrix (sTM) symmetry class, which classifies free-fermion dynamics at the single-particle level and connects to Anderson localization physics. In the free-fermion limit, we show that these two frameworks are equivalent via a novel dynamical bulk–boundary correspondence: the topology of the dynamical system's spacetime bulk determines the topology of the area-law entangled steady-state ensemble living on its temporal boundary. Next, we prove that symmetry-invariant, post-selection-free Gaussian measurements are realizable in only four of the ten mEO classes (A, AI, BDI, D); the remaining six require either post-selection or interacting (non-Gaussian) measurements. Building on these results, we construct general post-selection-free topological adaptive circuits that realize topological dynamical phases in any spatial dimension for the four admissible mEO classes. These circuits simultaneously provide a protocol for preparing and stabilizing free-fermion topological states in all ten symmetry classes. As a concrete demonstration, we construct and simulate 2+1d adaptive circuits that realize mEO-class-A topological dynamics, steering toward a steady-state ensemble of Chern insulators in ${\cal O}(1)$ circuit depth. Finally, we numerically characterize topological phase transitions, dynamical domain-wall modes, and robustness to coherent noise, identifying finite error thresholds at which trajectory-resolved and trajectory-averaged quantities undergo distinct phase transitions.

\end{abstract}
\maketitle

\section{Introduction}
Dynamical quantum systems driven by both unitary evolution and measurements provide a novel setting for exploring new realms of out-of-equilibrium quantum matter. Since the discovery of the measurement-induced entanglement phase transitions in dynamical qubit systems~\cite{skinner2019measurementinduced,li2018quantum,li2019measurementdriven,chan2019unitaryprojective,potter2022entanglement,fisher2023random}, the landscape of dynamical phases and transitions featuring novel entanglement scalings has been under extensive exploration (see, for example, Refs.~\cite{skinner2019measurementinduced,li2018quantum,li2019measurementdriven,chan2019unitaryprojective,vasseur2019entanglement,gullans2020dynamical,gullans2020scalable,zabalo2020critical,choi2020quantum,li2021conformal,jian2020measurementinduced,bao2020theory,lang2020entanglement,turkeshi2020measurementinduced,ippoliti2021entanglement,nahum2021measurement,zabalo2022operator,li2024statistical,potter2022entanglement,fisher2023random}). This landscape is further enriched by the interplay between global symmetry and many-body entanglement ~\cite{sang2021measurementprotected,han2022measurementinduced,lavasani2021measurementinduced,BaoSymmetryEnriched,agrawal2022entanglement,barratt2022field,jian2022criticality,jian2023measurementinduced,fava2023nonlinear,poboiko2023theory,fava2024monitored,majidy2023critical,chakraborty2024charge}.
 
In contrast, our understanding of the role of topology in dynamical systems with measurements is limited without assuming any spacetime periodicity. In qudit-based systems, previous works~\cite{BaoSymmetryEnriched,lavasani2021measurementinduced,lavasani2021topological} have identified examples of novel dynamical phases where the system can reach topologically nontrivial final states in the long-time limit. However, a systematic topology characterization in qubit-based dynamical systems has yet to be developed. Meanwhile, recent exploration of dynamical topological phenomena through the lens of fermionic systems has shown promise both in modeling new classes of topological dynamical systems~\cite{nahum2020entanglement,BeriSurfaceCode2024,pan2025topological,kells2023topological,KawabataTopoMonitored,oshima2025topology}, as well as advancing the general framework of topological classification~\cite{jian2022criticality,pan2025topological,KawabataTopoMonitored}. These recent developments are partly enabled by nontrivial topologies emerging even in the free-fermion limit. 

In free-fermion dynamical systems, both unitary evolution and measurements are implemented by Gaussian operators that preserve the non-interacting nature of fermionic states. In other words, a non-interacting fermionic state, described by a Slater determinant or its charge-unconserved generalization, remains non-interacting under both unitary evolution and measurement-induced wavefunction collapse. A general strategy to systematically study the topology of such free-fermion dynamics is based on the Altland-Zirnbauer (AZ) tenfold symmetry classification~\cite{altland1997nonstandard,jian2022criticality,pan2025topological,KawabataTopoMonitored}. Ref.~\cite{jian2022criticality} showed that free-fermion dynamics with measurements can be unified with the classic Anderson localization problem of disordered fermions in equilibrium under this AZ symmetry classification. The spacetime propagation of fermions in the dynamical system can be identified with the elastic scattering in space in the Anderson localization problem through a single-particle transfer matrix (sTM) framework. Consequently, the AZ symmetry classification, developed initially for disordered fermions~\cite{altland1997nonstandard}, also applies to free-fermion dynamics with measurements. In this paper, we refer to the resulting AZ symmetry class through the sTMs of the free-fermion dynamics as the sTM class ${\cal K}_{\rm sTM}$. Like in the Anderson localization problem, one generally expects the sTM class ${\cal K}_{\rm sTM}$ to govern the universal behaviors of free-fermion dynamics.

The unification between free-fermion dynamics and the Anderson localization problems further suggests that free-fermion dynamical phases exhibiting area-law entanglement entropy in the long-time limit share the same topological classification as disordered topological insulators or superconductors within the same sTM class ${\cal K}_{\rm sTM}$~\cite{ryu2010topological,pan2025topological,KawabataTopoMonitored}. For area-entangled free-fermion dynamics in 1+1d spacetime, concrete quantum circuit models that confirm this topological classification have been identified in Refs.~\cite{nahum2020entanglement,BeriSurfaceCode2024,pan2025topological,kells2023topological,KawabataTopoMonitored,oshima2025topology}.

A crucial and timely direction is to seek systematic microscopic models that realize topological free-fermion dynamics in higher-dimensional spacetime. The construction of the desired models, especially the ones with {\it post-selection-free} measurements, is not immediately clear from the abstract topological classification above and the existing 1+1d examples. Identifying such models will not only confirm the proposed abstract topological classification but also provide platforms for further investigating the phase transitions between dynamical phases with different topologies. Moreover, the models without post-selection will offer recipes for efficient simulations of the topological dynamical phases of matter on programmable quantum platforms~\cite{altmanQuantumSimulatorsArchitectures2021,monroeProgrammableQuantumSimulations2021,daleyPracticalQuantumAdvantage2022,bluvsteinLogicalQuantumProcessor2024, craneHybridOscillatorQubitQuantum2024}.

On a separate note, the formulation of the AZ symmetry class ${\cal K}_{\rm sTM}$ based on the sTMs has features that seem suboptimal and require deeper understanding. The time-reversal, particle-hole, and chiral symmetries associated with the sTM class ${\cal K}_{\rm sTM}$ constrain the structure of sTMs in free-fermion dynamics, but not in the conventional manner that global symmetries constrain the fermion propagator in Hamiltonian systems. That is because these symmetries are the explicit global symmetries of the Anderson localization problem that the free-fermion dynamics maps to in the aforementioned sTM framework. But they do not directly act on the Hilbert space of the dynamical fermion system.

Moreover, this current AZ symmetry classification based on sTMs is limited to free-fermion systems. Information about single-particle propagation is insufficient to describe interacting systems. Therefore, a new symmetry classification scheme is desirable based on the symmetries acting on the dynamical system and generalizable to interacting cases. As we show in this paper, developing this new symmetry classification scheme will also inform the general construction of topological free-fermion dynamics in higher spacetime dimensions. 

In this paper, we first introduce a new symmetry classification scheme for fermionic dynamics with measurements and feedforward. This scheme is directly based on the explicit symmetries of the dynamical system and the many-body evolution operators (mEOs) that capture the quantum evolution along different quantum trajectories. We will refer to the AZ symmetry class identified through the mEOs as the mEO class $\AZK{mEO}$. We show that, in the free-fermion limit, the mEO class $\AZK{mEO}$ is in one-to-one correspondence with an sTM class $\AZK{sTM}$. Also, mEO-based symmetry classification naturally extends to interacting fermion dynamics. Interestingly, by analyzing the effects of the time-reversal, particle-hole, and chiral symmetries associated with the mEO class $\AZK{mEO}$ in area-law entangled phases of the free-fermion dynamics, we uncover a novel {\it dynamical bulk-boundary correspondence} --- the nontrivial topology of the spacetime bulk implies a topological steady-state ensemble, which effectively resides on the temporal boundary of spacetime. In light of this dynamical bulk-boundary correspondence, constructing microscopic models of topological free-fermion dynamics amounts to designing dynamics that generate a topologically nontrivial steady-state ensemble within the same mEO class $\AZK{mEO}$.

The realization of area-law-entangled topological dynamics necessarily involves measurements. Notably, the symmetry class imposes nontrivial constraints on the allowed forms of measurement in the free-fermion limit. Specifically, we prove that in the free-fermion limit, measurements without post-selection are permitted only for the four mEO classes $\AZK{mEO} = \text{A, AI, BDI, and D}$. These four mEO classes correspond to sTM classes $\AZK{sTM} = \text{AIII, BDI, D, and DIII}$ respectively in the absence of interactions. In contrast, topological dynamics in the remaining six mEO class $\AZK{mEO} = \text{AIII, DIII, AII, CII, C, and CI}$ must involve post-selection in the free-fermion limit.

Building on the dynamical bulk-boundary correspondence and the measurement constraints imposed by symmetry classes, we develop a general construction of post-selection-free topological adaptive circuits that realize topological free-fermion dynamics in arbitrary spacetime dimensions for the mEO classes $\AZK{mEO} = \text{A, AI, BDI, and D}$. Each of these topological adaptive circuits implements measurements and feedforward operations to steer the system toward a target free-fermion topological state in one of these four symmetry classes. By the dynamical bulk-boundary correspondence, such adaptive circuits realize an ideal scenario of topological free-fermion dynamics, in which the steady-state ensemble consists of a single topological state. In our construction, the target topological state is formulated as the common eigenstate of a set of non-commuting stabilizers. The topological adaptive circuit repeatedly measures these stabilizers and applies ``error-correcting'' feedforward operations when the measurements yield undesired outcomes. Conceptually, our construction is similar to the measurement-induced steering protocol for the Affleck–Kennedy–Lieb–Tasaki (AKLT) state presented in Ref.~\cite{puente_quantum_2024,roy_measurement-induced_2020}, where the individual terms in the frustration-free Hamiltonian of the AKLT states play the role of the non-commuting stabilizers.

As a concrete example, we apply this general construction to topological free-fermion dynamics in mEO-class-A in 2+1d spacetime. The resulting adaptive circuits target and stabilize Chern insulator states in two spatial dimensions as their steady states. Furthermore, we use an effective Lindbladian description to estimate the timescale of the system's convergence toward the target Chern insulator states.  

The same construction of topological adaptive circuits is still applicable when we choose target states that belong to the remaining six mEO classes $\AZK{mEO} = \text{AIII, DIII, AII, CII, C, and CI}$. However, the resulting circuits should not be viewed as topological dynamics in these symmetry classes, as the measurements involved explicitly break the symmetries associated with these six symmetry classes. This does not contradict the dynamical bulk-boundary correspondence. In any case, these adaptive circuits can be treated as a novel method for preparing and stabilizing free-fermion topological states in all symmetry classes. This method is fundamentally different from the previously proposed method for preparing invertible topological states~\cite{barbarino2020preparing,AdaptiveMPSPrep}, which does not include mechanisms to correct errors in the preparation and to stabilize the target state after the preparation.

The topological adaptive circuits that steer the system precisely toward a single topological state generally require quantum gates with exponential tails. It is therefore essential to understand how these circuits behave when their gates are restricted to act within a finite range. As an example, we consider topological dynamics in mEO class A in $2+1$d spacetime. The finite-range version of the corresponding adaptive circuits will no longer steer the system toward a single state. Nevertheless, our numerical simulations show that they still produce nontrivial steady-state ensembles of Chern insulators with the same Chern number. Thus, these finite-range adaptive circuits remain valid models for mEO-class-A topological free-fermion dynamics in 2+1d spacetime. Notably, our simulations also show that the nontrivial steady-state ensemble is reached within $\mc{O}(1)$ circuit depth. This means that our finite-range topological adaptive circuits provide efficient recipes for preparing and stabilizing Chern insulator states in two dimensions. 

Our mEO-class-A topological adaptive circuits contain a parameter that allows us to adjust the Chern number in the steady-state ensemble. This parameter enables us to explore a phase diagram with different topological dynamics. Using numerical simulations, we show that the topologically distinct dynamics are separated by novel dynamical topological phase transitions. This transition further confirms the expectation that the Chern number of the steady-state ensemble reflects the topology in the spacetime bulk. Moreover, we construct a new dynamical system that interpolates between two topologically distinct adaptive circuits in real space. From the spacetime perspective, this dynamical system contains two topologically distinct $2+1$d spacetime domains. Notably, we find numerical evidence for nontrivial dynamical modes localized on the $1+1$d topological domain wall. This observation generalizes the previously discovered dynamical modes localized on the $0+1$d topological domain walls~\cite{pan2025topological}.

Lastly, we numerically investigate the stability of our topological adaptive circuits against coherent noise, modeled as additional local random unitary gates interspersed throughout the circuit. We find that the Chern numbers of the states in the steady-state ensemble remain nontrivial up to a finite error threshold. Interestingly, we find that the behavior of individual quantum trajectories and that of the ensemble-averaged density matrix undergo distinct quantum phase transitions at different error strengths.

\subsection*{Organization of Paper}

The organization of the paper is as follows:

\begin{itemize}[leftmargin=+0.2in]
    \item \textbf{Classification Framework and No-Go Theorems (Sec.~\ref{sec:topo_class}):} Introduces two complementary classification schemes: the single-particle transfer matrix (sTM) class and the many-body evolution operator (mEO) class. We show that, in the non-interacting limit, the two classification schemes are in one-to-one correspondence and give rise to equivalent topological classifications related by Bott periodicity. This correspondence underpins a novel dynamical bulk-boundary relation. Furthermore, we establish that post-selection-free Gaussian measurements are admissible in exactly four of the ten mEO classes; the remaining six require either post-selection or non-Gaussian (interacting) measurements.

    \item \textbf{Adaptive Circuit Construction (Sec.~\ref{sec:AdaptiveSteeringDynamics}):} Describes a general adaptive protocol for preparing arbitrary free-fermion topological states and for realizing free-fermion topological dynamics, with an explicit 2+1d construction that prepares Chern insulators. Also introduces an effective Lindbladian theory to estimate convergence timescales. 

    \item \textbf{Numerical Characterization (Sec.~\ref{sec:numerics}):} Presents a numerical study on dynamical topological phase transitions in our 2+1d adaptive circuit. We also numerically demonstrate and study the preparation of 1+1d dynamical topological domain walls. Furthermore, we characterize convergence and robustness to coherent errors, observing distinct error thresholds for trajectory-resolved and trajectory-averaged quantities.

    \item \textbf{Appendices:} These contain technical proofs, extended constructions, and supporting arguments. Topics include the sTM–mEO correspondence (App.~\ref{app: sTM-mEO_relation}), the relationship between symmetry class and admissibility of Gaussian positive-operator valued measures (POVMs) (App.~\ref{app: Gaussian POVM admissibility}), topological obstructions and overcomplete Wannier functions (App.~\ref{app: topological obstructions}), and the effective Lindbladian theory construction and convergence analysis (App.~\ref{app:lindblad}).
\end{itemize}

\section{Symmetry and Topological Classification of Fermionic Dynamics with Measurements}
\label{sec:topo_class} 

\begin{figure*}[t]
    \centering
    \includegraphics[width=\textwidth]{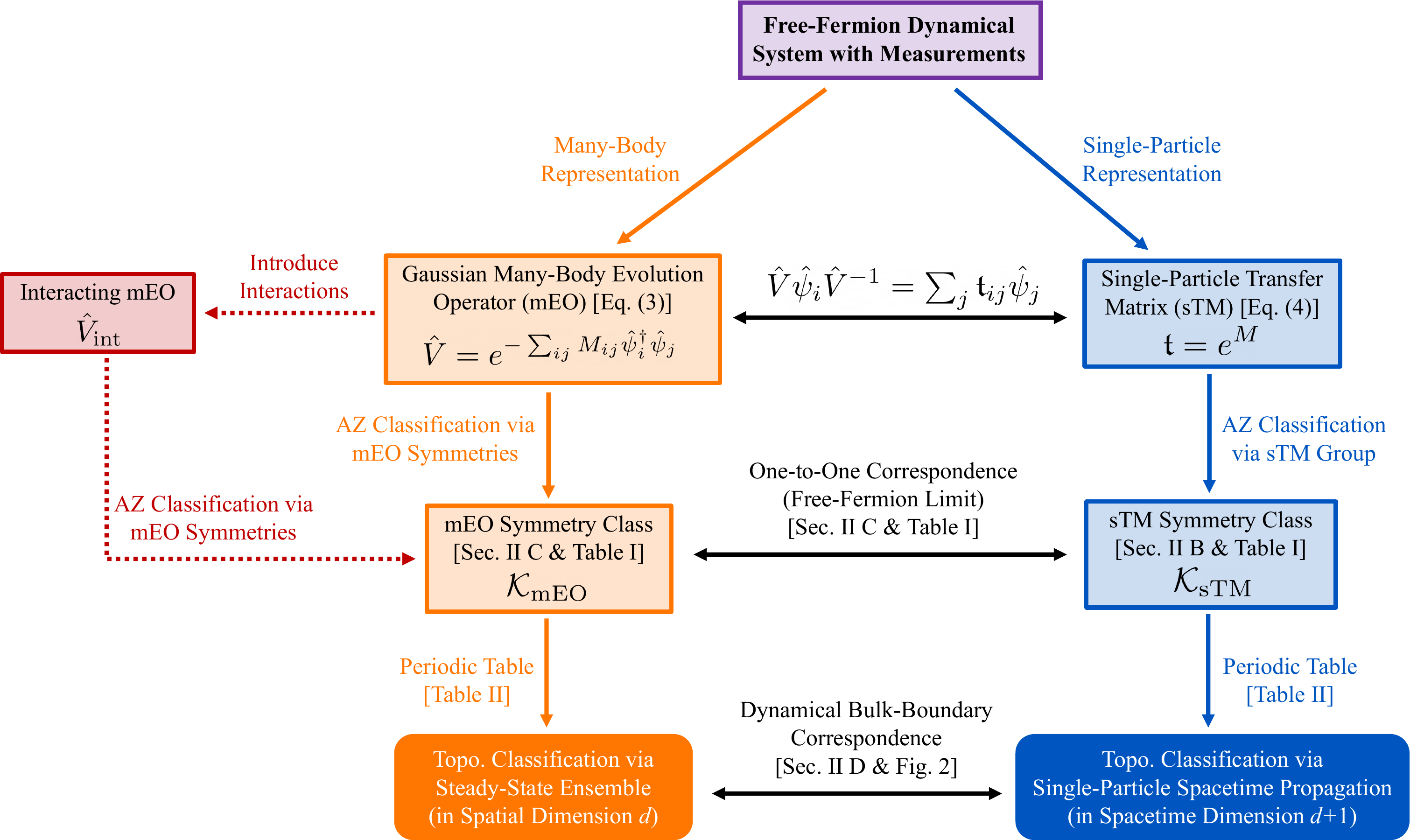}
    \caption{Conceptual framework for classifying fermionic dynamical systems with measurements using symmetry and topology. Here, $\hat\psi^\dag_i$ and $\hat\psi_i$ denote fermion operators satisfying $\{\hat\psi^\dag_i,\hat\psi_j\}=\delta_{ij}$. 
            A free-fermion dynamical system in $D = d+1$ spacetime dimensions can be described either at the many-body level by Gaussian many-body evolution {operators (mEOs)} or at the single-particle level by {single-particle} transfer matrices (sTMs). 
            Each representation admits an Altland–Zirnbauer (AZ) tenfold {symmetry} classification, with the respective AZ class denoted by $\mc{K}_{\mr{mEO}}$ and $\mc{K}_{\mr{sTM}}$. In the non-interacting limit, these two representations are related by a one-to-one correspondence. This correspondence further gives rise to two equivalent topological classifications: {one via the steady-state ensemble in $d$ spatial dimensions based on the mEO framework,}
            and {one via single-particle propagation in $d+1$ spacetime dimensions and its mapping to Anderson localization problems under the sTM framework.
            }
            The AZ classes $\mc{K}_{\mr{mEO}}$ and $\mc{K}_{\mr{sTM}}$ are related by dimensional reduction (as shown in Table~\ref{tab: bott periodicity table}), which underlies a dynamical bulk–boundary correspondence (shown in Fig.~\ref{fig: meo class stm class dim reduction}).
            Unlike the sTM representation, the mEO framework naturally accommodates interactions, enabling a direct extension of the classification to interacting fermionic dynamical systems.
    }
    \label{fig: secII-summary}
\end{figure*}

This section focuses on the symmetry and topological classification of fermion dynamics with measurements. We begin by reviewing the sTM symmetry class, introducing the newly identified mEO symmetry class, and showing their one-to-one correspondence in the free-fermion limit. Next, we study the topological classification of free-fermion dynamics with measurements within each symmetry class and introduce dynamical bulk-boundary correspondence. Finally, we examine how the symmetry classes impose constraints on the allowable forms of measurements. We provide a summary of each subsection next. Furthermore, a high-level overview of our classification schemes and the relationships between them is provided in Fig.~\ref{fig: secII-summary}.

In Sec.~\ref{sec: GaussianDynamicsWithMeasurements}, we introduce the basics of free-fermion dynamics with measurements, particularly the many-body evolution operators (mEO) that capture the system's evolution along different quantum trajectories. We will use the terminology free-fermion dynamics and Gaussian dynamics interchangeably. 

In Sec.~\ref{sec:stm}, we review the description of free-fermion dynamics using the single-particle transfer matrix (sTM). Following Ref.~\cite{jian2022criticality}, we review the universal group structure underlying the sTMs in free-fermion dynamics and the associated AZ symmetry classes. We refer to the AZ symmetry class identified via the sTM as the sTM class ${\cal K}_{\rm sTM}$ in this paper. In this sTM description, free-fermion dynamics is unified with the Anderson localization problem of disordered fermions in equilibrium. We discuss the suboptimal features of the sTM classification scheme, including the implicit role of the symmetries associated with ${\cal K}_{\rm sTM}$ in the dynamical systems and the difficulties in extending to interacting dynamics.  

In Sec.~\ref{sec:mEO class intro}, we introduce a new scheme for the AZ symmetry classification of fermion dynamics with measurements, which we refer to as the mEO class ${\cal K}_{\rm mEO}$. This new scheme is directly based on the mEOs of the dynamics and the symmetries that directly act on the Hilbert space of the dynamical systems. It applies to both free-fermion dynamics and interacting dynamics. In the case of free-fermion dynamics, we provide a one-to-one correspondence between the mEO class ${\cal K}_{\rm mEO}$ and the sTM class ${\cal K}_{\rm sTM}$. 

In Sec.~\ref{Sec:Bott_Clock}, we discuss the topology of the area-law entangled phases of free-fermion dynamics with measurements in all symmetry classes. A topological classification of area-law entangled free-fermion dynamics was proposed in Ref.~\cite{pan2025topological, KawabataTopoMonitored}. Concrete models exhibiting topologically nontrivial dynamics in 1+1d spacetime have been investigated in Ref. \cite{pan2025topological,KawabataTopoMonitored,oshima2025topology,nahum2020entanglement,BeriSurfaceCode2024}. In this work, we propose a dynamical bulk-boundary correspondence that connects the topology of area-law-entangled free-fermion dynamics with the topology of the steady-state ensemble generated by the dynamics. The relation between these two symmetry classification schemes, i.e., the sTM class and mEO class, plays a vital role in this dynamical bulk-boundary correspondence. This correspondence informs a general construction of microscopic models that realize the topological free-fermion dynamics, which will be presented in Sec. \ref{sec:AdaptiveSteeringDynamics}. 

In Sec.~\ref{Sec: POVMadmissibility}, we show that symmetry classes impose constraints on realizing free-fermion dynamics with measurements. In particular, in four of the ten symmetry classes ${\cal K}_{\rm mEO}=$ A, AI, BDI, and D, measurements without post-selection are permitted in the free-fermion limit. In the remaining six symmetry classes, measurements that maintain the free-fermion limit need post-selection.

\subsection{Free-Fermion Dynamics with Measurements and Many-Body Evolution Operators}
\label{sec: GaussianDynamicsWithMeasurements}
We review the basic formalism underlying free-fermion dynamics with measurements~\cite{bravyi2005lagrangian}. We will use the terminology free-fermion dynamics and Gaussian dynamics interchangeably. Free-fermion dynamics is a restricted class of quantum dynamics in which the fermionic quantum many-body states are always Gaussian throughout evolution. Pure Gaussian states can intuitively be thought of as ground states of free-fermion parent Hamiltonians, i.e., Slater determinant states (or their charge-unconserved counterparts). Mixed Gaussian states can be simply viewed as finite-temperature Gibbs states generated by free-fermion parent Hamiltonians. Note that when we use this intuition of a generic Gaussian state, we do not necessarily require the parent Hamiltonian to be local. The density matrix of a generic (charge-conserved) Gaussian state has the form
\begin{align}
    \hat{\rho} \propto e^{- \lambda\sum_{i,j} h_{ij} \hat\psi_i^\dag \hat\psi_j}
    \label{eq:GaussianDensityMatrix}
\end{align}
where $\hat\psi_i$ denotes the $i$th complex fermion operator in the system, $h$ is a Hermitian matrix, i.e., $h_{ij} = h_{ji}^*$, and $\lambda\in\mathbb{R}$. This density matrix describes a pure Gaussian state in the limit $\lambda \rightarrow \pm \infty$. One can straightforwardly generalize Eq.~\eqref{eq:GaussianDensityMatrix} to the cases without charge conservation. 

Another defining feature of Gaussian states is that they satisfy Wick's theorem: the two-point correlation functions encode any higher-point correlation functions in such states. As a result, Gaussian states are completely determined by their collection of two-point functions --- their correlation matrices. For example, in the case of charge-conserved complex fermions, the correlation matrix is given by $G_{ij} \equiv \Tr\left(\hat{\rho} \hat\psi_i^\dag \hat\psi_j\right)$. Efficient classical simulation of the Gaussian states, including their dynamics, can be performed using their correlation matrices.

Now, we introduce the notion of free-fermion or Gaussian dynamics with measurements. In the language of quantum circuits, such dynamics are generated by Gaussian unitary gates and Gaussian measurements. Their definition will be illustrated in the case of charge-conserved complex fermions and can be straightforwardly generalized to the charge-unconserved case using the Majorana fermions. Gaussian unitary gates take the general form of $e^{\ii \sum_{ij} {\cal J}_{ij}\hat\psi_i^\dag \hat\psi_j }$ with a Hermitian matrix ${\cal J}$. The Gaussianity of a fermionic quantum state remains after the evolution under a Gaussian unitary gate. 

Next, we introduce formalism for general measurements. A general quantum measurement can be defined using an ensemble of operators $\{\hat{K}_m\}_{m}$, called Kraus operators, and a collection of non-negative weights $\{w_m\geq0\}_{m}$~\cite{nielsen2012quantum} ($\{\cdot\}_m$ means the set including possible $m$ values). Both  Kraus operators $\hat{K}_m$ and the corresponding weight $w_m$ are labeled by the possible measurement outcomes $m$. The Kraus operators and their weights satisfy the positive-operator-valued-measure (POVM) condition:
\begin{align}
    \sum_{m} w_m\hat{K}^\dagger_m \hat{K}_m = \hat{\Id}. \label{eq:discrete POVM main text}
\end{align}
For any normalized incoming state $|\psi\rangle$, the probability of obtaining the measurement outcome $m$ is given by the Born-rule probability $p_m = w_m\langle\psi|\hat{K}^\dagger_m \hat{K}_m |\psi\rangle$. The POVM condition guarantees the normalization of the Born-rule probabilities $\sum_{ m} p_{ m} = 1$. Each outcome $m$ corresponds to a quantum trajectory where the wavefunction collapses according to $|\psi\rangle \rightarrow \frac{\hat{K}_m|\psi\rangle}{\norm{\hat{K}_m|\psi\rangle}}$. In principle, one can absorb the weights into the Kraus operators by redefining $\hat{K}_m \rightarrow \sqrt{w_m} \hat{K}_m$. Keeping the weight explicit is convenient for generalizing the case where the measurement outcomes form a continuous set. In this case, $w_m$ can be viewed as the measure for integration on this continuous set of outcomes $m$. 

For a Gaussian measurement, we require each Kraus operator $\hat{K}_m$ to be a Gaussian operator. More precisely,  $\hat{K}_m$ takes the form $e^{\sum {\mc{K}}^m_{ij} \hat\psi^\dag_i \hat\psi_j}$ with a general complex matrix $\mc{K}^m$ which depends on the measurement outcome. A simple example of a Gaussian measurement is the weak measurement of particle number $\hat\psi^\dag\hat\psi$ on a single fermion mode. The corresponding Kraus operator ensemble is $\{ e^{m \kappa (\hat\psi^\dag\hat\psi-1/2)}\}_{m=\pm}$ with $\kappa \in \mathbb{R}$ the measurement strength and weights $w_\pm = \frac{1}{2\cosh \kappa}$. This measurement recovers the standard projective measurement of $\hat\psi^\dag\hat\psi$ when the measurement strength diverges $\kappa\rightarrow+\infty$. The measurement outcomes $m=\pm$ correspond to finding $\hat\psi^\dag\hat\psi=$  1 or 0. One can readily see that a fermionic Gaussian state always evolves (or collapses) into a Gaussian state in each quantum trajectory of a Gaussian measurement. 

In a discrete-time formulation, Gaussian dynamics can be viewed as a quantum circuit comprising Gaussian unitary gates and Gaussian measurements. In every quantum trajectory, which is labeled by the outcomes ${\bf m}= (m_1, m_2,\dots)$ of all measurements in the dynamics, the evolution of the quantum state is given by a Gaussian (or free-fermion) many-body evolution operator (mEO):
\begin{align}
    \hat{V}_{\bf m} = \exp\left(-\sum_{i,j} M_{ij}^{\bf m} \hat{\psi}_i^\dagger \hat{\psi}_j\right), \label{eq:mEO def V}
\end{align}
where $M^{\bf m} $ is a trajectory-dependent complex matrix. Microscopically, $\hat{V}_{\bf m}$ is the time-ordered product of all the Gaussian unitaries and the Kraus operators associated with a quantum trajectory labeled by measurement outcomes $\mbf{m}$. The evolution of the system's quantum state within this trajectory is given by $|\psi\rangle \rightarrow \frac{\hat{V}_{\bf m}|\psi\rangle}{\norm{\hat{V}_{\bf m}|\psi\rangle}}$. Generically, $\hat{V}_{\bf m}$ is nonunitary when measurements are present. Generic Gaussian quantum dynamics (with both unitary gates and measurements) can thus be faithfully characterized by an ensemble of mEOs ${\cal M}_{\rm mEO} = \{\hat{V}_{\bf m}\}_{\bf m}$ and their corresponding weights. It is important to note that Gaussian quantum dynamics with measurements is distinct from generic nonunitary Gaussian dynamics due to the POVM condition for measurements when there is no post-selection. The POVM condition can be violated when we post-select a subset of measurement outcomes. 

When the mEO ensemble $\{\hat{V}_{\bf m}\}_{\bf m}$ includes operators beyond the Gaussian form Eq.~\eqref{eq:mEO def V}, we refer to the corresponding dynamics as interacting dynamics. For interacting dynamics with post-selection-free measurements, the POVM condition is still enforced on the interacting mEO ensemble. This paper focuses on the discrete-time formulation of the Gaussian quantum dynamics with measurements. For curious readers, the continuous-time counterpart can be studied using the framework of quantum state diffusion~\cite{devega2017dynamics,gisin1992quantumstate,diosi1998nonmarkovian}. The universal behavior of the Gaussian quantum dynamics should not depend on the choice of continuous-time or discrete-time formulations.

So far, we have not addressed the spatial structure of Gaussian dynamics. Later, when we consider the Gaussian dynamics of a quantum system in $d$ spatial dimensions, we will focus on the dynamics microscopically generated by local unitary gates and local measurements (with spatially local Kraus operators). 

\begin{table*}[t]
    \begin{ruledtabular}
    \caption{
    Classification of fermionic dynamics with measurements based on the discrete symmetries of many-body evolution operators (mEOs). 
    Each mEO class is labeled according to the Altland–Zirnbauer (AZ) classification scheme. For each mEO class $\mc{K}_{\mr{mEO}}$, we list the discrete symmetries, time-reversal symmetry (TRS), particle-hole symmetry (PHS), chiral symmetry (CS), the corresponding single-particle transfer matrix (sTM) class $\mc{K}_{\mr{sTM}}$ in the free-fermion limit, the sTM group $\mc{G}$, and whether the mEO class admits post-selection-free measurements.     
    Here, $\mc{K}_{\mr{mEO}}$ denotes the AZ or Cartan label associated with each class. See Ref.~\cite{ludwig2013lyapunov} for definitions of each sTM group.
    }
        \label{tab: mEO_sTM_symmetry_table}
        \begin{tabular}{cccc|cc|c}
            \textbf{mEO Class} $\mc{K}_\mr{mEO}$
            & \textbf{TRS}
            & \textbf{PHS}
            & ~~\textbf{CS}~~
            & \textbf{sTM Class} $\mc{K}_{\mr{sTM}}$
            & ~~$\mc{G}$~~
            & \makecell{ \textbf{Post-Selection-Free} \\  \textbf{Gaussian Measurements}}
            \\
            \hline
            A    & 0   & 0   & 0 & AIII & $\mathrm{GL}(N,\mathbb{C})$       & \cmark \\
            AIII & 0   & 0   & 1 & A    & $\mathrm{U}(N, M)$                 & \xmark \\
            \hline
            AI   & $+1$  & 0   & 0 & BDI  & $\mathrm{GL}(N, \mathbb{R})$       & \cmark \\
            BDI  & $+1$  & $+1$  & 1 & D    & $\mathrm{O}(N, M)$                 & \cmark \\
            D    & 0   & $+1$  & 0 & DIII & $\mathrm{O}(N, \mathbb{C})$        & \cmark \\
            DIII & $-1 $ & $+1$  & 1 & AII  & $\mathrm{SO}^*(2N)$                & \xmark \\
            AII  & $-1 $ & 0   & 0 & CII  & $\mathrm{U}^*(2N)$                 & \xmark \\
            CII  & $-1 $ & $-1 $ & 1 & C    & $\mathrm{Sp}(2N, 2M)$              & \xmark \\
            C    & 0   & $-1 $ & 0 & CI   & $\mathrm{Sp}(2N, \mathbb{C})$      & \xmark \\
            CI   & $+1$  & $-1 $ & 1 & AI   & $\mathrm{Sp}(2N, \mathbb{R})$      & \xmark \\
        \end{tabular}
    \end{ruledtabular}
\end{table*}

\subsection{Single-Particle Transfer Matrix and its Tenfold Symmetry Classification}
\label{sec:stm}

In this subsection, we briefly review the description of fermionic Gaussian quantum dynamics with measurements using the single-particle transfer matrix. The sTM description plays a crucial role in the unification of the Gaussian quantum dynamics of free-fermion systems in $d+1$ spacetime dimensions and the classical Anderson localization problem of disordered fermions in equilibrium in $(d+1)$ spatial dimensions under the AZ tenfold symmetry classification~\cite{altland1997nonstandard,jian2022criticality}. As explained later, this unification is an important stepping stone toward unveiling the nontrivial topology in fermionic quantum dynamics. 

As discussed above, the Gaussian dynamics with measurements can be characterized by an ensemble of Gaussian (or free-fermion) mEOs. A charge-conserving mEO of the form $\hat{V} =  \exp\left(-\sum_{i,j} M_{ij} \hat{\psi}_i^\dagger \hat{\psi}_j\right) $ with a complex matrix $M$ can be faithfully captured by its sTM $\mathfrak{t}$. The sTM $\mathfrak{t}$ is defined by how $\hat{V}$ evolves a single-fermion operator:
\begin{align}
    \hat{V} \hat{\psi}_i \hat{V}^{-1} = \sum_j \mathfrak{t}_{ij} \hat{\psi}_j, \quad \mathfrak{t} = e^{M}. 
    \label{eq:sTM_def1 main text}
\end{align}
Generalization of the sTM to cases without charge conservation is straightforward (see Ref.~\cite{jian2022criticality} for examples). Note that when $\hat{V}$ is not invertible, for example, in the presence of projective measurements, we should think of $\hat{V}$ as the limit of mEOs where the projective measurement is replaced by weak measurements of increasing strengths that asymptote toward the projective limit. Using Eq.~\eqref{eq:sTM_def1 main text}, one can describe any free-fermion mEO ensemble ${\cal M}_{\rm mEO}$ using the corresponding ensemble ${\cal M}_{\rm sTM}$ of sTMs. 

To understand the universal behavior of Gaussian dynamics in the limit of large spacetime volume, we should treat the sTM ensemble as an ensemble of elements of a certain group ${\cal G}$, an intrinsic characteristic of the dynamics. The group structure of the sTM emerges due to the simple fact that the product of the sTMs $\mathfrak{t}_{1}$ and $\mathfrak{t}_{2}$ for two consecutive periods of time should produce the sTM $\mathfrak{t} = \mathfrak{t}_2 \mathfrak{t}_1$ for the combined evolution. Ref.~\cite{jian2022criticality} showed that the sTM group ${\cal G}$ of a generic Gaussian dynamics with measurements, due to the randomness of measurement outcomes, should be one of the ten classical groups~\cite{weyl1946classical} listed in the 6th column of Table~\ref{tab: mEO_sTM_symmetry_table}. Each of these groups is associated with an AZ symmetry class~\cite{schnyder2008classification,ludwig2013lyapunov,jian2022criticality}, which we now refer to as the sTM class ${\cal K}_{\rm sTM}$ in this paper. The definition of each group can be found in Ref.~\cite{ludwig2013lyapunov}.

Now, we provide some heuristics on how the sTM group ${\cal G}$ and the sTM class $\mc{K}_{\mr{sTM}}$ emerge. Microscopically, it was shown that the sTMs of Gaussian dynamics with measurements in a fermionic system of $d+1$ \textit{spacetime} dimensions can be mapped to the transfer matrices that describe the elastic scattering amplitude of free fermions across a static disordered system in $(d+1)$ \textit{spatial} dimensions~\cite{jian2022criticality}. The latter setting is the classical Anderson localization problem of disordered non-interacting fermions in equilibrium. The measurement outcomes in the Gaussian dynamics correspond to the disorder pattern in the corresponding Anderson localization problem. Therefore, there is a general correspondence between Gaussian dynamics with measurements in $(d+1)$-dimensional spacetime and the Anderson localization problem in $(d+1)$-dimensional space.

For the Anderson localization problem, it is known that the transfer matrices that compute the elastic-scattering amplitudes (with generic disorder) should be treated as the elements of one of the ten classical groups. The time-reversal symmetry (TRS), the particle-hole symmetry (PHS), and chiral symmetry (CS) of the Hamiltonian underlying the elastic disorder scattering dictate which of the ten classical groups emerges~\cite{caselle2006symmetric,schnyder2008classification,schnyder2009classification,ludwig2013lyapunov}. Depending on whether TRS, PHS, and CS are present, and whether their single-particle actions square to $\pm1$ if present, there are exactly ten AZ symmetry classes, with each class leading to one of the ten classical groups. The symmetry class (and its corresponding transfer matrix group) controls the universal behavior of the disordered free fermions at long wavelengths. 

Based on the correspondence between Gaussian dynamics and the Anderson localization problems, the AZ symmetry classification of the latter applies to the former systems as well~\cite{jian2022criticality}. This classification is crucial in understanding the phase diagram of Gaussian dynamics with measurements~\cite{jian2022criticality,fava2023nonlinear,jian2023measurementinduced,poboiko2023theory,chahine2024entanglement,fava2024monitored,poboiko2024measurementinduced,pan2025topological,xiao2024topology}. Because this symmetry classification of Gaussian dynamics with measurements is based on sTMs, we refer to this symmetry class as the sTM class ${\cal K}_{\rm sTM}$ of the Gaussian dynamics. This terminology is to be contrasted with a related but different symmetry classification scheme, namely the mEO class ${\cal K}_{\rm mEO}$, which is generalizable to interacting dynamics (see Sec.~\ref{sec:mEO class intro}). This tenfold sTM classification can alternatively be obtained from the perspective of nonunitary evolution generated by non-Hermitian Hamiltonians. Combining the symmetry classification of generic non-Hermitian Hamiltonians and the group structure of the sTM inherited from time evolution, one can also obtain the same tenfold symmetry classification of sTM ensembles~\cite{bernard2002classification,kawabata2019symmetry,xiao2024topology}. 

Note that, despite the aforementioned correspondence, Gaussian dynamics with measurements and the classical Anderson localization problems are \textit{not} the same problem. In the former, the probability distribution of the random measurement outcomes is controlled by the Born rule (which depends on the system's quantum state). In contrast, the probability distribution of static disorder in the Anderson localization problem is predetermined by the model or the material of interest. In replica-trick-based analytical treatments, this difference in the probability distributions amounts to different replica limits~\cite{jian2023measurementinduced,poboiko2023theory,fava2023nonlinear}. Refs.~\cite{jian2023measurementinduced,poboiko2023theory} have also provided examples where this difference can lead to properties of critical phases and transitions in the Gaussian dynamics that differ from their counterparts in the Anderson localization problem. However, it is possible that the correspondence between area-law-entangled phases in the Gaussian dynamics and their Anderson-localized counterparts is more robust. We will elaborate on this point in Sec.~\ref{Sec:Bott_Clock}.

Moreover, we want to address the relation between the POVM measurement conditions and the symmetry classification. As discussed above, Ref.~\cite{jian2022criticality} introduced the symmetry classification of Gaussian dynamics with measurements based on the sTM ensemble ${\cal M}_{\rm sTM}$ associated with the ensemble of quantum trajectories in the dynamics. This previous treatment did not focus on possible relations between the symmetry classes and the POVM conditions, which can have nontrivial implications on the possible microscopic realizations of the Gaussian dynamics in different symmetry classes. One of the intriguing new results we obtained in this work is the nontrivial constraint between Gaussianity, symmetry class, and the POVM conditions. In fact, mEO ensembles in certain symmetry classes cannot satisfy the POVM condition unless either interactions are included (i.e., non-Gaussian measurements) or the ensemble consists of purely unitary Gaussian operators (up to multiplicative constants).

These results are summarized in the last column of Table~\ref{tab: mEO_sTM_symmetry_table} and are extensively discussed in Sec.~\ref{Sec: POVMadmissibility}. Here, we briefly highlight the following points. In the four sTM classes ${\cal K}_{\rm sTM} = \text{AIII, BDI, D, and DIII}$, the POVM condition can be satisfied by Gaussian mEO ensembles in the presence of measurements. Physically, this means that there exist Gaussian measurements compatible with the relevant symmetry constraints \textit{without} post-selecting quantum trajectories. Many previously studied models, including those in Refs.~\cite{alberton2021entanglement,nahum2020entanglement,turkeshi2021measurementinduced,kells2023topological,jian2022criticality,jian2023measurementinduced,merritt2023entanglementa,klocke2023majorana,poboiko2023theory,chahine2024entanglement,fava2023nonlinear,fava2024monitored,poboiko2024measurementinduced}, already provide examples of fermionic Gaussian dynamics with post-selection-free measurements in these four sTM classes (even though the symmetry aspects were not always explicitly emphasized). 

In contrast, for the remaining six sTM classes ${\cal K}_{\rm sTM} = \text{A, AII, CII, C, CI, and AI}$, the POVM condition \textit{cannot} be satisfied by any Gaussian mEO ensemble in the presence of measurements (see Sec.~\ref{Sec: POVMadmissibility} for more details). Note that an mEO ensemble with each mEO proportional to a unitary operator is considered not to contain any measurements, as it physically describes random unitary evolutions. This result implies that the realization of Gaussian measurements in these six sTM classes requires post-selection. 

However, one can introduce interactions to eliminate constraints between symmetry classes and the POVM condition. In other words, post-selection can be avoided in non-Gaussian measurements (see Sec.~\ref{Sec: POVMadmissibility}). 

Despite the significance of the sTM-based symmetry classification in understanding the universal behavior and phase diagrams of dynamical free-fermion systems with measurements, it has suboptimal features. First, the symmetries associated with the sTM class $\AZK{sTM}$ should not be viewed as symmetries acting on the Hilbert space of the dynamical fermion system in the conventional sense. Instead, they are the explicit symmetries of the corresponding Anderson localization problem and, hence, determine the sTM group ${\cal G}$. Second, this sTM-based symmetry classification is limited to free-fermion systems, as there is no faithful single-particle description of interacting dynamics. In Sec.~\ref{sec:mEO class intro}, we introduce a new symmetry classification scheme—the mEO class—that overcomes these two shortcomings of the sTM-based approach.

\subsection{Symmetry Classification of mEOs and its Correspondence with the sTM-based Classification}
\label{sec:mEO class intro}

We have reviewed the sTM-based symmetry classification ${\cal K}_{\rm sTM}$ of Gaussian dynamics with measurements and its limitations. In the following, we introduce a new symmetry classification scheme, the mEO class ${\cal K}_{\rm mEO}$, directly based on the mEOs of the dynamical systems. This new scheme makes the roles of the symmetries, including TRS, PHS, and CS, more explicit and accommodates interacting dynamical systems. In the free-fermion limit, there is a one-to-one correspondence between the mEO class ${\cal K}_{\rm mEO}$ and the sTM class ${\cal K}_{\rm sTM}$. This new symmetry classification scheme also lays the foundation for the discussion of the  
topologies of fermionic dynamics in Sec. \ref{Sec:Bott_Clock}.

To analyze the AZ symmetry classification of fermion dynamics at the many-body level, we directly examine TRS, PHS, and CS of the mEO ensemble ${\cal M}_{\rm mEO}$ that describes the quantum trajectories. We denote the many-body TRS action as $\hat{\mc{T}}$, the PHS action as $\hat{\mc{C}}$, and the CS action as $\hat{\mc{S}}$. The action $\hat{\mathcal{C}}$ is {\it unitary}, while both $\hat{\mc{T}}$ and $\hat{\mc{S}}$ are {\it anti-unitary} many-body operators. The actions of TRS, PHS, and CS can be generally written as 
\begin{align}
    &\text{TRS:} && \hat{\mathcal{T}} \hat{\psi}_i \hat{\mathcal{T}}^{-1} = \sum_{j}(\mathcal{U}_T)_{ij} \hat{\psi}_j, && \hat{\mathcal{T}} \ii \hat{\mathcal{T}}^{-1} = -\ii, \label{eq:trs_transform}  \\ 
    &\text{PHS:} && \hat{\mathcal{C}} \hat{\psi}_i \hat{\mathcal{C}}^{-1} =  \sum_{j}(\mathcal{U}_C)_{ij} \hat{\psi}_j^\dagger, && \hat{\mathcal{C}} \ii \hat{\mathcal{C}}^{-1} = +\ii, \label{eq:phs_transform} \\ 
    &\text{CS:}  && \hat{\mathcal{S}} \hat{\psi}_i \hat{\mathcal{S}}^{-1} =  \sum_{j}(\mathcal{U}_S)_{ij} \hat{\psi}_j^\dagger, && \hat{\mathcal{S}} \ii \hat{\mathcal{S}}^{-1} = -\ii, \label{eq:cs_transform}
\end{align}
where the first-quantized actions of TRS, PHS, and CS are denoted by the unitary matrices $\mc{U}_{T,C,S}$, respectively. Furthermore, there are two types of TRS and PHS corresponding to
\begin{align}
    \hat{\cT}^2 = (\pm 1)^{\hat N_F}, \quad \text{and} \quad \hat{\cC}^2 = (\pm 1)^{\hat N_F},
    \label{eq:TwoTypesTP main text}
\end{align}
where $(-1)^{\hat N_F}$ represents the (many-body) fermion parity operator. These two types of TRS and PHS in Eq.~\eqref{eq:TwoTypesTP main text} correspond to the following constraints on their first-quantized actions:
\begin{align}
    \cU_T \cU_T^* = \pm \Id, \quad \text{and} \quad \cU_C \cU_C^* = \pm \Id.
\end{align}  
CS is the product of TRS and PHS. When both TRS and PHS are present, CS is also preserved. When both are absent, CS might still be a symmetry of the system. Depending on which types of TRS, PHS, and CS are present, there are in total ten possible symmetry patterns, known as the ten AZ symmetry classes, which are summarized in the first four columns of Table \ref{tab: mEO_sTM_symmetry_table}. In conventional notation, $0$ represents the absence of the symmetry and $\pm 1 $ refers to the two types of TRS and PHS.

Next, we establish the criteria to determine if TRS, PHS, and CS are present in the dynamics with measurements. Let ${\cal M}_{\rm mEO}$ denote the mEO ensemble associated with the dynamics. If TRS, PHS, or CS is present in the dynamics, we require every mEO $\hat{V} \in {\cal M}_{\rm mEO}$ to commute with the corresponding symmetry action. Physically, this means that a symmetry, if present, should be preserved in every quantum trajectory.

Take TRS as an example. For the dynamics to respect TRS, each mEO $\hat{V} \in \mathcal{M}_{\rm mEO}$ must satisfy
\begin{align}
    \hcT \hat{V} \hcT^{-1} = \hat{V}.
    \label{eq:TRSofV1}
\end{align}
This condition ensures closure under mEO multiplication: if $\hat{V}$ and $\hat{V}'$ both satisfy TRS, then so does their product $\hat{V}\hat{V}'$. Consequently, TRS remains preserved under coarse-graining of time scales and governs the universal behavior in the limit of large spacetime volume. Moreover, if the dynamics preserves TRS, any TRS-invariant state $|\psi\rangle = \hcT |\psi\rangle$ evolves into a TRS-invariant state in every quantum trajectory, i.e., $\hat{V}|\psi\rangle = \hcT \hat{V}|\psi\rangle$.

Importantly, this notion of TRS does not involve reversing the direction of time. Instead, the anti-unitary nature of $\hcT$ acts as a ``realness'' constraint on each mEO. The criteria for particle-hole symmetry (PHS) and chiral symmetry (CS) naturally generalize Eq.~\eqref{eq:TRSofV1}, and are likewise stable under mEO multiplication. We caution that one might view the mEO as an exponentiated non-Hermitian operator and apply the 38-fold non-Hermitian symmetry classification. However, those non-Hermitian symmetry classes incompatible with mEO multiplication are irrelevant to the behavior of the dynamics after coarse-graining. Our formulation of TRS, PHS, and CS is robust under coarse-graining. A similar single-particle consideration appears in Ref.~\cite{xiao2024topology}.

With the general definition above, we are now ready to define the symmetry classification for the free-fermion nonunitary dynamics, based on whether TRS, PHS, and CS are present in ${\cal M}_{\rm mEO}$, using the AZ tenfold symmetry classification (see the first four columns of Table \ref{tab: mEO_sTM_symmetry_table}).
We call this symmetry classification the mEO class ${\cal K}_{\rm mEO}$. This definition does not rely on the specific form of the mEOs and therefore applies to both Gaussian and interacting fermionic dynamics with measurements. Later, we will argue that in area-law-entangled phases of fermionic dynamics, the information of the initial state is quickly washed out by measurements. Consequently, the symmetries in the mEO class ${\cal K}_{\rm mEO}$ will also emerge as symmetries of the system’s area-law-entangled steady-state ensemble in the long-time limit.

For Gaussian dynamics with measurements, one can define both the sTM class ${\cal K}_{\rm sTM}$ and the mEO class ${\cal K}_{\rm mEO}$. However, they are not necessarily equal for a given system: $\mc{K}_{\mr{sTM}} \neq \mc{K}_{\mr{mEO}}$. Instead, as shown below, there is a one-to-one correspondence between them. In this sense, the two symmetry classifications are equivalent in the free-fermion limit.

To establish this one-to-one correspondence between ${\cal K}_{\rm sTM}$ and ${\cal K}_{\rm mEO}$, we examine the TRS constraint Eq.~\eqref{eq:TRSofV1} and its PHS and CS counterparts on a Gaussian mEO 
$\hat{V} = \exp\left(-\sum_{i,j} M_{ij} \hat{\psi}_i^\dagger \hat{\psi}_j\right)$: 
\begin{align}
    & \text{TRS: } && \h{\mc{T}}\hat{V}\h{\mc{T}}^{-1} = \hat{V} \Longleftrightarrow M = \mc{U}_T^\dagger M^* \mc{U}_T, \label{eq:trs_constraint} \\
    & \text{PHS: } && \h{\mc{C}}\hat{V}\h{\mc{C}}^{-1} = \hat{V} \Longleftrightarrow M = -(\mc{U}_C^\dagger M \mc{U}_C)^\intercal, \label{eq:phs_constraint} \\
    & \text{CS: } && \h{\mc{S}}\hat{V}\h{\mc{S}}^{-1} = \hat{V} \Longleftrightarrow M = -(\mc{U}_S^\dagger M^* \mc{U}_S)^\intercal. \label{eq:cs_constraint}
\end{align}
Recall that $M$ is the generator of the sTM, i.e., $\mathfrak{t} = e^M$. One can show that the constraints on $M$ in each mEO class ${\cal K}_{\rm mEO}$ match exactly the requirements for $\mathfrak{t} = e^M$ to be an element of one of the ten sTM groups ${\cal G}$. Since the sTM class ${\cal K}_{\rm sTM}$ is defined according to the sTM group ${\cal G}$ (see Sec.~\ref{sec:stm}), this implies a one-to-one correspondence between the mEO classes ${\cal K}_{\rm mEO}$ and the sTM classes ${\cal K}_{\rm sTM}$, listed row-by-row in Table~\ref{tab: mEO_sTM_symmetry_table}. A full proof is given in App.~\ref{app: sTM-mEO_relation}.

Interestingly, this correspondence between the mEO and sTM classes manifests as a permutation: within the eight real symmetry classes (AI, BDI, D, DIII, AII, CII, C, and CI), the sTM and mEO classes are related by a cyclic permutation. Likewise, the two complex classes (A and AIII) are permuted under this correspondence. A similar permutation of symmetry classes also appears in the context of Anderson localization problems~\cite{ludwig2013lyapunov,schnyder2008classification,ludwig2016topological}.

Another remark is that, in light of the one-to-one correspondence between the mEO and sTM classes in Gaussian dynamics, the former can be viewed as a way to extend the sTM classes to interacting dynamics. Strictly speaking, the sTM class is not well-defined in interacting dynamics due to the lack of a faithful single-particle description. However, recent works~\cite{guo2025field,poboiko2025measurementinduced} show that for interacting fermionic dynamics continuously connected to non-interacting limits, the sTM class of the non-interacting system still strongly influences the universal behavior when interactions are turned on. With the mEO classification developed here, we argue that the universal behavior of the interacting dynamics studied in Refs.~\cite{guo2025field,poboiko2025measurementinduced} should be understood as governed by the mEO class of the system, which maps to the corresponding sTM class in the non-interacting limit. More generally, we expect the mEO-based symmetry classification to serve as a general tool for predicting universal behavior in interacting fermionic dynamics in the large-spacetime-volume limit, which we will pursue in the future.

\subsection{Topological Dynamics, Steady-State Ensembles, and Dynamical Bulk-Boundary Correspondence}
\label{Sec:Bott_Clock}
In this subsection, we focus on the classification of area-law-entangled topological Gaussian dynamics in the ten AZ symmetry classes. We will show that the topology of the dynamics in the spacetime bulk can be manifested by the topology of the steady-state ensemble effectively living on the temporal boundary of the spacetime. This relationship between the two types of topologies can be viewed as a {\it dynamical bulk-boundary correspondence}.

First, we provide a general argument for the topological classification of the Gaussian dynamics with measurements. In Sec.~\ref{sec:stm}, we reviewed the unification of the Gaussian dynamics with measurements in $d+1$ spacetime dimensions and the Anderson localization problem of disordered fermions in $(d+1)$ spatial dimensions. This unification further suggests a correspondence between the phases of Gaussian dynamics with area-law entanglement and disordered localized phases in the Anderson localization problem in the same sTM class ${\cal K}_{\rm sTM}$~\cite{pan2025topological, xiao2024topology}. Within each symmetry class, the latter can be further subdivided into insulators or superconductors with different topologies (see reviews Refs.~\cite{qi2011topological, hasan2010colloquium, ludwig2016topological}). These topological insulators and superconductors are expected to have corresponding area-law entangled Gaussian dynamics with the same topological classification.
This reasoning leads to a topological classification of area-law-entangled Gaussian dynamics summarized in Table~\ref{tab: bott periodicity table}, with the first column interpreted as the sTM class ${\cal K}_{\rm sTM}$ and the dimension $D$ referring to the spacetime dimensions of the Gaussian dynamics (as well as the spatial dimension of the corresponding disordered Anderson-localized phase). 

This classification provides a general principle for exploring topological Gaussian dynamics with measurements. However, there has not been a general recipe to construct microscopic models for these topological phases of Gaussian dynamics. A particularly interesting question is how to realize a topological phase of Gaussian dynamics {\it without post-selecting the quantum trajectories}. Avoiding post-selection is crucial for efficiently simulating the dynamics of interest on programmable quantum platforms. Refs.~\cite{pan2025topological, xiao2024topology, BeriSurfaceCode2024, nahum2020entanglement, kells2023topological} have identified examples of $\mathbb{Z}_2$ and $\mathbb{Z}$-classified topological Gaussian dynamics in 1+1d spacetime realized in the form of post-selection-free monitored Gaussian circuits in sTM class ${\cal K}_{\rm sTM} = \text{DIII and D}$. However, sTM-class-A topological Gaussian dynamics in 1+1d has only been found using post-selection~\cite{pan2025topological}. Concrete post-selection-free realizations of topological Gaussian dynamics in higher dimensions remain unexplored. 

An important result of this paper is a general construction of Gaussian circuits with post-selection-free measurements that realize topological Gaussian dynamics in general dimensions in sTM classes ${\cal K}_{\rm sTM} = \text{AIII, BDI, D, and DIII}$. For the remaining six classes ${\cal K}_{\rm sTM} = \text{A, AII, CII, C, CI, AI}$, we show that the topological dynamics classified in Table~\ref{tab: bott periodicity table} requires post-selection in the free-fermion limit.

To better understand the nature of the topological dynamical phases classified in Table~\ref{tab: bott periodicity table}, and to motivate our construction of the corresponding Gaussian circuit, it is informative to examine the symmetries and topologies of the steady-state ensemble produced by the mEOs. For this purpose, using the mEO symmetry class ${\cal K}_{\rm mEO}$ to characterize the dynamics is helpful. By definition, if the initial state $|\psi_0\rangle$ respects the symmetries in ${\cal K}_{\rm mEO}$, the evolved state $\hat V_{\bf m}|\psi_0\rangle$ must respect those same symmetries in every quantum trajectory. More interestingly, in the area-law-entangled topological phase, we argue that symmetry preservation should asymptotically hold for every long-time trajectory \emph{regardless} of the initial state. In other words, the steady-state ensemble should consist of states respecting the symmetries in ${\cal K}_{\rm mEO}$. This expectation is motivated by prior works showing that, in the area-law entangled phases, the frequent measurement-induced wavefunction collapse can efficiently erase information about the initial state~\cite{gullans2020dynamical,choi2020quantum,sang2021measurementprotected,agrawal2022entanglement}. As a result, in the long-time limit, the states in the steady-state ensemble inherit the global symmetries of the dynamics regardless of the initial state (unless those symmetries are spontaneously broken).

Next, we argue that the steady-state ensemble of the area-law-entangled topological phase of Gaussian dynamics is an ensemble of short-range-correlated Gaussian states. The single-particle spacetime propagation in the topological Gaussian dynamics maps to the spatial propagation of a particle in the corresponding Anderson-localized topological phase (see Sec.~\ref{sec:stm} for review). Therefore, we expect that the steady-state ensemble of the area-law-entangled topological Gaussian dynamics is an ensemble of short-range-correlated Gaussian states, each satisfying the symmetries in the ${\cal K}_{\rm mEO}$ class of the dynamics. This steady-state ensemble resembles the localized ground states of disordered free-fermion Hamiltonians in the symmetry class ${\cal K}_{\rm mEO}$. Therefore, the topology of the steady-state ensemble should also follow the topological classification shown in Table~\ref{tab: bott periodicity table}, where the AZ class should now be interpreted as the mEO class ${\cal K}_{\rm mEO}$ and $D$ as the spatial dimension. Here, we assume that for dynamics that respect locality, the area-law-entangled states in the same steady-state ensemble should share the same topological invariant.

\begin{figure}[ht]
    \centering
    \includegraphics[width=0.8\linewidth]{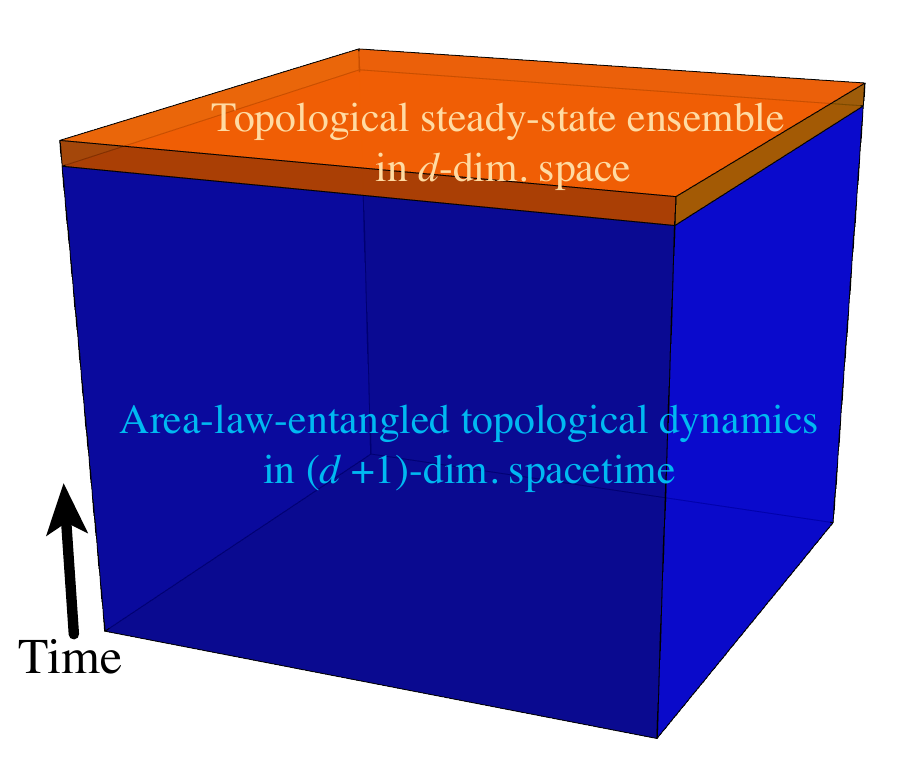}
    \caption{Schematic illustration of dynamical bulk-boundary correspondence between the area-law-entangled topological dynamics in $(d+1)$-dimensional spacetime and the topolgical steady-state ensemble in the $d$-dimensional space on the spacetime boundary.}
    \label{fig: meo class stm class dim reduction}
\end{figure}

Now, we have two different ways to interpret the classification in Table~\ref{tab: bott periodicity table}, depending on whether the AZ class refers to ${\cal K}_{\rm sTM}$ or ${\cal K}_{\rm mEO}$ and whether $D$ refers to the spacetime or spatial dimensions of the dynamical system. We argue that the two interpretations reflect the following {\it dynamical bulk-boundary correspondence}. For the topological classification of dynamics in the spacetime bulk, we use the sTM class ${\cal K}_{\rm sTM}$ and spacetime dimension. When we switch to the equivalent mEO class (see Sec.~\ref{sec:mEO class intro}) and replace the spacetime dimension by the dimension of the space located at the boundary of spacetime (see Fig.~\ref{fig: meo class stm class dim reduction}), we move diagonally in Table~\ref{tab: bott periodicity table} as indicated by the arrows. Notice that the topological invariants connected by the arrows are identical. Mathematically, this phenomenon is known as Bott periodicity. It has appeared previously in the classification of topological insulators and superconductors~\cite{kitaev2009periodic,teo2010topological,ryu2010topological}. In the context of Gaussian dynamics, this observation strongly suggests a novel dynamical bulk-boundary correspondence: 
\begin{mdframed}
{\it The topology of the area-law-entangled Gaussian dynamics in the spacetime bulk can be manifested by the topology of the steady-state ensemble effectively living on the time slice at the spacetime boundary.}
\end{mdframed}
This dynamical bulk-boundary correspondence exactly matches the examples identified in Ref.~\cite{pan2025topological,xiao2024topology}. For instance, the two area-law-entangled topological Gaussian dynamics in 1+1d spacetime in sTM class ${\cal K}_{\rm sTM}=\text{DIII}$ were found to produce the two topologically-distinct steady-state ensembles in one spatial dimension in mEO class ${\cal K}_{\rm mEO}=\text{D}$.

In light of this dynamic bulk-boundary correspondence, designing microscopic models for topological Gaussian dynamics with measurements is equivalent to constructing dynamical systems that generate a topological steady-state ensemble. Guided by this observation, we obtain a systematic construction of Gaussian circuits with post-selection-free measurements that represent the topological Gaussian dynamics in mEO classes ${\cal K}_{\rm mEO}=$ A, AI, BDI, and D (equivalent to sTM classes ${\cal K}_{\rm sTM}=$ AIII, BDI, D, and DIII). This construction is presented in Sec.~\ref{sec:AdaptiveSteeringDynamics}. 

For the remaining six symmetry classes ${\cal K}_{\rm mEO}=$ AIII, DIII, AII, CII, C, and CI, we show in App.~\ref{app: Gaussian POVM admissibility} that Gaussian dynamics in these symmetry classes must involve post-selection if measurements are included. Since fully unitary dynamics cannot realize any area-law-entangled dynamical phase, the topological dynamics in these remaining six symmetry classes require post-selection when realized in the free-fermion limit, consistent with the sTM-class-A example found in Ref.~\cite{pan2025topological}. However, symmetry-compatible interacting measurement operations can relax the need for post-selection, leading to genuine symmetry-constrained fermionic dynamics with measurements. The next subsection will provide an example for the mEO class AII. The possibility of realizing topological dynamics in these six symmetry classes without post-selection in the interacting limit is left for future study.

\begin{table}[t]
\begin{ruledtabular}
\caption{Periodic table of topological insulators and superconductors~\cite{ludwig2016topological,chiu2016classification}, organized by the Bott clock~\cite{kitaev2009periodic, teo2010topological, ryu2010topological}. The arrows illustrate the ``dimensional reduction" from the Gaussian circuits in sTM class $\mc{K}_{\mathrm{sTM}}$ in $D=(d{+}1)$-dimensional spacetime to their steady-state ensembles in mEO class $\mc{K}_{\mathrm{mEO}}$ in $D=d$ spatial dimensions. The map between AZ labels $\mc{K}_{\mathrm{sTM}}\rightarrow \mc{K}_{\mathrm{mEO}}$ obeys a shift in the Bott clock. The topological invariants connected by the arrow are identical, reflecting a dynamical bulk-boundary correspondence: the topology in the spacetime of the Gaussian dynamics with measurements determines the topology of the steady-state ensemble at the spacetime boundary. For example, a $1{+}1$d circuit in sTM class DIII yields a 1d steady state in mEO class D, with both exhibiting a $\mathbb{Z}_2$ classification.}
\label{tab: bott periodicity table}
    \begin{tabular}{ccccc }
        \textbf{AZ Class } & \textbf{$D=1$} & \textbf{$D=2$} & \textbf{$D=3$} &  \textbf{$D=4$}\\
        \hline
        A    & 0 & \tikzmark{A_d2}\textcolor{red}{$\mathbb{Z}$} & \tikzmark{A_d3} 0 &  \tikzmark{A_d4} \textcolor{red}{$\mathbb{Z}$} \\
        AIII & \tikzmark{AIII_d1}\textcolor{red}{$\mathbb{Z}$} & 0 & \tikzmark{AIII_d3}\textcolor{red}{$\mathbb{Z}$} & 0\\
        \hline
        AI   & 0 & 0 & 0 & \tikzmark{AI_d4}\textcolor{brown}{$\mathbb{Z}$}\\
        BDI  & \tikzmark{BDI_d1}\textcolor{blue}{$\mathbb{Z}$} & 0 & 0 & 0\\
        D    & \tikzmark{D_d1}\textcolor{green}{$\mathbb{Z}_2$} & \tikzmark{D_d2}\textcolor{blue}{$\mathbb{Z}$} & 0 & 0\\
        DIII & \tikzmark{DIII_d1}\textcolor{magenta}{$\mathbb{Z}_2$} & \tikzmark{DIII_d2}\textcolor{green}{$\mathbb{Z}_2$} & \tikzmark{DIII_d3}\textcolor{blue}{$\mathbb{Z}$} & 0\\
        AII  & 0 & \tikzmark{AII_d2}\textcolor{magenta}{$\mathbb{Z}_2$} & \tikzmark{AII_d3}\textcolor{green}{$\mathbb{Z}_2$} & \tikzmark{AII_d4}\textcolor{blue}{$\mathbb{Z}$} \\
        CII  & \tikzmark{CII_d1}\textcolor{brown}{$\mathbb{Z}$} & 0 & \tikzmark{CII_d3}\textcolor{magenta}{$\mathbb{Z}_2$} &  \tikzmark{CII_d4}\textcolor{green}{$\mathbb{Z}_2$} \\
        C    & 0 & \tikzmark{C_d2}\textcolor{brown}{$\mathbb{Z}$} & 0 & \tikzmark{C_d4}\textcolor{magenta}{$\mathbb{Z}_2$}\\
        CI   & 0 & 0 & \tikzmark{CI_d3}\textcolor{brown}{$\mathbb{Z}$} & 0\\
    \end{tabular}
\end{ruledtabular}

\begin{tikzpicture}[overlay, remember picture, 
    shorten >=20pt, shorten <=10pt]
\draw[->, red, thick] ([yshift=2pt,xshift=-1.5pt]pic cs:A_d4) -- ([yshift=2pt,xshift=-1.5pt]pic cs:AIII_d3);
\draw[->, blue, thick] ([yshift=4pt,xshift=-1.5pt]pic cs:AII_d4) -- ([yshift=4pt,xshift=-1.5pt]pic cs:DIII_d3);
\draw[->, green, thick] ([yshift=4pt]pic cs:CII_d4) -- ([yshift=4pt]pic cs:AII_d3);
\draw[->, magenta, thick] ([yshift=4pt]pic cs:C_d4) -- ([yshift=4pt]pic cs:CII_d3);
\draw[->, brown, thick] ([yshift=4pt,xshift=-1.5pt]pic cs:AI_d4) -- ([yshift=-10pt,xshift=10pt]pic cs:CI_d3);

\draw[->, red, thick] ([yshift=4pt,xshift=-1.5pt]pic cs:AIII_d3) -- ([yshift=4pt,xshift=-1.5pt]pic cs:A_d2);
\draw[->, blue, thick] ([yshift=4pt,xshift=-1.5pt]pic cs:DIII_d3) -- ([yshift=4pt,xshift=-1.5pt]pic cs:D_d2);
\draw[->, green, thick] ([yshift=4pt]pic cs:AII_d3) -- ([yshift=4pt]pic cs:DIII_d2);
\draw[->, magenta, thick] ([yshift=4pt]pic cs:CII_d3) -- ([yshift=4pt]pic cs:AII_d2);
\draw[->, brown, thick] ([yshift=4pt,xshift=-1.5pt]pic cs:CI_d3) -- ([yshift=4pt,xshift=-1.5pt]pic cs:C_d2);

\draw[->, red, thick] ([yshift=2pt,xshift=-1.5pt]pic cs:A_d2) -- ([yshift=2pt,xshift=-1.5pt]pic cs:AIII_d1);
\draw[->, blue, thick] ([yshift=4pt,xshift=-1.5pt]pic cs:D_d2) -- ([yshift=4pt,xshift=-1.5pt]pic cs:BDI_d1);
\draw[->, green, thick] ([yshift=4pt]pic cs:DIII_d2) -- ([yshift=4pt]pic cs:D_d1);
\draw[->, magenta, thick] ([yshift=4pt]pic cs:AII_d2) -- ([yshift=4pt]pic cs:DIII_d1);
\draw[->, brown, thick] ([yshift=4pt,xshift=-1.5pt]pic cs:C_d2) -- ([yshift=4pt,xshift=-1.5pt]pic cs:CII_d1);
\end{tikzpicture}
\end{table}

\subsection{Gaussian POVM Admissibility in Different Symmetry Classes}
\label{Sec: POVMadmissibility}
\label{sec:POVM}

As discussed in Sec.~\ref{sec: GaussianDynamicsWithMeasurements}, without post-selecting quantum trajectories, the mEO ensemble $\mathcal{M}_{\rm mEO}=\{ \hat V_{\bf m}\}_{\bf m}$ must satisfy the POVM condition
\begin{align}
    \sum_{\bf m} w_{\bf m}\hat V^\dag_{\bf m} \hat V_{\bf m} = \hat \Id,
    \label{eq: POVM_V}
\end{align}
where $w_{\bf m} \geq 0$ is the weight for the quantum trajectory with measurement outcome ${\bf m}$. If an mEO ensemble does not satisfy this condition, its realization as dynamics with measurements must involve post-selection. We now ask: for which symmetry classes can Gaussian mEO ensembles with measurements satisfy Eq.~\eqref{eq: POVM_V}?

\begin{definition}[Gaussian POVM Admissibility]
\label{def:GPOVM}
An mEO class $\mathcal{K}_{\mathrm{mEO}}$ is said to \emph{admit a Gaussian POVM} if there exists an ensemble of Gaussian Kraus operators $\{\hat{K}_m\}_m$, invariant under the symmetries of $\mathcal{K}_{\mathrm{mEO}}$, together with non-negative weights $\{w_m \geq 0\}_m$, satisfying the POVM condition $\sum_m w_m \hat{K}_m^{\dagger} \hat{K}_m = \hat{\Id}$, such that the ensemble is \emph{not} reducible to a set of Gaussian unitary operators (up to multiplicative constants). This definition is independent of the spatial dimension and the number of fermion modes.
\end{definition}

The exclusion of unitary-only ensembles is essential: any symmetry class trivially satisfies the POVM condition if only unitary mEOs are considered, which corresponds to random unitary dynamics without any measurements. With this definition in hand, we state the main result of this subsection.

\begin{theorem}[Gaussian POVM Admissibility Constraint]
\label{thm:GPOVM}
In the free-fermion limit, exactly four of the ten mEO classes admit symmetry-invariant Gaussian POVM:
\begin{equation}
  \mathcal{K}_{\mathrm{mEO}} \in \{\mathrm{A},\; \mathrm{AI},\; \mathrm{BDI},\; \mathrm{D}\}.
\end{equation}
For the remaining six mEO classes, $\mathcal{K}_{\mathrm{mEO}} \in \{\mathrm{AIII},\; \mathrm{DIII},\; \mathrm{AII},\; \mathrm{CII},\; \mathrm{C},\; \mathrm{CI}\}$, no Gaussian mEO ensemble containing measurements can satisfy the POVM condition. The proofs of admissibility (by explicit construction) and inadmissibility (by contradiction using the Cartan decomposition of the log-singular-value spectrum) are given in Appendix~\ref{app: Gaussian POVM admissibility}.
\end{theorem}

\begin{remark}[Role of interactions]
\label{rem:interactions}
The Gaussian constraint is essential to the above Theorem. Introducing interactions can restore POVM admissibility in all six excluded classes. For instance, in mEO class AII, measuring the non-Gaussian operator $\hat{n}_{\uparrow}\hat{n}_{\downarrow}$ yields the Kraus operator ensemble $\{\hat{n}_{\uparrow}\hat{n}_{\downarrow},\; \hat{\Id} - \hat{n}_{\uparrow}\hat{n}_{\downarrow}\}$, both elements of which commute with the time-reversal symmetry $\hat{\mc{T}}$ with $\hat{\mc{T}}^2 = (-1)^{\hat{N}_F}$ and satisfy the POVM condition Eq.~\eqref{eq: POVM_V}. Therefore, the realizability constraints identified here apply strictly to the free-fermion limit and motivate the search for interacting, post-selection-free realizations of topological dynamics in the six excluded mEO classes.
\end{remark}

These results are summarized in the last column of Table~\ref{tab: mEO_sTM_symmetry_table}. In the following, we provide heuristics regarding the constraints between the mEO class $\mathcal{K}_{\mathrm{mEO}}$ and the POVM condition.

For the mEO classes A, AI, BDI, and D, one can find fermion modes that are singlets under the corresponding symmetries, so measuring their occupation numbers yields symmetry-compatible Kraus operators. For example, consider mEO class ${\cal K}_{\rm mEO}=\text{AI}$ (equivalent to sTM class ${\cal K}_{\rm sTM}=\text{BDI}$), which features TRS with $\hat{\mc{T}}^2 = +1$. One can find a basis where each fermion mode $\h \psi$ is invariant under $\hat{\mc{T}}$, i.e., $\hat{\mc{T}} \h \psi \hat{\mc{T}}^{-1} = \hat \psi$. Weakly measuring the occupation number $\hat{n}=\hat\psi^\dag \hat\psi$ leads to two Kraus operators
\begin{equation}
    \hat{K}^{\mr{AI}}_{m=\pm} = \frac{1}{\sqrt{2\cosh \kappa}}e^{m \kappa (\hat n-1/2)}
\end{equation}
associated with the two measurement outcomes $m=\pm$. Both Kraus operators commute with $\hat{\mc{T}}$. In the limit $\kappa \rightarrow \infty$, this measurement becomes the projective measurement of $\hat n$, with outcomes $m=\pm$ corresponding to $\hat n =1$ and 0.

In contrast, the symmetries in mEO classes AIII, DIII, AII, CII, C, and CI require a certain doublet structure in the fermion modes. Take the mEO class AII as an example. The TRS with $\h {\cal T}^2 = (-1)^{\h N_F}$ requires the fermion operators to form Kramers doublets. Let $\h\psi_\uparrow$ and $\h\psi_\downarrow$ denote a pair of fermion operators in a Kramers doublet. Measuring the number operators $\hat n_\uparrow$ and $\hat n_\downarrow$ independently without post-selection breaks TRS when their outcomes mismatch, because TRS requires the measurement outcome of $\hat n_\uparrow$ to be the same as that of $\hat n_\downarrow$, which can only be achieved by post-selecting the measurement outcomes. In App.~\ref{app: Gaussian POVM admissibility}, we present a systematic proof that no Gaussian mEO ensemble with measurements in the six mEO classes AIII, DIII, AII, CII, C, and CI can satisfy the POVM condition. Therefore, their realizations require measurements with post-selection. Note that this type of post-selection essentially rejects all but an exponentially small fraction of quantum trajectories of the entire quantum circuit. Direct application of quantum-classical cross-correlation methods \cite{GarrattEhudPostMeasurement2024,LiCrossEntropy2023} does not reduce the cost of this form of post-selection.

\section{Topological Adaptive Dynamics and Steering Toward Free-Fermion Topological States}
\label{sec:AdaptiveSteeringDynamics}

In this section, we focus on a general microscopic construction of free-fermion dynamical systems consisting of unitary evolution and measurements that can realize distinct dynamical topological area-law-entangled phases. These microscopic models are formulated as adaptive fermionic Gaussian circuits with post-selection-free measurements and feedforward operations. Based on the dynamical bulk-boundary correspondence (Sec.~\ref{Sec:Bott_Clock}), if the adaptive Gaussian circuit of the mEO class ${\cal K}_{\rm mEO}$ can steer the quantum system toward a topological steady-state ensemble in the same mEO class, the topology of the circuit dynamics can be classified by identifying the mEO class and spatial dimension of the dynamics. The corresponding topological classification can then be obtained from Table~\ref{tab: bott periodicity table}. In an ideal scenario, the topological steady-state ensemble contains only one free-fermion topological state. 

In Sec.~\ref{section: general protocol}, we present a general construction for post-selection-free topological adaptive Gaussian circuits that can steer the system toward a target free-fermion topological state. This construction applies to target topological states in any dimension and symmetry class. However, its implication differs depending on the target state's (traditionally defined) symmetry class. When the chosen target topological state belongs to the symmetry classes $\{\text{A, AI, BDI, D}\}$, our construction leads to an adaptive Gaussian circuit whose mEO class matches the target's symmetry class and, therefore, realizes the topological Gaussian dynamics classified in Table~\ref{tab: bott periodicity table}. When the chosen target topological state belongs to classes $\{\text{AIII, DIII, AII, CII, C, CI}\}$, the corresponding adaptive circuit contains gates that do \textit{not} respect the target state's symmetry class. Hence, it does not realize the topological Gaussian dynamics in these classes, which is consistent with the fact that topological dynamics in these classes require post-selection in the free-fermion limit. Nevertheless, we can still treat the constructed adaptive circuit as a novel protocol for preparing and stabilizing free-fermion topological states.

In Sec.~\ref{sec: adaptive-dynamics-chern-insulators}, we provide a concrete demonstration of mEO-class-A topological dynamics in 2+1d spacetime. The corresponding topological adaptive circuit targets a Chern insulator in two spatial dimensions. In addition, we estimate the convergence timescale to the target state by analyzing an effective Lindbladian description of the trajectory-averaged density matrix dynamics.

\subsection{General Construction of Topological Adaptive Circuit}\label{section: general protocol}
In the following, we introduce the general construction of a topological adaptive circuit that steers the physical system toward a target free-fermion topological state in any symmetry class and any spatial dimension {using local operations}. As mentioned earlier, the topological adaptive circuits realize the topological Gaussian dynamics in mEO classes $\AZK{mEO} =$ A, AI, BDI, and D when we choose the target topological state to respect one of these four symmetry classes. For target states in all ten symmetry classes, our adaptive circuit construction serves as a novel protocol for preparing and stabilizing free-fermion topological states.

Our construction is based on a stabilizer formalism of free-fermion topological states, which we introduce in Sec.~\ref{sec: stabilizer}. Measuring these stabilizers can determine whether the state of the system matches the target topological state. Our topological adaptive circuits are designed to repeatedly measure these stabilizers and perform error-correcting feedforward operations to steer the system toward the target topological state. The detailed construction is presented in Sec.~\ref{sec: AdaptiveDynamicsGeneralStrategy}. Importantly, our adaptive circuit is post-selection-free.

\subsubsection{Stabilizer Formalism for Free-Fermion Topological States} 
\label{sec: stabilizer}
We outline a general stabilizer-type formalism for free-fermion topological states {in all AZ symmetry classes}.
In this formalism, the free-fermion topological state of interest will be described as the unique state that satisfies a set of stabilizing conditions with spatially localized stabilizers. The symmetry properties of the stabilizers depend on the symmetry class. This formalism will serve as the foundation for the topological adaptive dynamics that steer the system toward the desired free-fermion topological states using a measurement-and-feedforward protocol.

A free-fermion topological state $\TS$ can be generally treated as the insulating ground state of a non-interacting parent Hamiltonian $\hat{\cal H}_{\rm TS}$. Here, we will use the double ket notation $| ~\cdot ~\rangle\!\rangle$ to denote the many-body wavefunction, while single kets $| ~\cdot ~\rangle$ denote single-particle states. To demonstrate the main idea of the stabilizer formalism, we will focus on the case where $\TS$ (and its parent Hamiltonian) are realized in a translation-invariant lattice system. We will later discuss the generalization to non-translation-invariant systems. A translation-invariant free-fermion topological state $\TS$ is essentially a non-interacting band insulator (of either charged complex fermions or Bogoliubov quasiparticles). We will introduce the localized stabilizers for the topological state $\TS$ using the overcomplete Wannier (OW) basis~\cite{rashba1997orthogonal} of the single-particle band structure of the non-interacting parent Hamiltonian $\hat{\cal H}_{\rm TS}$. Note that the OW basis has been referred to as the \textit{coherent state} basis in past literature~\cite{rashba1997orthogonal,qi2011generic,jian2013crystalsymmetry,li2024constraints}. As will become clear, the overcompleteness of this basis circumvents possible obstructions~\cite{brouder2007exponential,li2024constraints} to finding a local orthonormal basis for topological bands, enabling the description of $\TS$ using localized stabilizers. 

Concretely, let us consider the free-fermion topological state $\TS$ on a $d$-dimensional lattice with $N$ fermion modes per unit cell, where $\hat{c}_{\bfr, \mu}$ and $\hat{c}^\dag_{\bfr, \mu}$ denotes the annihilation and creation operators of the $\mu$th fermion modes in the unit cell at location ${\bf r}$. Using a properly chosen band structure given by the parent Hamiltonian $\hat{\cal H}_{\rm TS}$, $\TS$ can be defined as a band insulator (or superconductor) that fully occupies the $M\leq N$ lowest bands (and leaves the rest of the bands empty). Depending on the nature of $\TS$, the relevant band structure is either formed by regular charged complex fermions or chargeless Bogoliubov quasiparticles. 

For a given band with index $n$ ($1\le n \le N$), the projection of the fermion operators centered at $\bfr$ with the flavor $\mu$, $\hat{c}_{\bfr,\mu}$, onto this band, produces a localized fermion operator $\hat{\chi}_{\bfr,\mu,n}$ associated with the single-particle OW wavefunction $W_{\bfr, \mu, n}(\bfr',\mu')$ (with arguments $\bfr'$ and $\mu'$) after normalization: 
\begin{align}
    \hat{\chi}_{\bfr,\mu,n} = \sum_{\bfr',\mu'} W_{\bfr, \mu, n}(\bfr',\mu') \hat{c}_{\bfr', \mu'}. 
\end{align}
The normalization of this OW wavefunction is such that $\sum_{\bfr',\mu'} | W_{\bfr, \mu,n}(\bfr',\mu') |^2=1$. We denote the single-particle OW state with the wavefunction $W_{\bfr, \mu, n}(\bfr',\mu')$ as $|W_{\bfr, \mu, n} \rangle$. The single-particle OW state $|W_{\bfr, \mu, n} \rangle$ consists of degrees of freedom only in the $n$th band and is generically exponentially localized around $\bfr$ in real space~\cite{rashba1997orthogonal}. $|W_{\bfr, \mu, n} \rangle$'s with different values of $\bfr$ and $\mu$ are not necessarily orthogonal to each other. However, the set $\{|W_{\bfr, \mu, n} \rangle \}_{\bfr, \mu}$ of OW states (for all values of unit cell location $\bfr$ and mode index $\mu$) forms an {\it overcomplete localized} single-particle basis for the $n$th band despite the band topology~\cite{rashba1997orthogonal, qi2011generic, jian2013crystalsymmetry, li2024constraints}~\footnote{If $\hat{c}_{\bfr,\mu}$ with a certain mode index $\mu$ vanishes after the projection to the $n$th band, this mode index will not appear in the overcomplete OW basis $\{|W_{\bfr, \mu, n} \rangle \}_{\bfr, \mu}$ of the $n$th band. This circumstance will not qualitatively alter the subsequent discussions.}. In Sec. \ref{sec: ChernStabilizers}, we will demonstrate the technical details for the construction of the OW states $|W_{\bfr, \mu, n} \rangle$ and the associated fermion operators $\hat{\chi}_{\bfr, \mu, n}$ in a concrete example of a Chern insulator. Here, we will not dive into the technicalities and focus on the key ideas for establishing the stabilizer description of $\TS$.

The number operators  
\begin{align}
    \hat{\cal N}_{\bfr, \mu, n} \equiv \hat{\chi}_{\bfr, \mu, n}^\dag\hat{\chi}_{\bfr, \mu, n}
    \label{eq:LocalizedNumOp}
\end{align}
of the localized OW modes $\hat{\chi}_{\bfr, \mu, n}$ can serve as the stabilizers for the free-fermion topological state $\TS$. The fact that $\TS$ fully occupies the bands $n=1,...,M$ and leaves the bands $n=M+1,...,N$ completely empty is equivalent to the following stabilizing conditions: 
\begin{align}
    \hat{\cal N}_{\bfr, \mu, n} \TS & = \TS,~~ \text{for}~n = 1,...,M, \nonumber \\
    (1-\hat{\cal N}_{\bfr, \mu, n})\TS & = \TS,~~ \text{for}~n = M+1,...,N,
    \label{eq:TsStabilizerCondition}
\end{align}
for all choices of $\bfr$ and $\mu$. Note that the stabilizers $\hat{\cal N}_{\bfr, \mu, n}$ and $(1-\hat{\cal N}_{\bfr, \mu, n})$ with different subscripts $\bfr$, $\mu$ and $n$ do not always commute with each other (unlike those in standard quantum error-correcting stabilizer codes). 

Nevertheless, these stablizers enjoy the following three crucial properties: (1) each of the stabilizers $\hat{\cal N}_{\bfr, \mu, n}$ and $(1-\hat{\cal N}_{\bfr, \mu, n})$ is exponentially localized in real space; (2) each stabilizer eigenvalues $0$ and $1$; (3) the free-fermion topological state $\TS$ is the {\it unique} common eigenstate of all the stabilizers with eigenvalue being $1$. In Sec.~\ref{sec: AdaptiveDynamicsGeneralStrategy}, we will see that these properties underpin the topological adaptive circuit that steers the system toward the target topological state $\TS$. In particular, property (1) will ensure the locality of the dynamics. Measuring the number operator $\hat{\cal N}_{\bfr, \mu, n}$ will allow us to examine whether the system locally resembles the target state $\TS$. The topological adaptive dynamics will be designed to gradually steer the system toward the state that obeys the stabilizing conditions Eq.~\eqref{eq:TsStabilizerCondition}. The overcompleteness of the OW basis guarantees the uniqueness of the solution to Eq. \eqref{eq:TsStabilizerCondition} [see property (3)] and, hence, ensures that the topological adaptive circuit eventually reaches the desired target state $\TS$. 

Now, we argue that the stabilizer formalism introduced above can be generalized to free-fermion topological states $\TS$ without translation symmetry. In this case, we can treat such $\TS$ as the insulating ground state of a disordered non-interacting parent Hamiltonian. The single-particle Hilbert space can be divided into two orthogonal subspaces, one occupied by $\TS$ and the other unoccupied by $\TS$. Every local fermion operator $\h c_{\bfr, \mu}$ can be projected onto these two single-particle sub-Hilbert spaces, leading to a generalization of the localized OW modes $\h \chi_{\bfr, \mu, n}$. Subsequently, one can write down the number operators associated with the projected modes and the stabilizing conditions for $\TS$ similar to Eq.~\eqref{eq:TsStabilizerCondition}. 

Moreover, we would like to discuss the relation between the stabilizers and the symmetry classes of the topological state $\TS$. If TRS or PHS is present and squares to $1$, one can choose the OW modes $\hat{\chi}_{\bfr, \mu, n}$ so that each of them is a singlet under all the symmetries. One can work in the Majorana fermion basis to find such singlet OW modes in some cases. The corresponding number operator $\hcN_{\bfr, \mu, n}$ commutes with the symmetries. The mEO symmetry class of the evolution generated by measuring $\hcN_{\bfr, \mu, n}$ is the same as the symmetry class of the free-fermion topological state $\TS$ in the conventional definition. This is the case in (mEO) symmetry classes A, AI, BDI, and D.

In the remaining six symmetry classes (AIII, DIII, AII, CII, C, and CI), the stabilizers do not commute with the symmetry operations. For example, a topological state with spinful electrons in symmetry class AII, with TRS $\hat{\cal T}^2 = (-1)^{N_F}$, requires all the single-particle fermion modes to form Kramers doublets. A stabilizer associated with a spin-up fermion mode must transform into one associated with a spin-down fermion mode. Measuring these stabilizers independently without post-selection will lead to dynamics with TRS-broken mEOs.  

\subsubsection{Topological Adaptive Circuit that Steers Toward a Free-Fermion Topological State}
\label{sec: AdaptiveDynamicsGeneralStrategy}
Now, we present the general construction of the topological adaptive circuit that steers the quantum system toward a target free-fermion topological state $\TS$ (regardless of the initial state). The steering is achieved by repeated stabilizer measurements and error-correcting feedforward operations. We will discuss the timescale for the convergence toward $\TS$. This circuit can realize the topological Gaussian dynamics in mEO classes $\AZK{mEO} = \text{A, AI, BDI, and D}$ classified in Table~\ref{tab: bott periodicity table}. At the same time, it provides a general protocol for preparing and stabilizing free-fermion topological states in all symmetry classes, which we will elaborate on.

Concretely, the topological adaptive circuit targets the free-fermion topological state $\TS$ by iterating the following cycle of measurements and feedforward operations. In each cycle, the adaptive circuit measures the localized number operator $\hat{\cal N}_{\bfr, \mu, n}$ [See Eq.~\eqref{eq:LocalizedNumOp}] of the OW modes for all values of $\bfr$, $\mu$, and $n$ one after another. The measurement outcome for each $\hcN_{\bfr, \mu, n}$ is $0$ or $1$. After each measurement, the feedforward operation depends on whether the outcome matches the stabilizing condition Eq.~\eqref{eq:TsStabilizerCondition} for $\TS$. If the outcome matches the stabilizing condition, the adaptive circuit will proceed to measure the next localized number operator. If the measurement outcome of $\hcN_{\bfr, \mu, n}$ does not match Eq.~\eqref{eq:TsStabilizerCondition}, the circuit will apply an ``error-correcting" unitary gate that attempts to correct the undesired outcome with a finite probability before moving on to the next localized number operator. A construction of this error-correcting unitary will be presented below. A cycle of operation completes after every operator $\hcN_{\bfr, \mu, n}$ is measured and the subsequent feedforward operation is carried out. The so-defined cycle is iteratively implemented in the proposed topological adaptive circuit. All the operations involved in this circuit are Gaussian and exponentially localized. 

It is straightforward to conclude that once the system reaches the target topological state $\TS$, it will remain in the same state under the measurement and feedforward operations above. Because $\TS$ is the unique solution to the stabilizing condition Eq.~\eqref{eq:TsStabilizerCondition}, $\TS$ is the only steady state of this topological adaptive circuit. If an error occurs on the state $\TS$, this circuit will remove the error and restore $\TS$. In this sense, this topological adaptive circuit not only prepares but also stabilizes the topological state $\TS$.

Now, we describe the error-correcting unitary gate that corrects the undesired measurement outcomes of $\hcN_{\bfr, \mu, n}$ that violate the stabilizing conditions formulated in Eq.~\eqref{eq:TsStabilizerCondition}. To implement such error correction, we first introduce an extra ancillary system that shares the same lattice structure as the original physical system where the target state $\TS$ is defined. We denote the fermion operators in the ancillary systems as $\hat{d}_{\bfr,\mu}$. Throughout the dynamics, the ancillary system will always be in a product state with a finite particle density and trivial topology. We should view the total system as a \textit{bilayer structure}, where the top layer represents the physical system $\mathcal{P}$ and the bottom layer represents the ancillary system $\mathcal{A}$. 

We can choose the error-correcting unitary gates in the form of fermionic SWAP (fSWAP) gates:
\begin{equation}
    \mr{fSWAP}(\hat{c} , \hat{d}) \equiv \exp\left(\ii \frac{\pi}{2} (\hat{c}^\dagger   - \hat{d}^\dagger )(\hat{c}  - \hat{d})\right),
\end{equation}
which interchanges any given pair of physical fermion mode $\hat{c}$ and ancillary fermion mode $\hat{d}$, under the conjugatation by the fSWAP gate~\cite{cervera-lierta2018exact}. This unitary gate moves the particles between the physical and the ancillary system. When the measurement of $\hcN_{\bfr, \mu, n}$ yields an undesired outcome, the subsequent error-correcting unitary gate is given by the local fSWAP gate $\mr{fSWAP}(\hat{\chi}_{\bfr, \mu, n}, \hat{d}_{\bfr, \mu})$. Conceptually, the ancillary system plays the role of a particle bath. When the undesired occupation $\hcN_{\bfr, \mu, n}$ occurs, the fSWAP gate can (probabilistically) correct it by transferring particles between the physical system and the bath. One way to make it local is to imagine that the physical and ancillary systems are arranged in a bilayer geometry, which ensures that the error-correcting fSWAP operations remain local. The use of an ancillary system to pump particles in and out of the physical system is similar in spirit to a recently proposed error correction scheme proposed for fault-tolerant fermionic quantum processors~\cite{ottErrorcorrectedFermionicQuantum2024}.

For example, consider the case where $\hcN_{\bfr, \mu, n}= 0$ matches the stabilizing condition Eq.~\eqref{eq:TsStabilizerCondition} 
as an example. The undesired outcome $\hcN_{\bfr, \mu, n}= 1$ can be corrected by $\mr{fSWAP}(\hat{\chi}_{\bfr, \mu, n}, \hat{d}_{\bfr, \mu})$ which moves the particle from the mode $\hat{\chi}_{\bfr, \mu, n}$ to the ancillary mode $\hat{d}_{\bfr, \mu}$ when the latter mode is empty. When the ancillary mode $\hat{d}_{\bfr, \mu}$ is occupied, the action of $\mr{fSWAP}(\hat{\chi}_{\bfr, \mu, n}, \hat{d}_{\bfr, \mu})$ is trivial, leaving the undesired measurement outcome of $\hcN_{\bfr, \mu, n} $ uncorrected. Therefore, the probability of successful error correction depends on the probability of finding the mode $\hat{d}_{\bfr, \mu}$ empty. To ensure that the error correction can happen with a finite probability, we include the random incoherent local redistribution of particles in the ancillary system as a part of each cycle of the adaptive dynamics. This redistribution dynamics can be implemented as a unitary gate that generates random particle hopping amongst the ancillary modes, followed by the measurements of all density operators $\hat{d}^\dag_{\bfr,\mu}\hat{d}_{\bfr,\mu}$. Because of the incoherent redistribution dynamics, any mode $\hat{d}_{\bfr, \mu}$ has a finite probability to be empty (or occupied). In the case where $\hcN_{\bfr, \mu, n}= 1$ matches the stabilizing condition Eq.~\eqref{eq:TsStabilizerCondition}, one can perform a parallel analysis to show that $\mr{fSWAP}(\hat{\chi}_{\bfr, \mu, n}, \hat{d}_{\bfr, \mu})$ can correct the undesired measurement outcome $\hcN_{\bfr, \mu, n}= 0$ with a finite probability.

Regarding the locality of this topological adaptive circuit, both the measurements and the error-correcting gates are localized with exponentially decaying tails due to the localization property of the OW modes $\hat{\chi}_{\bfr,\mu,n}$. The exponential tails are present because this circuit precisely targets one single topological state $\TS$. A finite-range topological adaptive circuit can be obtained by replacing all of the OW modes $\hat{\chi}_{\bfr,\mu,n}$ by their finite-range-truncated (and normalized) versions. Every operation in this circuit becomes strictly finite-ranged. Even though Eq.~\eqref{eq:TsStabilizerCondition} no longer holds for the number operators of the truncated OW modes, the finite-range topological adaptive circuit can still steer the system toward a steady-state ensemble consisting of states with the same topology as the original target state $\TS$ {for mEO classes A, AI, BDI, and D}.
We provide numerical evidence for this statement in concrete examples presented in Sec.~\ref{sec:numerics}. 

The symmetry properties of this topological adaptive circuit depend on the target topological state $\TS$. If $\TS$ belongs to symmetry classes A, AI, BDI, and D (in the conventional sense), one can see that all the operations, including the measurements and the unitary gates, respect (following the definition of Sec.~\ref{sec:mEO class intro}) the symmetries in the relevant symmetry class (upon choose the right basis). In other words, the mEO class of the circuit dynamics is the same as the symmetry class of the target state. By the dynamical bulk-boundary correspondence, we conclude that our topological adaptive circuits realize the topological Gaussian dynamics in these four symmetry classes as classified in Table~\ref{tab: bott periodicity table}. When we choose a target topological state $\TS$ in the symmetry classes AIII, DIII, AII, CII, C, and CI, the post-selection-free measurements in the adaptive circuit do not respect the relevant symmetries. Hence, the topological adaptive circuits with a target state in these six symmetry classes should {\it not} be treated as the topological Gaussian dynamics in these six mEO symmetry classes. This is consistent with our earlier conclusion (see Sec.~\ref{Sec: POVMadmissibility}) that the realization of the topological dynamics in the mEO classes AIII, DIII, AII, CII, C, and CI must involve post-selection in the free-fermion limit.
Regardless of the symmetry aspects, our topological adaptive circuit can always be treated as a novel protocol to both prepare and stabilize free-fermion topological states $\TS$ in any dimension and any symmetry class.

Now, we discuss the timescale for the system to converge to the target topological state $\TS$ under the proposed topological adaptive circuit {focusing on the symmetry classes A, AI, BDI, and D}. Because different localized number operators $\hat{\cal N}_{\bfr, \mu, n}$ do not always commute, and because the error-correcting unitary only corrects the undesired measurement outcomes with a nonzero probability, the system generically does not reach the target topological state $\TS$ after one cycle. Nevertheless, as the adaptive circuit repeats the cycle described above, the system will be steered toward $\TS$ in $\mc{O}(1)$ cycles, which will be shown in concrete examples in Sec.~\ref{sec: adaptive-dynamics-chern-insulators} and Sec.~\ref{sec:numerics}. Within each cycle, the number of measurements scales with the total system size. In the original (untruncated) adaptive circuit, the measurements of different $\hat{\cal N}_{\bfr, \mu, n}$ do not commute and, hence, need to be implemented \textit{sequentially}, rendering the total circuit depth per cycle to scale with the total system size. In contrast, in the finite-range adaptive circuit, the number operators of the finite-range-truncated OW modes commute when their supports do not overlap. This feature enables the parallelization of many measurement and feedforward operations, reducing the circuit depth of each cycle to $\mc{O}(1)$ and is thus independent of the total system size. Crucially, this truncation does not change the symmetry properties of the circuit.

In addition, we would like to compare our protocol with other methods for preparing topological states. Notably, Ref.~\cite{barbarino2020preparing} proposed a protocol to prepare an invertible topological state together with its conjugate via unitary operations, and discards the unwanted conjugate copy in the end. This method lacks a mechanism to eliminate or mitigate errors that may occur during the preparation process. Additionally, it does not protect the target state against noise after preparation. In contrast, our adaptive protocol can not only steer the systems toward the target state but also protect the target state by persistent error correction. The stability of our adaptive protocol is studied numerically in a concrete example in Sec.~\ref{sec:robustness}.

Conceptually, our construction is similar to the previous measurement-induced steering protocol for the 1d AKLT state presented in Ref.~\cite{puente_quantum_2024,roy_measurement-induced_2020}. The AKLT state enjoys a frustration-free Hamiltonian. That means individual terms in the Hamiltonian annihilate the AKLT state. In the measurement-induced steering protocol, these individual Hamiltonian terms play a similar role to our non-commuting stabilizers.

It is also helpful to compare our adaptive circuits with the existing dissipative schemes used to prepare topological states. There has been extensive work on engineering Lindbladians that dissipatively prepare target quantum states~\cite{diehlTopologyDissipationAtomic2011a, bardynTopologyDissipation2013a, budichDissipativePreparationChern2015a, reiterScalableDissipativePreparation2016, goldsteinDissipationinducedTopologicalInsulators2019a, chenQuantumThermalState2023, yangDissipativeBoundaryState2023b, guoDesigningOpenQuantum2025, pocklingtonAcceleratingDissipativeState2025, linDissipativePreparationManyBody2025, zhanRapidQuantumGround2026}. In particular, Refs.~\cite{budichDissipativePreparationChern2015a, goldsteinDissipationinducedTopologicalInsulators2019a} study the dissipative preparation of Chern insulators in two spatial dimensions. In general, our adaptive circuit protocol avoids engineered dissipation entirely, relying instead on local measurement dynamics combined with unitary feedforward (and an ancillary system) to prepare topological states at the level of individual quantum trajectories. In Sec. \ref{sec: adaptive-dynamics-chern-insulators}, we will apply our adaptive circuit construction to mEO class A in 2+1d. The corresponding adaptive circuit prepares and stabilizes two-dimensional Chern insulators. As we will see, the finite-range version of our adaptive circuit can produce \textit{pure-state} Chern insulators in individual quantum trajectories in \textit{finite time}. There are trajectory-to-trajectory variations arising from an intrinsic measurement randomness and the truncation of OW modes. In contrast, the previous dissipative schemes only prepare a single mixed-state analogue of a Chern insulator when using finite-range jump operators \cite{budichDissipativePreparationChern2015a,goldsteinDissipationinducedTopologicalInsulators2019a}. 

The technical details skipped in the discussion above will be illustrated in a concrete example in Sec. \ref{sec: adaptive-dynamics-chern-insulators}.

\subsection{Topological Adaptive Steering Dynamics that Targets a Chern Insulator}\label{sec: adaptive-dynamics-chern-insulators}

In the following, we discuss a concrete example of a Gaussian topological adaptive circuit that realizes the mEO class A (and equivalently, sTM class AIII) topological Gaussian dynamics in 2+1d spacetime dimensions. As required by the mEO symmetry class A, the circuit conserves the total charge (or total particle number) in the system. It steers the system toward a Chern insulator in two spatial dimensions, which is a manifestation of the topology in the dynamics via the dynamical bulk-boundary correspondence (see Sec.~\ref{Sec:Bott_Clock}). This example follows the general construction outlined in Sec.~\ref{section: general protocol}.

Section~\ref{sec: ChernStabilizers} presents the technical details of the stabilizer formalism for a target Chern insulator state $\CI$.  
Section~\ref{sec: ChernInsulator_SterringCircuit} presents the topological adaptive circuit that steers the system toward $\CI$.
In Sec. \ref{sec:ChernInsulator_QuantumChannel}, we estimate the timescale for the system's convergence to $\CI$ using an effective Lindbladian description of the evolution of the trajectory-averaged density matrix under the proposed circuit dynamics.

\subsubsection{A Chern Insulator in 2d and its Stabilizer Formalism}
\label{sec: ChernStabilizers}

\begin{figure*}[t]
    \centering
    \includegraphics[width=\textwidth]{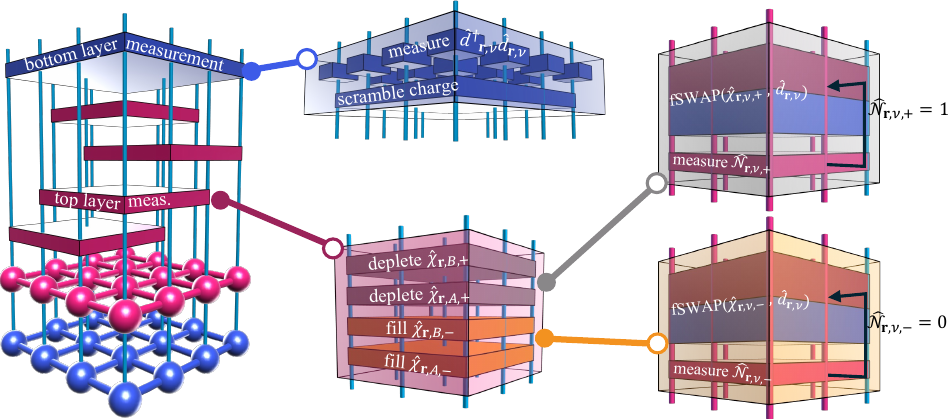}
    \caption{Schematic for the topological adaptive circuit as described in Algorithm~\ref{algo:chern}. The physical system (red on the top layer) and the ancillary system (blue on the bottom layer) are both 2d square lattices of the same size with two fermion modes per unit cell. 
    In each cycle, the circuit first measures the localized number operators $\hcN_{\bfr,\nu,\pm}$ in the physical system and applies feedforward operations to correct the measurement outcomes (red boxes).
    The circuit then redistributes the particles in the ancillary system (blue box) to prepare for the next cycle.}
    \label{fig: schematic}
\end{figure*}

We present the stabilizer formalism for a 2d Chern insulator state $\CI$, setting the stage for the construction of the topological adaptive circuit that targets this state. Concretely, we focus on a Chern insulator state $\CI$ given by the ground state of a non-interacting two-band parent Hamiltonian $\h{\mc{H}}_{\rm CI}$. In real space, the system is defined on a 2d square lattice with two fermion modes $\h c_{\bfr,1}$ and $\h c_{\bfr,2}$ per unit cell. The unit cell coordinate $\bfr= (x,y)\in \mathbb{Z}^2$ is given by a pair of integers. 
For convenience, we start with the momentum-space resolved parent Hamiltonian of the target state $\CI$ as
\begin{align}
    \h{\mc{H}}_{\rm CI} &= \int\frac{d^2\mbf{k}}{(2\pi)^2}\h{ c}^\dagger({\mbf{k}})(\vec{n}(\mbf{k})\cdot\vec{\sigma}) \h c(\mbf{k})\label{eq:toymodel} 
\end{align}
with 
\begin{align}
    \vec{n}(\mbf{k}) &= (\sin k_x, \ \sin k_y, \ -\alpha-\cos k_x - \cos k_y)\label{eq:spectral vector for CI}.
\end{align}
$\vec{\sigma} =(\sigma^x, \sigma^y, \sigma^z)$ represents the vector of Pauli matrices and $\h c(\mbf{k}) = (\h c_1(\mbf{k}), \ \h c_2(\mbf{k}))^\intercal$ is the momentum-space fermion operator with its two components corresponding to the two modes $\h c_1$ and $\h c_2$ of a unit cell. Here, {$\alpha\in\mathbb{R}$} is a tuning parameter for the target state $\CI$. The parent Hamiltonian $\h{\mc{H}}_{\rm CI}$ gives rise to two single-particle bands, and the Chern insulator $\CI$ is defined as the insulating ground state, with the lower band fully occupied {and the upper band empty}. The band gap is finite as long as $\alpha \neq 0, \pm2$, which we will assume unless stated otherwise. The topological properties of $\CI$ can be tuned by the parameter $\alpha$: $\CI$ has Chern number $1$ for $\alpha \in (-2,0)$, $-1 $ for $\alpha \in (0,2)$, and $0$ for $|\alpha|>2$. 

The single-particle projectors {onto} the upper and lower bands of $\h{\mc{H}}_{\rm CI}$ are denoted by the subscript $\pm$, respectively, as
\begin{align}
    P_{\pm}(\mbf{k}) = \frac{1}{2} \left(\Id \pm \frac{\vec{n}(\mbf{k})}{|\vec{n}(\mbf{k})| }\cdot\vec{\sigma} \right).
    \label{eq:CI_ParentHamiltonian_Projectors}
\end{align}
Using this projector, we can write down the (normalized) localized fermion operators $\hat{\chi}_{\bfr, A/B, \pm}$ whose single-particle OW wavefunction forms an overcomplete basis of the upper ($+$) and the lower ($-$) bands:
\begin{align}
    \hat{\chi}_{\bfr, A/B, \pm} \propto \int\frac{d^2\mbf{k}}{(2\pi)^2} e^{\ii \mbf{k}\cdot\bfr}  \tau_{A/B}^\dag P_{\pm}(\mbf{k}) \h c(\mbf{k}), \label{eq:chi_CI}
\end{align}
where $\tau_A = (1, \ 1)^\intercal/\sqrt{2}$ and  $\tau_B = (1, \ -1)^\intercal/\sqrt{2}$. Up to normalization factors implicit in Eq.~\eqref{eq:chi_CI}, the fermion operators $\hat{\chi}_{\bfr, A/B, \pm}$ are the projection of the local fermion operators $\h c_{\bfr,A} = \frac{1}{\sqrt{2}}(\h c_{\bfr,1} + \h c_{\bfr,2})$ and $\h c_{\bfr,B} = \frac{1}{\sqrt{2}}(\h c_{\bfr,1} - \h  c_{\bfr,2})$ onto the upper ($+$) and the lower ($-$) band of $\h{\mc{H}}_{\rm CI}$. 
In principle, any choice of two orthogonal vectors $\tau_{A/B}$ suffices for our construction. For example, if we choose $\tau_A'= (1,0)^\intercal$ and $\tau_B'=(0,1)^\intercal$ instead, the corresponding localized OW modes will be the projections of $c_{\bfr,1} $ and $c_{\bfr,2}$. Here, under our current choice of $\tau_{A/B}$, the normalization factors for all the OW modes remain $\mc{O}(1)$ for all values of the tuning parameter $\alpha$, which is convenient for the subsequent numerical studies of the adaptive dynamics.

The single-particle OW state $|W_{\bfr, A/B, \pm}\rangle$ associated with the localized fermion operator $\hat{\chi}_{\bfr, A/B, \pm}$ has the wavefunction:
\begin{align}
     W_{\bfr, A/B, \pm}(\bfr',\mu') \propto \int\frac{d^2\mbf{k}}{(2\pi)^2} e^{\ii \mbf{k}\cdot(\bfr-\bfr')} \left[\tau_{A/B}^\dag P_{\pm}(\mbf{k}) \right]_{\mu'},
\end{align}
where $[\dots]_\mu$ denotes the $\mu$th component of the OW function.
This wavefunction is exponentially localized around $\bfr$ as long as the band gap in $\hat{\cal H}_{\rm CI}$ is nonzero, i.e., $\alpha \neq 0, \pm 2$, {leading to a tunable localization length via the parameter $\alpha$. }

{The set of OW states $\{|W_{\bfr, \nu, \pm} \rangle \}_{\bfr\in \mathbb{Z}^2,\nu=A,B}$ forms an overcomplete localized basis for the upper band $+$ (lower band $-$) of $\h{\mc{H}}_{\rm CI}$, though neither $\{|W_{\bfr, A, \pm} \rangle \}_{\bfr\in \mathbb{Z}^2}$ nor $\{|W_{\bfr, B, \pm} \rangle \}_{\bfr\in \mathbb{Z}^2}$ is a complete basis of the upper (lower) band on its own. }
(See App.~\ref{app: topological obstructions} for more discussion on the completeness of the basis in Chern bands.) 
As discussed in Sec.~\ref{sec: stabilizer}, the (over)completeness of the basis will guarantee that the target Chern insulator state $\CI$ is the {unique} many-body state {satisfying} all the stabilizing conditions presented below in Eqs.~\eqref{eq:CIStabilizerCondition1} and \eqref{eq:CIStabilizerCondition2}.---a crucial property that will ensure the convergence of the proposed topological adaptive circuit in Sec.~\ref{sec: ChernInsulator_SterringCircuit}.

The stabilizing condition for the Chern insulator state $\CI$ is given by
\begin{align}
    \hat{\cal N}_{\bfr, \nu, -} \CI & = \CI,  \ \forall \ \bfr\in\mathbb{Z}^2, \ \nu=A, B,  \label{eq:CIStabilizerCondition1}\\ 
    (1-\hat{\cal N}_{\bfr, \nu, +})\CI & = \CI, \ \forall \ \bfr\in\mathbb{Z}^2, \ \nu=A, B, 
    \label{eq:CIStabilizerCondition2}
\end{align}
where the number operators $\hat{\cal N}_{\bfr, A/B, \pm}$ are those of the (normalized) localized OW modes $\hat{\chi}_{\bfr, \nu, \pm}$:
\begin{align}
   \hat{\cal N}_{\bfr, \nu, \pm} \equiv \hat{\chi}_{\bfr,\nu, \pm}^\dag \hat{\chi}_{\bfr, \nu, \pm},
\end{align}
for $\nu = A, B$. 
When $\alpha \neq 0, \pm2$, namely when the band gap in $\hat{\cal H}_{\rm CI}$ is nonzero, each $\hat{\cal N}_{\bfr, A/B, \pm}$ is exponentially localized around $\bfr$. The localization length diverges as the gap closes. With finite localization length, the adaptive circuit constructed from the recipe in Sec. \ref{section: general protocol} will respect the locality of the system and steer the system precisely to the target Chern insulator state $\CI$. As we will see later, the parameter $\alpha$ will offer a tuning knob that interpolates mEO class A Gaussian dynamics with different topologies, manifested by the different steady-state Chern numbers.

\subsubsection{Topological Adaptive Circuit that Steers toward a Chern Insulator}
\label{sec: ChernInsulator_SterringCircuit}

Equipped with the stabilizer formalism of the Chern insulator $\CI$, we are ready to introduce the topological adaptive circuit that steers the physical system to $\CI$ following the recipe outlined in Sec. \ref{sec: AdaptiveDynamicsGeneralStrategy}.

The total system includes the physical system that will host the Chern insulator $\CI$ and an ancillary system. Both systems are 2d square lattices with two modes per unit cell. The two auxiliary fermion modes at each unit cell $\bfr$ in the ancillary system are denoted as $\hat d_{\bfr, A}$ and $\hat d_{\bfr, B}$. Conceptually, the desired topological adaptive circuit steers the physical system to $\CI$ by filling (depleting) the lower (upper) band of $\hat{\cal H}_{\rm CI}$ [Eq.~\eqref{eq:CI_ParentHamiltonian_Projectors}] via measurement and feedforward operations. The fully filled lower band corresponds to the stabilizing condition Eq.~\eqref{eq:CIStabilizerCondition1}, while the completely depleted upper band corresponds to Eq.~\eqref{eq:CIStabilizerCondition2}. 

The topological adaptive circuit for $\CI$ implements the measurement-and-feedforward quantum algorithm summarized in Algorithm~\ref{algo:chern} {(which is also schematically shown in Fig.~\ref{fig: schematic})}. Below, we provide the details and reasons for each step. First, we initialize the physical system {(red dots in Fig.~\ref{fig: schematic})} and the ancillary system {(blue dots in Fig.~\ref{fig: schematic})} in two decoupled pure Gaussian states. Each of the initial Gaussian states has a fixed charge (or particle number), and the total charge $Q$ of the entire system must be in the range $Q\in (N_{\rm uc},  \ 3N_{\rm uc}) $. Here, $N_{\rm uc}$ is the number of unit cells in the physical system as well as in the ancillary system. Our topological adaptive circuit conserves the total charge (or particle number). Regardless of the details of the initial state, this circuit will steer the physical system toward $\CI$, which consists of $N_{\rm uc}$ particles, and bring the ancillary system to a trivial $(Q-N_{\rm uc})$-particle product state. The constraint on $Q$ is simply due to the range of charge that the ancillary system can hold, i.e., $Q - N_{\rm uc}\in (0, \ 2N_{\rm uc})$.

After initialization, the topological adaptive circuit iterates a two-step cycle. The first step involves stabilizer measurements and error-correcting feedforward operations (see the right column in Fig.~\ref{fig: schematic}). Concretely, the circuit goes through the unit cell label $\bfr$ one by one. For each unit cell location $\bfr\in \mathbb{Z}^2$, four local density operators $\hcN_{\bfr, A/B, \pm}$ will be measured. After each measurement, the feedforward operation is a do-nothing operation if the measurement outcome matches the stabilizing conditions Eq.~\eqref{eq:CIStabilizerCondition1} and Eq.~\eqref{eq:CIStabilizerCondition2}. If the outcome does not match, the feedforward operation is an error-correcting fSWAP gate. 

More specifically, when a measurement finds $\hcN_{\bfr, A/B, -} = 0$, it means that the localized OW mode $\h \chi_{\bfr, A/B, -}$ in the lower band of $\hat{\cal H}_{\rm CI}$ is missing a particle. The local unitary gate ${\rm fSWAP}(\hat{\chi}_{\bfr, A/B, -},\h d_{\bfr, A/B})$ probabilistically corrects the error by taking a particle from the ancillary mode $d_{\bfr, A/B}$ and filling it in the mode $\hat \chi_{\bfr, A/B, -}$. The success rate of this error correction depends on the probability of having an available particle in the ancillary mode $d_{\bfr, A/B}$, which is finite.  When a measurement finds $\hcN_{\bfr, A/B, +} = 1$, the localized mode $\h \chi_{\bfr, A/B, +}$ of the upper band contains an unwanted particle. The local error-correcting unitary gate ${\rm fSWAP}(\hat{\chi}_{\bfr, A/B, +},\h d_{\bfr, A/B})$ probabilistically depletes the mode $\h \chi_{\bfr, A/B, +}$ by moving the unwanted particle to the ancillary mode $d_{\bfr, A/B}$. The success probability is tied to the probability of finding the mode $d_{\bfr, A/B}$ empty (before the fSWAP gate), which is again finite. 

The second step in the cycle randomly redistributes local charge within the ancillary system (see the top blue box in Fig.~\ref{fig: schematic}). An implementation of this redistribution begins with a charge-conserved random unitary gate:
\begin{align}
    \prod_{\langle \bfr, \bfr' \rangle}& e^{\ii\theta_\bfr \sum_{\nu = A,B}(\h d_{\bfr, \nu}^\dag \h d_{\bfr', \nu}+h.c.)} \prod_\bfr e^{\ii\theta_\bfr' (\h d_{\bfr, A}^\dag \h d_{\bfr, B}+h.c.)}  \label{eq:random unitary on bottom layer},
\end{align}
where $\langle \bfr, \bfr' \rangle$ denotes the nearest-neighbor pairs of unit cells at $\bfr$ and $\bfr'$. Furthermore, $\theta_\bfr'$ and $\theta_\bfr$ are independent random phase angles between $0$ and $2\pi$ for every $\bfr$. After the random unitary gate, the circuit measures all the particle number operators $\hat{d}_{\bfr, \nu}^\dag \hat{d}_{\bfr, \nu} $ for $\bfr \in \mathbb{Z}^2$ and $\nu = A,B$. These measurements collapse the ancillary system into a product state with a random charge distribution. The randomness of the charge distribution in the ancillary system guarantees finite success probabilities for the aforementioned fSWAP gates in correcting the undesired measurement outcomes of $\hcN_{\bfr,A/B,\pm}$. Conceptually, the ancillary system acts as a bath that supplies the particles needed to fill the lower bands in the physical system and provides space to dump the unwanted particles detected in the upper band of the physical system. 

The iteration of the two-step cycle described above gradually fills (depletes) all of OW modes $\hat{\chi}_{\bfr, A/B, -}$ ($\hat{\chi}_{\bfr, A/B, +}$) in the lower (upper) band of the physical system. The overcompleteness of $\{|W_{\bfr, \nu, +} \rangle \}_{\bfr\in \mathbb{Z}^2,\nu=A,B}$ and $\{|W_{\bfr, \nu, -} \rangle \}_{\bfr\in \mathbb{Z}^2,\nu=A,B}$ as a basis of the upper ($+$) and the lower ($-$) bands guarantees the physical system's convergence to the desired Chern insulator state $\CI$. Here, we have assumed that the band gap in $\hat{\cal H}_{\rm CI}$ is nonzero, i.e., $\alpha \neq 0, \pm2$. This convergence timescale is $\mc{O}(1)$ in the number of cycles, which we can estimate using an effective Lindbladian description for the evolution of the trajectory-averaged density matrix (see Sec.~\ref{sec:ChernInsulator_QuantumChannel}).

\begin{algorithm}[hptb]
    \SetAlgoLined
    \vspace{5pt}
    \textbf{Initialization:} Arbitrary free-fermion states in both the physical and the ancillary system, with definite total charge $Q \in (N_{\rm uc}, 3N_{\rm uc})$.\\ 

    \textbf{Output:} The physical system converges to the Chern insulator state $\ket{\mr{CI}}$, while the ancillary system hosts a random product state. \\ [10pt] 

    \textit{\# Adaptive Protocol}\\
    \textbf{Step 1: Stabilizers measurements and error-correcting feedforward operations}\\ 
    \For{$\bfr \in  \text{\rm unit cell coordinates}$}{
        \For{$\nu = A,B  $}{
            \vspace{5pt}
            \textit{\# Fill lower band of }$\hat{\cal H}_{\rm CI}$\\
            Measure $\hcN_{\bfr,\nu,-}$\\
            \If{$\hcN_{\bfr,\nu,-} = 0$}{
                Apply $\mathrm{fSWAP}\left(\hat{\chi}_{\bfr, \nu, -} , \ \hat{d}_{\bfr, \nu} \right)$
            }\vspace{10pt}
            
            \textit{\# Deplete upper band of }$\hat{\cal H}_{\rm CI}$\\
            Measure $\hcN_{\bfr,\nu, +}$\\
            \If{$\hcN_{\bfr,\nu, +}= 1$}{
                Apply $\mathrm{fSWAP}\left(\hat{\chi}_{\bfr, \nu, +} , \ \hat{d}_{\bfr, \nu} \right)$
            }
        }
    }\vspace{10pt}

    \textbf{Step 2: Particle redistribution and measurement in the ancillary system}\\
    \indent Apply random local charge-conserving Gaussian unitary gates Eq.~\eqref{eq:random unitary on bottom layer} to the ancillary system to redistribute particles. Measure all the number operators $\h d^\dag_{\bfr, A/B}d_{\bfr, A/B}$ to collapse the ancillary system to a product state. 
    \vspace{10pt}

    \textbf{Step 3: Iterate}\\
    Repeat steps 1--2 until the physical system converges to the Chern insulator state $\ket{\mr{CI}}$.

    \caption{Topological adaptive protocol for steering toward 2d Chern Insulator}
    \label{algo:chern}
\end{algorithm}

Recall that each number operator $\hcN_{\bfr, A/B,\pm}$ is localized around $\bfr$ with an exponential tail.  The corresponding localization length, controlled by the parameter $\alpha$, is finite (except at $\alpha = 0, \pm2$). Since the operators $\hcN_{\bfr, A/B,\pm}$ do not always commute with each other, the Algorithm~\ref{algo:chern} implements their measurements sequentially in each cycle. That means the circuit depth of each cycle is ${\cal O}(N_{\rm uc})$. 

We can modify the topological adaptive circuit into a finite-range version by truncating and normalizing each OW mode $\hat{\chi}_{\bfr, A/B, \pm}$ within a finite-range neighbor of $\bfr$. Then, we replace $\hcN_{\bfr, A/B, \pm}$ by the number operator of the corresponding truncated OW modes. This modification ensures that all operations in the circuit are strictly local (i.e., finite-ranged with no exponential tails). Since the number operators of the truncated OW modes commute when their supports do not overlap, their measurements and feedforward operations can be parallelized. This parallelization can reduce the circuit depth of each cycle to $\mc{O}(1)$.

Note that the number operators of the truncated OW modes no longer strictly obey the stabilizing conditions Eqs.~\eqref{eq:CIStabilizerCondition1} and \eqref{eq:CIStabilizerCondition2} for $\CI$. Nevertheless, our numerical studies in Sec. \ref{sec:numerics} show that the finite-range topological adaptive circuit can still steer the physical system toward a steady-state ensemble of 2d area-law entangled states sharing the same nontrivial Chern number. That means the finite-range topological adaptive circuits also provide examples of the topological Gaussian dynamics in mEO class A (which is equivalent to sTM class AIII). The parameter $\alpha$ can be used to tune the topology of the circuit dynamics, manifested by the steady-state Chern numbers. As will be shown in Sec.~\ref{sec:numerics}, the change of the topology of the dynamics leads to a dynamical topological phase transition.

\subsubsection{Effective Lindbladian Description}
\label{sec:ChernInsulator_QuantumChannel}
Under the topological adaptive circuit dynamics for $\CI$, the evolution of the system's density matrix averaged over all quantum trajectories can be captured using quantum channels. In the following, we will write this quantum channel and obtain its effective Lindbladian description. We will analyze this effective Lindbladian description to gain insight into the system's convergence to the target Chern insulator state $\CI \langle\!\langle {\rm CI}|$. Our analysis here corroborates our numerical studies presented later in Sec.~\ref{sec:numerics}. In the following, we present the main results of our analysis. The technical details are summarized in App.~\ref{app:lindblad}.

In the topological adaptive circuit, the most important ingredients that achieve the steering toward the Chern insulator are the measurements of the localized number operators $\hcN_{\bfr, A/B, \pm}$ and the subsequent feedforward operations. Each of the measurement-and-feedforward operations induces a quantum channel acting on the trajectory-averaged density matrix $\h\rho_{\rm tot}$ for the entire system, including both the physical and ancillary systems:
\begin{widetext}
    \begin{align}
    \mc{E}_{\bfr,\nu,-}(\h \rho_{\rm tot})&=\hcN_{\bfr,\nu,-} \h\rho_{\rm tot} \hcN_{\bfr,\nu,-}  + {\rm fSWAP}(\h\chi_{\bfr,\nu,-},\h d_{\bfr,\nu} ) \cdot
    (1-\hcN_{\bfr,\nu,-}) \h \rho_{\rm tot} (1-\hcN_{\bfr,\nu,-}) \cdot {\rm fSWAP}(\h\chi_{\bfr,\nu,-},\h d_{\bfr,\nu} )^\dag, \label{eq:filling channel}\\
     \mc{E}_{\bfr,\nu,+}(\h \rho_{\rm tot})&=(1 -\hcN_{\bfr,\nu,+} ) \h\rho_{\rm tot} (1 -\hcN_{\bfr,\nu,+} )  + {\rm fSWAP}(\h\chi_{\bfr,\nu,+},\h d_{\bfr,\nu} ) \cdot
    \hcN_{\bfr,\nu,+}  \h\rho_{\rm tot}  \hcN_{\bfr,\nu,+} \cdot  {\rm fSWAP}(\h\chi_{\bfr,\nu,+},\h d_{\bfr,\nu} )^\dag, \label{eq:depletion channel}
\end{align}
\end{widetext}
with $\nu = A, B$. Here, $\mc{E}_{\bfr,\nu,\pm}$ corresponds to the measurement-and-feedforward operation associated with $\hcN_{\bfr,\nu,\pm}$. The tensor product of $\CI \langle\!\langle {\rm CI}|$ in the physical system and any density matrix $\h \rho_{\rm a}$ in the ancillary system, i.e., $\h\rho_{\rm tot} = \CI \langle\!\langle {\rm CI}| \otimes \h \rho_{\rm a}$, stays invariant under the action of these two quantum channels. This conclusion is the direct consequence of the stabilizing conditions Eqs.~\eqref{eq:CIStabilizerCondition1} and \eqref{eq:CIStabilizerCondition2}.

Recall that the topological adaptive circuit randomly redistributes particles in the ancillary system after each cycle. To approximate the ancillary system as an incoherent particle bath, we treat it as a Markovian environment that is instantaneously refreshed after each elementary measurement-feedforward step. Specifically, we assume the state of the total system remains in a product form throughout, $\h\rho_{\rm tot} = \h\rho_{\mathrm{ph}} \otimes \h\rho_{\mathrm{a}}$, where $\h\rho_{\mathrm{ph}}$ is the physical system's density matrix, and $\h\rho_{\mathrm{a}}$ is approximated as a product of Gaussian mixed states on each ancillary mode with a fixed particle number $\bar{n}_{\mathrm{a}}$ per ancillary fermion mode. Formally, we enforce $\mathrm{Tr}[\h\rho_{\mathrm{a}} d^\dagger_{\mbf{r},\nu}d_{\mbf{r},\nu}] = \bar{n}_{\mathrm{a}}$ for each $\nu$ and unit cell $\mathbf{r}$. To avoid correlations between distinct elementary measurement-feedforward operations that reuse the same ancillary modes, we minimally enlarge the ancillary Hilbert space by introducing two additional modes per unit cell, allowing each elementary feedforward operation (fSWAP) to act on an independent ancillary mode. Tracing out the ancillary system under such approximations yields the effective quantum channels acting solely on the physical system:
\begin{widetext}
\begin{align}
    \tilde{\mathcal{E}}_{\mathbf{r},\nu,-}(\h\rho_{\mathrm{ph}})&=\hat{\mathcal{N}}_{\mathbf{r},\nu,-}\h\rho_{\mathrm{ph}}\hat{\mathcal{N}}_{\mathbf{r},\nu,-}+(1-\bar{n}_{\mr{a}})(1-\hat{\mathcal{N}}_{\mathbf{r},\nu,-})\h\rho_{\mathrm{ph}}(1-\hat{\mathcal{N}}_{\mathbf{r},\nu,-})+\bar{n}_{\mr{a}}\hat{\chi}_{\mathbf{r},\nu,-}^\dagger\h\rho_{\mathrm{ph}}\hat{\chi}_{\mathbf{r},\nu,-},\label{eq:effective filling channel} \\
    \tilde{\mathcal{E}}_{\mathbf{r},\nu,+}(\h\rho_{\mathrm{ph}})&=(1-\hat{\mathcal{N}}_{\mathbf{r},\nu,+})\h\rho_{\mathrm{ph}}(1-\hat{\mathcal{N}}_{\mathbf{r},\nu,+})+\bar{n}_{\mr{a}}\hat{\mathcal{N}}_{\mathbf{r},\nu,+}\h\rho_{\mathrm{ph}}\hat{\mathcal{N}}_{\mathbf{r},\nu,+}+(1-\bar{n}_{\mr{a}})\hat{\chi}_{\mathbf{r},\nu,+}\h\rho_{\mathrm{ph}}\hat{\chi}_{\mathbf{r},\nu,+}^\dagger.\label{eq:effective depletion channel}
\end{align}
\end{widetext}
Here, the local particle density $\bar{n}_{\mr{a}}$ in the Markovian-approximated ancillary system is in the range $\bar{n}_{\mr{a}}\in[0,1]$. The set of channels $\{\tilde{\mathcal{E}}_{\mathbf{r},\nu,-}\}_{\mbf{r}\in\mathbb{Z}^2,\nu=A,B}$ generate processes that fill the lower band of the parent Hamiltonian $\hat{\mathcal{H}}_{\mathrm{CI}}$, while the set of channels $\{\tilde{\mathcal{E}}_{\mathbf{r},\nu,+}\}_{\mbf{r}\in\mathbb{Z}^2,\nu=A,B}$ deplete the upper band. We readily verify that the target density matrix $\ket{\mathrm{CI}}\!\rangle\langle\!\bra{\mathrm{CI}}$ remains invariant under these channels.

To understand how the physical system's average density matrix $\h \rho_{\rm ph}$ converges to $\ket{\mathrm{CI}}\!\rangle\langle\!\bra{\mathrm{CI}}$, we approximate the effects of the discrete-time channels $\tilde{\mc{E}}_{\bfr,\nu,\pm}$ 
by a continuous-time Lindbladian description by assigning infinitesimal probabilities $dt$ to each measurement-feedforward operation~\cite{paz1998continuous,ahn2003quantum,oreshkov2013continuoustime,ippoliti2015perturbative}. This yields an effective Lindblad generator for each cycle:
\begin{equation}
    \mathcal{L}^{\mathrm{cycle}} = \sum_{\mbf{r}\in\mathbb{Z}^2}\sum_{\nu=A,B}\left[\mathcal{L}_{\mbf{r},\nu}^{\mathrm{gain}} + \mathcal{L}_{\mbf{r},\nu}^{\mathrm{loss}} + \mathcal{L}_{\mbf{r},\nu}^{\mathrm{decoh.}}\right],\label{eq:Lindblad cycle}
\end{equation}
with components explicitly defined as
\begin{align}
&\mathcal{L}_{\mbf{r},\nu}^{\mathrm{gain}} = \bar{n}_\mathrm{a}\mathcal{D}[\hat{\chi}_{\bfr,\nu,-}^\dagger]\label{eq:lindblad gain}, \\ 
&\mathcal{L}_{\mbf{r},\nu}^{\mathrm{loss}} = (1-\bar{n}_\mathrm{a})\mathcal{D}[\hat{\chi}_{\bfr,\nu,+}]\label{eq:lindblad deplete}, \\ 
&\mathcal{L}_{\mbf{r},\nu}^{\mathrm{decoh.}} = (2-\bar{n}_\mathrm{a})\mathcal{D}[\hat{\mathcal{N}}_{\bfr,\nu,-}] + (1+\bar{n}_\mathrm{a})\mathcal{D}[\hat{\mathcal{N}}_{\bfr,\nu,+}]\label{eq:lindblad decoh},
\end{align}
where $\mathcal{D}[\hat{L}](\h \rho)=\hat{L}\h \rho\hat{L}^\dagger - \frac{1}{2}\{\hat{L}^\dagger\hat{L},\h \rho\}$ is the standard dissipator. Formally, the Lindbladian evolution of the average density matrix over time $t$ is given by $e^{t\mc{L}^\mr{cycle}}$. 
Crucially, the continuous parameter $t$ corresponds to the number of cycles in the discrete time formulation of the adaptive protocol.

We analyze how the physical system's density matrix $\hat\rho_{\rm ph}$ converges to $\CI \langle\!\langle {\rm CI}|$ in the effective Lindbladian description through the evolution of the two-point correlators. Despite the non-Gaussian terms present in the Lindbladian~\eqref{eq:Lindblad cycle}, we can obtain a closed-form differential equation that controls the dynamics of the two-point correlator (see App.~\ref{app:lindblad}), allowing us to estimate the convergence timescale toward the Gaussian Chern insulator steady-state. Due to the translational invariance of $\mc{L}^\mr{cycle}$, it is convenient to formulate the Lindbladian in momentum space. Accordingly, we will study the dynamics of the momentum space two-point function, which is given by:
\begin{equation}
    G_{\mu\mu'}(\mathbf{k}) = \langle\hat{c}^\dagger_{\mu}(\mbf{k})\hat{c}_{\mu'}(\mbf{k})\rangle
\end{equation}
with mode orbital indices $\mu,\mu'=1,2$, and where the expectation value $\langle\dots\rangle$ is taken with respect to the physical system's density matrix $\h\rho_{\mr{ph}}$.

While the non-Gaussian terms limit complete integrability of the two-point function dynamics, we can analytically determine the convergence time scale of the lower and the upper band \textit{occupation densities}. We define the occupation densities via
\begin{equation}
 \bar{g}_{\pm}(t)
= \int \frac{d^2\mathbf{k}}{(2\pi)^2} \bra{\psi_\pm(\mbf{k})}G^*(\mbf{k})\ket{\psi_{\pm}(\mbf{k})}
\end{equation}
where $\ket{\psi_{\pm}(\mbf{k})}$ are upper $(+)$ and lower $(-)$ band Bloch states at momentum $\mbf{k}$, and $(\dots)^*$ denotes complex conjugation. We show in App.~\ref{app:lindblad} that the occupation densities converge to their steady states $\bar{g}_{+}(t\to\infty)\to0$, $\bar{g}_{-}(t\to\infty)\to1$ exponentially, with a convergence time $T_{\mr{conv.}}$ determined by
\begin{equation}
    1/T_{\mr{conv.}} = \underset{n,\mbf{k}}{\mr{min}} \sum_{\nu=A,B} \frac{1}{Z_{\nu,n}}\tau_{\nu}^\dagger P_{n}(\mbf{k})\tau_\nu\label{eq: main text convg timescale}
\end{equation}
where $Z_{\nu,n}\equiv\int\frac{d^2\mbf{k}}{(2\pi)^2}\tau_{\nu}^\dagger P_{n}(\mbf{k})\tau_\nu$ is a normalization factor.
While $\tau_{\nu}^\dagger P_{n}(\mbf{k})\tau_\nu/Z_{\nu,n}$ can vanish at certain $\mbf{k}$'s for a given $\nu$ due to nontrivial band topology, these zeroes occur at distinct momenta for different orbital choices $\nu$---see App.~\ref{app: topological obstructions} for a detailed discussion. As a result, the sum $\sum_{\nu=A,B} \tau_{\nu}^\dagger P_{n}(\mbf{k})\tau_\nu/Z_{\nu,n}$ remains nonzero for all $\mathbf{k}$. Consequently, the convergence time must be finite, i.e., $T_{\mr{conv.}} \sim {\cal O}(1)$. 

The fact that the convergence time remains finite is a concrete reflection of how redundancy in the (overcomplete) set of local jump processes circumvents topological zeros (the zeros of $\tau_{\nu}^\dagger P_{n}(\mbf{k})\tau_\nu/Z_{\nu,n}$ occur at distinct $\mathbf{k}$ for $\nu=A,B$, so the sum over $\nu$ stays strictly positive). 
Topological obstructions in the dissipative preparation of Chern insulators have been noted previously~\cite{budich2015dissipative, goldsteinDissipationinducedTopologicalInsulators2019a}. 
Our use of an overcomplete family of exponentially localized jump operators stabilizes a \textit{pure-state} Chern insulator steady-state with a finite, system-size independent convergence time.

Furthermore, the analysis presented here (and elaborated on in App.~\ref{app:lindblad}) shows that our adaptive circuit can also be viewed as a dissipative topological ground-state preparation protocol with exponentially-localized jump operators, in which the ancillary layer is replaced by a memoryless supply of fresh ancillas, in close analogy with recent ancilla-based dissipative state-preparation protocols~\cite{dingSingleancillaGroundState2024, zhanRapidQuantumGround2026}. 
While the effective Lindbladian description developed here was derived specifically for the $2{+}1$D mEO-class-A adaptive circuit, it is not hard to see that analogous effective Lindbladians can be constructed for adaptive circuits in other mEO classes and in different spatial dimensions.

We caution that in the effective Lindbladian description, the filling of all the lower band OW modes and the depletion of the upper band OW modes effectively occur simultaneously. Therefore, the result $T_{\mr{conv.}} \sim {\cal O}(1)$ should be interpreted as the conclusion that our discrete-time adaptive protocol reaches the steady-state Chern insulator after $T_{\mr{conv.}}\sim\mathcal{O}(1)$ cycles of the measurement-and-feedforward operations on all of the OW modes, as discussed earlier. This Lindbladian description does not address the circuit depth needed to implement each cycle. 

In the finite-range versions of the topological adaptive circuits, each cycle can be implemented in ${\cal O}(1)$ circuit depth. Our numerical simulation in Sec.~\ref{sec:numerics} will show that these circuits steer the physical system toward the topological steady-state ensemble in ${\cal O}(1)$ circuit depth, consistent with our analysis using the approximate Lindbladian description.

\section{Numerical Simulations: Adaptive Topological Dynamics for Chern Insulators and Phase Transitions}
\label{sec:numerics}

\begin{figure*}[t]
    \centering
    \includegraphics[width=\textwidth]{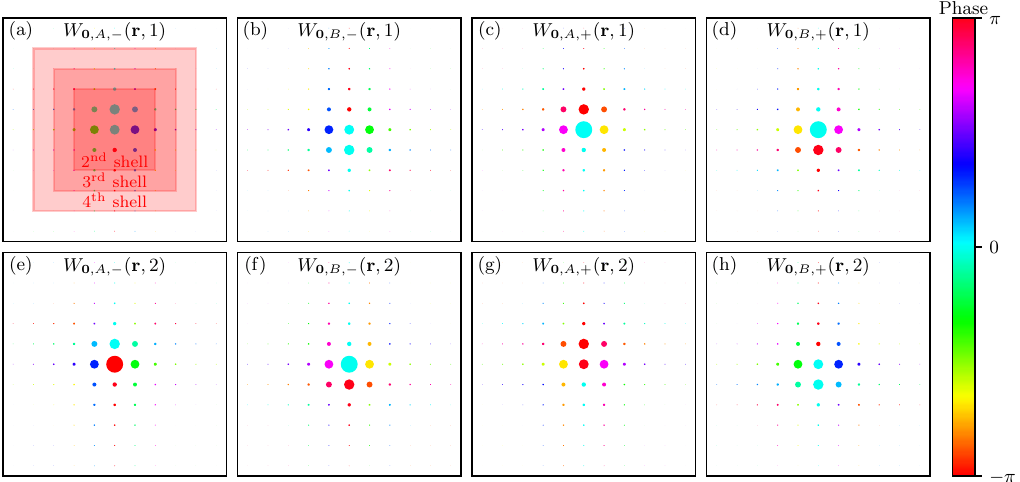}
    \caption{The real-space profile of the four local modes $\hat{\chi}_{\bm{0}, \nu, \pm} = \sum_{\bfr,\mu} W_{\bm{0},\nu,\pm}(\bfr,\mu) c_{\bfr,\mu}$ centered at the origin (from the left to the right column are $\hat{\chi}_{\bm{0}, A, -}, \hat{\chi}_{\bm{0}, B, -}, \hat{\chi}_{\bm{0}, A, +}, \hat{\chi}_{\bm{0}, B, +}$) on sublattice 1 (top) and 2 (bottom), corresponding to the topological phase at $\alpha=1$ in $\h{\mc{H}}_{\rm CI}$ in Eq.~\eqref{eq:toymodel}. The radius of the circle indicates the amplitude of each localized mode, and the color indicates the phase. The top-left panel shows how the finite-range OW functions are defined, where the truncation range is specified by the number of `shells'. These truncated OW functions are what are used for numerical simulation of our adaptive protocol, as discussed in Sec.~\ref{sec:trajectory_resolved}}.
    \label{fig: OW truncation}
\end{figure*}

In the following, we numerically study the finite-range version of the topological adaptive protocol, which steers toward the Chern insulator state $\CI$, as described in Algorithm~\ref{algo:chern}. We characterize its steering ability, convergence timescale, and phase transitions.
In Sec.~\ref{sec:trajectory_resolved}, we examine the steering ability of the adaptive circuit as a function of the tuning parameter $\alpha$ that controls the target-state topology. We also investigate the $\alpha$-tuned topological phase transition between different topological dynamical phases.
In Sec.~\ref{sec:trajectory_resolved}, away from the transitions, we observe that the dynamics of the adaptive protocol converges exponentially fast, reaching a topological state in $\mc{O}(1)$ cycles regardless of the initial state and for various spatial truncations of the OW modes.
This exponential convergence motivates us to study the dynamical topological modes localized on the topological domain wall created by spatially varying $\alpha$. Regarding the purification dynamics, we find that the topological domain-wall mode purifies algebraically slowly, unlike the bulk modes, which purify exponentially fast, as shown in Sec.~\ref{sec:domain_wall}.
Finally, we test the robustness of the adaptive protocol against local perturbations in Sec.~\ref{sec:robustness}, where we add a local unitary perturbation periodically between every cycle. We find that the adaptive protocol is robust against symmetry-invariant noise below a critical noise strength, beyond which the protocol fails to steer the system toward the nontrivial Chern insulator states. Two different transitions in both the trajectory-averaged and trajectory-resolved quantities are found as a function of the noise strength.

\begin{figure*}[ht]
    \centering
    \includegraphics[width=6.8in]{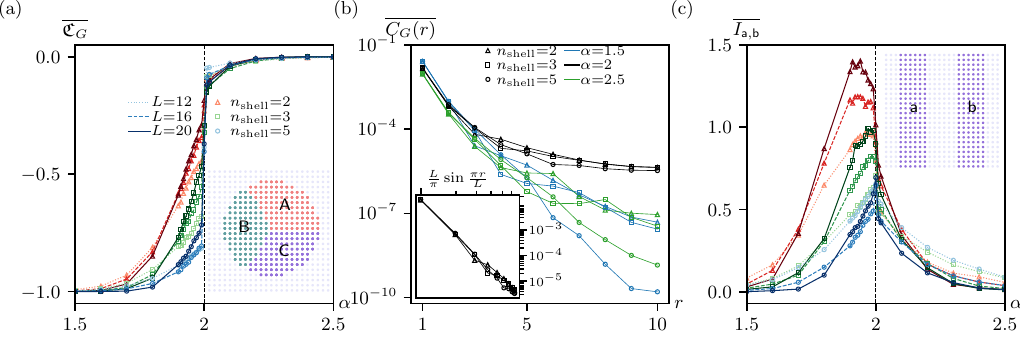}
    \caption{
        (a) Trajectory-resolved real-space Chern number in Eq.~\eqref{eq:Chern} as a function of $\alpha$ on the square lattice for the physical system $\mathcal{P}$ with $L=12$ (dotted lines), $L=16$ (dashed lines), and $L=20$ (solid lines). The localized modes $\hat{\chi}_{\bfr,\nu,\pm}$ in Eq.~\eqref{eq:chi_CI} are truncated to a finite range upto the 2nd shell (red triangles), 3rd shell (green squares), and 5th shell (blue circles). The ensemble size is 100 with self-averaging over different partitions of the triple regions $\mathsf{A}, \mathsf{B}, \mathsf{C}$, as indicated in the inset.
        (b) Averaged two-point correlation function $\overline{C_{G}(r)}$ in Eq.~\eqref{eq:Cr} in an $L=20$ square lattice in the topological phase ($\alpha=1.5$, blue lines), at critical point ($\alpha=2$, black lines), and trivial phase ($\alpha=2.5$, green lines) with localized modes $\hat{\chi}_{\bfr,\nu,\pm}$ truncated upto the 2nd shell (triangles), 3rd shell (squares), and 5th shell (circles).
        The inset shows the correlation function at $\alpha=2$ with an algebraic decay as a function of the chord distance $\frac{L}{\pi}\sin(\frac{\pi r}{L})$ in the log-log scale.
        (c) Averaged bipartite mutual information $\overline{I_{\mathsf{a},\mathsf{b}}}$ between two strip regions $\mathsf{a}$ and $\mathsf{b}$ (see inset) as a function of $\alpha$ in a square lattice of $L=12$ (dotted lines), $L=16$ (dashed lines), and $L=20$ (solid lines) with the localized mode truncated to a finite range of the 2nd shell (red triangles), 3rd shell (green squares), and 5th shell (blue circles).
    }
    \label{fig:trajectory_resolved_alpha}
\end{figure*}

\begin{figure}[ht]
    \centering
    \includegraphics[width=3.4in]{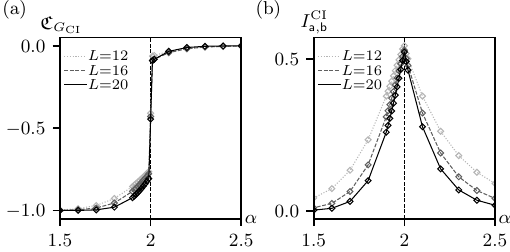}
    \caption{
        (a) Real-space Chern number in Eq.~\eqref{eq:Chern} $\mathfrak{C}_{G_{\text{CI}}}$ of the ground state $\CI$ of the Hamiltonian $\h{\mc{H}}_{\rm CI}$ in Eq.~\eqref{eq:toymodel} for $L=12$ (dotted lines), $L=16$ (dashed lines) and $L=20$ (solids lines).   
        (b) Ideal bipartite mutual information $I_{\mathsf{a},\mathsf{b}}^{\text{CI}}$ of $\CI$ between two regions $\mathsf{a}$ and $\mathsf{b}$ (see the inset in Fig.~\ref{fig:trajectory_resolved_alpha}(c)).
       }
    \label{fig:chern_I2_raw}
\end{figure}

\subsection{Dynamical Topological Phases and Transitions}
\label{sec:trajectory_resolved}
In the following, we present a numerical study of the topological adaptive circuit introduced in Sec.~\ref{sec: adaptive-dynamics-chern-insulators}. This numerical investigation focuses on the \textit{finite-range} version of the topological adaptive circuit (the adaptive circuit is shown in Fig.~\ref{fig: schematic}, while the finite-range OW functions illustrated in Fig.~\ref{fig: OW truncation}). By tuning the parameter $\alpha$ [see Eq.~\eqref{eq:toymodel}], we obtain a phase diagram of distinct area-law-entangled topological dynamical phases and dynamical topological phase transitions. 

To investigate the phase diagram, we study both ``trajectory-resolved" and ``trajectory-averaged" quantities. Trajectory-resolved quantities refer to quantities that require the resolution of individual quantum trajectories. They are typically the averages of nonlinear functions, such as the real-space Chern number (see Eq.~\eqref{eq:Chern}) and entanglement entropy, of the density matrices in individual quantum trajectories. In contrast, trajectory-averaged quantities refer to quantities that can be directly
obtained using the average density matrix over all trajectories. The resolution of individual quantum trajectories is not required for these quantities.

We start with a physical system on a square lattice $\mathcal{P}$ of size $L\times L$ with periodic boundary conditions, with two complex fermions per site, as shown in the red dots in Fig.~\ref{fig: schematic}.
The corresponding ancillary system, {effectively acting as a bath defined on a square lattice $\mathcal{A}$ with the same size}, is shown in the blue dots.
We initialize both systems in a random product state labeled by real-space charge distribution, and evolve the system under the protocol in Algorithm~\ref{algo:chern} for $L$ cycles.

To ensure the circuit consists of only finite-range operations, we truncate the OW fermionic mode $\hat{\chi}_{\bfr, \nu,\pm}$ up to different ranges parameterized by $n_{\text{shell}}$, as illustrated in Fig.~\ref{fig: OW truncation}.
Here, $n_{\text{shell}}$ means that the truncated OW mode $\hat{\chi}_{\bfr, \nu,\pm}$ is supported on $(2n_{\text{shell}}+1) \times (2n_{\text{shell}}+1)$ sites. Our finite-range adaptive circuit is built from the truncated (normalized) OW modes and their associated number operators. As we later show in Sec.~\ref{sec:dynamics}, these finite-range topological adaptive circuits can steer the physical system toward a topological steady-state ensemble in ${\cal O}(1)$ circuit depth for a range of $n_{\text{shell}}$ values.

We first focus on the trajectory-resolved quantities, which can help us understand the properties in individual quantum trajectories. To verify the steering ability of the adaptive protocol, we compute the real-space Chern number $\mathfrak{C}$ for the Gaussian state $\h\rho$ appearing in a quantum trajectory from its correlation matrix~\cite{kitaev2008anyons,fan2023generalized} as 
\begin{equation}\label{eq:Chern}
    \mathfrak{C}_G \equiv 12 \pi \ii  \left[ \tr(G P_{\mathsf{A}}G P_{\mathsf{B}}G P_{\mathsf{C}}) - \tr(G P_{\mathsf{C}}G P_{\mathsf{B}}G P_{\mathsf{A}})  \right]
\end{equation}
where $P_{\mathsf{A},\mathsf{B},\mathsf{C}}$ are the spatial projectors onto the three disjoint regions $\mathsf{A}$, $\mathsf{B}$, and $\mathsf{C}$ on the physical system, as shown in the inset of Fig.~\ref{fig:trajectory_resolved_alpha}(a), and 
\begin{equation}\label{eq:correlation_matrix}
    [G]_{(\bfr,\mu), (\bfr',\mu')} \equiv \tr(\h\rho c_{\bfr,\mu}^\dagger c_{\bfr',\mu'}),
\end{equation} 
is the two-point correlation matrix of the Gaussian state $\h\rho$ within the physical system $\mathcal{P}$~\footnote{We ensure the ancillary system $\mathcal{A}$ is always decoupled from the physical system $\mathcal{P}$ after each cycle by the strong projective local charge measurements $\hat{d}_{\bfr,\nu}^\dagger \hat{d}_{\bfr,\nu}$, enforcing a trivial state in the ancillary system. Therefore, even including the ancillary system in the partition does not change the putative topology at all.}.
Ideally, we require the length scale of the triple-region, $\mathsf{A} \cup \mathsf{B} \cup \mathsf{C} $, to be larger than the correlation length but smaller than the system size to ensure the quantization of the Chern number.
Numerically, we choose the radius of the triple-region to be $0.4L$, which ensures the stability of the estimated Chern number $\mathcal{C}$ away from the critical point, namely $\alpha$ is away from 0 and $\pm 2 $. 
To save numerical resources, for each state $\h\rho$ from a specific trajectory, we average over different ways of partitioning the triple-regions by sweeping their intersection point over the entire lattice $\mathcal{P}$ for physical systems.

By averaging the Chern number over different trajectories, we obtained the trajectory-resolved Chern number $\overline{\mathfrak{C}_G}$, shown in Fig.~\ref{fig:trajectory_resolved_alpha}(a) as a function of $\alpha$ for various system sizes $L=$12, 16, and 20 (dotted, dashed, and solid lines). Here, the notation $\overline{(\dots)}$ denotes the average over trajectories properly weighted by the Born-rule probabilities.
We find a dynamical topological phase transition at $\alpha=2$, separating the two area-law-entangled dynamical phases whose steady-state ensembles possess distinct topologies. We clarify that the numerical study here is restricted to the parameter range $\alpha\in (1.5,2.5)$. When $\alpha$ is smaller than and away from 2, the adaptive circuit generates a topological steady-state ensemble that exhibits an average Chern number $\overline{\mathfrak{C}_G}$ that converges to $-1$, while $\overline{\mathfrak{C}_G}$ vanishes in the topologically-trivial dynamical phase for $\alpha >2$.
The topological phase transition at $\alpha=2$, indicated by the jump of $\overline{\mathfrak{C}_G}$, is robust against the truncation of the localized mode $\hat{\chi}_{\bfr,\nu,\pm}$, even the truncation is only upto the 2nd shell (triangles in Fig.~\ref{fig:trajectory_resolved_alpha}(a)), i.e., $n_{\text{shell}}=2$.
Near the critical point $\alpha=2$, $\overline{\mathfrak{C}_G}$ becomes non-quantized due to the diverging correlation length that exceeds the size of $\mathsf{A}\cup \mathsf{B}\cup \mathsf{C}$.
Therefore, we do not observe a single-parameter scaling behavior of the Chern number. 

The topological phase transition induced by $\alpha$, the parameter that controls the target state (before truncation), is also accompanied by a change in the squared two-point correlation function
\begin{equation}\label{eq:Cr}
    \begin{split}
        C_{G}(r) &\equiv \frac{1}{2L^2}\sum_{\mu,\mu',\bfr',i=\left\{ x,y \right\}} \Big|\tr(\h\rho {c_{\bfr',\mu}^\dag c_{\bfr'+r \bf{e}_{i},\mu'}})\Big|^2 \\
        &= \frac{1}{2L^2}\sum_{\mu,\mu',\bfr',i=\left\{ x,y \right\}}  \big| \left[G \right]_{(\bfr',\mu),(\bfr'+r \bf{e}_{i},\mu')}\big|^2.
    \end{split}
\end{equation} 
We take the squared modulus to remove random phases. The average over $\bf r'$ and the directions $\bf{e}_{i}$ is a technical step for saving numerical resources.
After averaging over the trajectories, we obtain the trajectory-resolved second moment of the two-point correlation function $\overline{C_{G}(r)}$, which exponentially decays inside the topological phase at $\alpha=1.5$ in blue lines, and the trivial phase at $\alpha=2.5$ in green lines, respectively, as shown in Fig.~\ref{fig:trajectory_resolved_alpha}(b). 
This exponential decay is robust against the truncation of the localized mode.
At $\alpha=2$, it remains a long-range algebraic decay, manifesting a characteristic feature of the criticality in the continuous topological phase transition.

In Fig.~\ref{fig:trajectory_resolved_alpha}(c), we show the trajectory-resolved bipartite mutual information $\overline{I_{\mathsf{a},\mathsf{b}}} $ between two strips
$\mathsf{a}$ and $\mathsf{b}$ of widths both being $L/4$
in the physical system $\mathcal{P}$ as sketched in cyan in the inset (here $I_{\mathsf{a},\mathsf{b}} = S_{\mathsf{a}} + S_{\mathsf{b}} - S_{\mathsf{a} \cup \mathsf{b}}$, where $S_{\mathsf{a}}, S_{\mathsf{b}}$, and $S_{\mathsf{a} \cup \mathsf{b}}$ are the von Neumann entanglement entropies for the subsystem $\mathsf{a}, \mathsf{b}$, and $\mathsf{a}\cup \mathsf{b}$, respectively). 
We find that the mutual information asymptotically vanishes in both the topological phase and the trivial phase, implying area-law-entangled phases in a 2d lattice for $\alpha$ away from 2.
At the $\alpha=2$, the mutual information manifests a singular jump decreasing as the system size and $n_{\text{shell}}$ increase.

This steady-state ensemble generated by our $\alpha$-tuned, finite-range, adaptive circuit protocol
shares some similarities with the quantum state in conventional topological quantum phase transitions in equilibrium systems.
In Fig.~\ref{fig:chern_I2_raw}, we show the Chern number $\mathfrak{C}_{G_{\text{CI}}}$, and the bipartite mutual information $I_{\mathsf{a},\mathsf{b}}^{\text{CI}}$, corresponding to the ideal ground state $\CI$ of the Hamiltonian $\h{\mc{H}}_{\rm CI}$ in Eq.~\eqref{eq:toymodel} with the same triple-region partitioning as in Fig.~\ref{fig:trajectory_resolved_alpha}(a) and strip regions in Fig.~\ref{fig:trajectory_resolved_alpha}(c).
We find that the real-space Chern number $\mathfrak{C}_{G_{\text{CI}}}$ in the equilibrium system does show a jump at $\alpha=2$, qualitatively consistent with the dynamical system as in Fig.~\ref{fig:trajectory_resolved_alpha}(a).
For the bipartite mutual information $I_{\mathsf{a},\mathsf{b}}^{\text{CI}}$, both area-law entangled phase shows a vanishing $I_{\mathsf{a},\mathsf{b}}^{\text{CI}}$, and the critical point shows a peak around 0.5, consistent with the dynamical system (see Fig.~\ref{fig:trajectory_resolved_alpha}(c)) in the thermodynamic limit where $L\rightarrow \infty$ and $n_{\text{shell}}\rightarrow \infty$.

\begin{figure}[ht]
    \centering
    \includegraphics[width=3.4in]{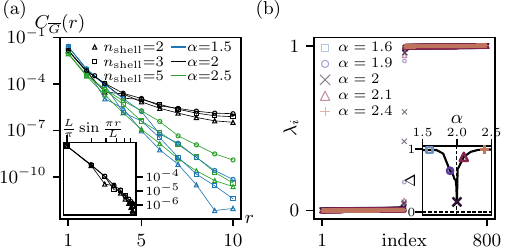}
    \caption{
        (a) Two-point correlation function $C_{\overline{G}}(r)$ associated with the trajectory-averaged $\mathfrak{C}_{ \overline{G}}$ in an $L=20$ square lattice in the topological phase ($\alpha=1.5$, blue lines), at critical point ($\alpha=2$, black lines), and trivial phase ($\alpha=2.5$, green lines) with localized modes $\hat{\chi}_{\bfr,\nu,\pm}$ truncated upto the 2nd shell (triangles), 3rd shell (squares), and 5th shell (circles).
        The inset shows the correlation function at $\alpha=2$ with an algebraic decay as a function of the chord distance $\frac{L}{\pi}\sin(\frac{\pi r}{L})$ in the log-log scale.
        (b) Eigenvalue spectrum $\lambda_i$ of the trajectory-averaged correlation matrix $\overline{G}$ for the physical system $\mathcal{P}$ under various $\alpha$ = 1.6 (squares), 1.9 (circles), 2 (crosses), 2.1 (triangles), and 2.4 (plus sign). 
        Inset shows the spectral gap $\Delta$ (Eq.~\eqref{eq:spectral_gap}) as a function of $\alpha $ using the same color and marker, with dashed lines indicating the vanishing of the spectral gap around $\alpha=2$.
    }
    \label{fig:trajectory_average_alpha}
\end{figure}
As we tune the target state by the parameter $\alpha$, there is also a singular change in the trajectory-averaged density matrix of the system, which should be responsible for some of the singular features shown in Fig.~\ref{fig:trajectory_resolved_alpha}. This singular change in the trajectory-averaged density matrix is manifested by
a change in the spatial scaling of the trajectory-averaged correlation matrix $C_{\overline{G}}(r)$ defined as
\begin{equation}\label{eq:Cr2}
        C_{\overline{G}}(r) \equiv  \frac{1}{2L^2}\sum_{\mu,\mu',\bfr',i=\left\{ x,y \right\}}  \big| \left[\overline{G} \right]_{(\bfr',\mu),(\bfr'+r \bf{e}_{i},\mu')}\big|^2,
\end{equation} 
as shown in Fig.~\ref{fig:trajectory_average_alpha}(a). $\overline{G}$ is the trajectory-averaged correlation matrix.
Here, we again observe that long-range algebraic decay emerges around $\alpha=2$, implying that most of the contribution of the criticality is already contained in the averaged correlation matrix $\overline{G}$.

At this stage, our numerical evidence demonstrates the existence of two topologically distinctive phases. Regarding the nature of the phase transitions, there are still two possible scenarios: (1) the trajectory-averaged and trajectory-resolved quantities both experience a transition at the same value $\alpha =2$ and (2)  trajectory-resolved quantities experience a different transition at a smaller $\alpha$ in addition to the singularities at $\alpha = 2$ caused by the trajectory-averaged transition. In previously investigated non-topological adaptive dynamics, the trajectory-averaged and trajectory-resolved quantities are often found to undergo different transitions \cite{KhemaniAbsorbing,Iadoecola2023,XiaoChenFeedback,PiroliTriviality,XhekControlling}, suggesting the possibility of scenario (2) in our setting. However, with the current numerical precision, we cannot yet conclusively determine which scenario occurs. Further numerical investigation will be conducted in future work to distinguish these two scenarios.

Meanwhile, we do have an example where the trajectory-averaged and trajectory-resolved transitions split once we include the random unitaries, as discussed in Sec.~\ref{sec:robustness}.

Another interesting feature of our adaptive circuit is the similarity in magnitude between the trajectory-resolved and trajectory-averaged correlation matrices, implying a small degree of trajectory fluctuation.
This can be probed by a spectral gap $\Delta$ defined the eigenvalue spectrum $\{\lambda_i \}$ of $\overline{G}$ in the physical system $\mathcal{P}$~\footnote{Since the ancillary system is always decoupled from the physical system after every complete cycle, and is in a maximally mixed state due to random measurement outcomes, all eigenvalues are trivially 0.},
\begin{equation}\label{eq:spectral_gap}
    \Delta\equiv \min_{{\lambda_i}>0.5} {\lambda_i} - \max_{{\lambda_i}<0.5} {\lambda_i} .
\end{equation}
For a pure Gaussian state, the eigenvalue of the correlation matrix must be either 0 or 1. For a mixed or non-Gaussian state, there are eigenvalues in between, i.e., ${\lambda_i}\in (0,1)$. 
Therefore, this spectral gap quantifies how close the system is compared to a pure Gaussian state. 
Figure~\ref{fig:trajectory_average_alpha}(b) shows that the final steady-state ensemble is very close to a single pure Gaussian state when $\alpha$ is away from 2, implying a very small trajectory-to-trajectory fluctuation. This is because the adaptive circuit effectively steers most trajectories toward the same final state. Intuitively, we can think of the steady-state ensemble as the ideal target state with different disorders caused by the truncation of the OW wave function. Around $\alpha=2$, the spectral gap closes, and more disorder appears in the ensemble, eventually causing a phase transition.

\subsection{Dynamics of Converging to the Target State}\label{sec:dynamics}
\begin{figure}[ht]
    \centering
    \includegraphics[width=2.2in]{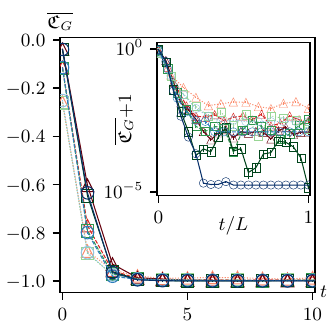} 
    \caption{
        Dynamics of the trajectory-resolved Chern number converging to a target topological state at $\alpha=1.5$ averaged over 100 different trajectories.
        The system sizes are $L=12$ (dotted), $L=16$ (dashed), and $L=20$ (solid), with localized modes truncated up to the 2nd shell (red triangles), 3rd shell (green squares), and 5th shell (blue circles).
        The inset shows the trajectory-resolved Chern number offset by one on the log scale.
    }
    \label{fig:dynamics}
\end{figure}

We also study the dynamics of the adaptive protocol to reach the steady state for various system sizes (dotted, dashed, and solid lines for $L=$ 12, 16, and 20) and various truncation ranges (triangles, squares, and circles for truncation up to the second shell, third shell, and fifth shell), as shown in Fig.~\ref{fig:dynamics}. We confirm that the average trajectory-resolved Chern number $\overline{\mathfrak{C}_G}$ reaches the steady state with Chern number being $-1$ in $\mc{O}(1)$ cycles for the topological phase at $\alpha=1.5$, for all truncations $n_{\rm shell}=2,3,5$ of the localized mode $\hat{\chi}_{\bfr,\nu,\pm}$. Furthermore, it approaches the putative topological phase with the Chern number $-1$ exponentially in the early stage, as shown in the inset.

\subsection{Dynamical Topological Domain-Wall Modes}\label{sec:domain_wall}
We have established that the (finite-range) topological adaptive circuit can be tuned between different topological dynamical phases in 2+1 spacetime dimensions using the parameter $\alpha$. We now introduce a 1+1d spacetime domain wall between different topological dynamical phases and study the dynamical domain-wall modes. This study generalizes the investigation of dynamical topological modes on a 0+1d domain wall between 1+1d topological dynamical phases in Ref.~\cite{pan2025topological}. 

\begin{figure}[ht]
  \centering
   \includegraphics[width=3.4in]{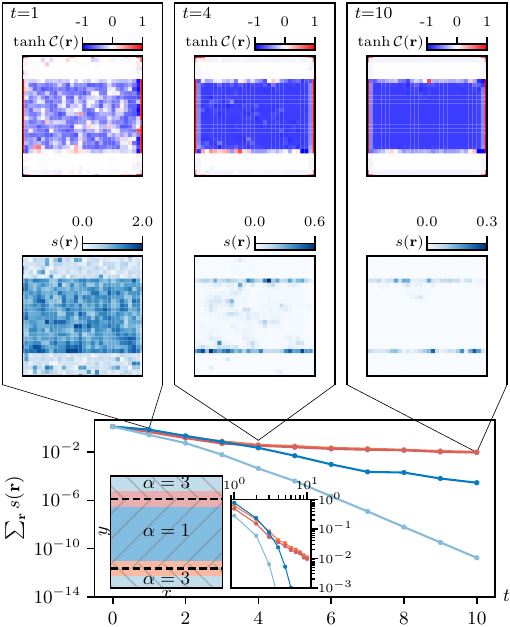}
    \caption{
        Topological edge modes induced by a line-defect domain wall in a 2d square lattice of $L=30$ with the periodic boundary condition for a single trajectory, separating the topological phase ({dark blue} region in the left inset) and trivial phase ({light blue} region) parameterized by $\alpha=1$ and $\alpha=3$, respectively.
        The bottom panel shows the dynamics of the integrated entanglement contours (dark blue line) integrated in the topological region  (dark blue region with $\alpha=1$), trivial region (light blue line and light blue region with $\alpha=3$), and the domain wall region (orange line and orange region) in the semi-log scale, with the smaller inset showing the log-log scale.
        The top row shows the snapshots of the local Chern marker $\mathcal{C}(\bfr)$ [Eq.~\eqref{eq:chern_marker}] (rescaled by $\tanh$ for visibility) and the middle row shows the entanglement contour $s(\bfr)$ [Eq.~\eqref{eq: EntanglementContour}] for the physical system $\mathcal{P}$ at three typical time steps, $t=$1, 4, and 10 (from left to right).
    }
    \label{fig: domain walls}
\end{figure}

Concretely, we create the domain wall by considering the topological adaptive circuit with a spatially varying $\alpha$ as shown in the inset of Fig.~\ref{fig: domain walls}. We choose $\alpha = 1$ in the darker blue region for the topologically nontrivial dynamics (with steady-state Chern number $-1$) and $\alpha=3$ for the topologically trivial dynamics (with steady-state Chern number 0) in the light blue region on a 2d square lattice of $L=30$ with periodic boundary conditions. The dynamical domain wall lies along the horizontal dashed lines.

To detect the topological edge mode, we introduce two metrics, the local Chern marker~\cite{bianco2011mapping,kitaev2008anyons} and the entanglement contour~\cite{chen2014entanglement}, which can be considered as the real-space-resolved Chern number and real-space-resolved entanglement entropy. Note that these quantities are specialized to Gaussian states. The definitions of these quantities are as follows.

The local Chern marker is computed for a pure Gaussian state described by the correlation matrix $G$ as~\cite{bianco2011mapping,kitaev2008anyons}
\begin{equation}\label{eq:chern_marker}
    \mathcal{C}(\bfr)\equiv2\pi \ii \sum_{\mu}\left[ G X G Y G - G Y G X G\right]_{\left( \bfr,\mu \right),\left( \bfr,\mu \right)},
\end{equation}
where $X$ and $Y$ are position operators in the $x$ and $y$ directions, i.e., $\left[  X \right]_{\left( \bfr,\mu \right), \left( \bfr',\mu' \right) }=x\delta_{\bfr,\bfr'}\delta_{\mu,\mu'} $ and $\left[  Y \right]_{\left( \bfr,\mu \right), \left( \bfr',\mu' \right) }=y\delta_{\bfr,\bfr'}\delta_{\mu,\mu'} $ with $\bfr=\left( x,y \right)$. This local Chern marker tests the non-commutativity of the two spatial directions, and is nonzero only in a topological state. Therefore, it can be used to directly detect the topology in the bulk of the physical system $\mathcal{P}$.

The entanglement contour~\cite{chen2014entanglement} provides the spatial distribution of entanglement that contributes to the total entanglement entropy, defined as
\begin{equation}
\begin{split}
    s(\bfr) &\equiv -\sum_{\mu} \big[ G_{\mathcal{P}\cup\mathcal{A}}\log G_{\mathcal{P}\cup\mathcal{A}} \\
    &+  \left(\mathbb{1}-G_{\mathcal{P}\cup\mathcal{A}}\right)\log \left(\mathbb{1}-G_{\mathcal{P}\cup\mathcal{A}}\right) \big]_{(\bfr,\mu),(\bfr,\mu)},
\end{split}
    \label{eq: EntanglementContour}
\end{equation}
which can be used to visualize the spatial position of the emergent topological modes. Note that the entanglement contour $s(\bfr)$ spatially resolves the von Neumann entanglement entropy between the region where $\bfr$ is defined (i.e., $\mathcal{P}\cup\mathcal{A}$) and its complement.
Here, we introduce an additional ``reference system'' with the same number of sites $\mathcal{R}$ customary for studying the purification dynamics~\cite{gullans2020dynamical}. 

We initialize the reference system $\mathcal{R}$ to be maximally entangled with the physical and ancillary system $\mathcal{P}\cup\mathcal{A}$, and then track the time scale of the purification process at each site via $s(\bfr)$. The expectation is that the degrees of freedom away from the domain wall should purify quickly, as they reside in the bulk of the area-law-entangled topological dynamics. In contrast, the degrees of freedom on the domain wall are expected to exhibit distinct purification dynamics, indicating the existence of dynamical topological domain-wall modes. This purification behavior has been demonstrated for the 0+1d domain walls between 1+1d topological dynamical phases in Ref.~\cite{pan2025topological}. Here, we extend it to 2+1d spacetime dimensions.

We choose a single trajectory of the adaptive protocol as shown in Fig.~\ref{fig: domain walls}, and present the local Chern marker $\mathcal{C}(\bfr)$ (top row) and entanglement contour $s(\bfr)$ (middle row)  in the physical system $\mathcal{P}$ at three typical snapshots at $t=$1, 4, and 10.~\footnote{For the ancillary system, the local Chern marker simply vanishes everywhere and the entanglement contour immediately vanishes after the first cycle.}
The three curves are the entanglement contour (dark blue) integrated over the topological phase $\alpha=1$ (the dark blue region in the inset), integrated over the trivial region $\alpha=3$ (light blue regions), and integrated over the crossover regions (orange stripes), respectively.

In the very early stage $t=1$, the local Chern marker already shows a topological bulk region with a nonzero Chern marker, whose support spatially overlaps with the topological region in the bulk, as seen in the dark blue region between the two dashed lines in the inset of Fig.~\ref{fig: domain walls}. After a few cycles $t=4$, the local Chern marker saturates in the bulk topological region (blue), and remains zero in the trivial region (white)~\footnote{The positive contour near the edge of $x=1$ and $x=30$ is because the position operator $X$ and $Y$ is ill-defined within the periodic boundary condition.}.
At the same time, the entanglement contour develops a peak at the boundary between the topologically different regions, suggesting the presence of nontrivial modes localized at the boundary (see middle row of Fig.~\ref{fig: domain walls}). The integrated entanglement contour around the boundary decays algebraically (orange lines in the bottom panel of Fig.~\ref{fig: domain walls}), much slower than the exponential decay of the entanglement contour in both the bulk region for the trivial phase (light blue lines) and topological phase (dark blue lines). 

Our numerical results provide strong evidence for the existence of nontrivial dynamical modes localized on the topological domain wall. Conceptually, we expect these dynamical domain-wall modes to be a non-equilibrium version of the chiral modes between two different Chern insulators. A detailed theoretical characterization of these modes is left for future investigation. 

\subsection{Robustness of Topological Adaptive  Circuit and Noise-Induced Transitions}\label{sec:robustness}

Finally, we demonstrate the robustness of the adaptive protocol in steering the dynamics to the target topological state against local coherent noise.

Here, we model the local coherent noise as random charge-preserving U(1) unitaries on each site $(\bfr,\mu)$, which are periodically added between each cycle of the adaptive protocol, i.e.,
\begin{equation}\label{eq:perturbation}
    \prod_{\bfr,\mu}U(\theta_{\bfr,\mu}) \equiv \prod_{\bfr,\mu} e^{4\pi\ii \theta_{\bfr,\mu}  \left( \hat{c}_{\bfr,\mu}^\dagger \hat{c}_{\bfr,\mu} -\frac{1}{2}\right)},
\end{equation}
where $\theta_{\bfr,\mu}$ is a random parameter independently sampled from a uniform distribution in $[0,\sigma)$, and $\sigma\in\left[ 0,1 \right]$ is a site-independent global tuning parameter controlling the strength of the perturbation (all previous results can be understood as the limit of $\sigma=0$).

A single application of Eq.~\eqref{eq:perturbation} acting on any state $G$ does not change the calculated Chern number defined in Eq.~\eqref{eq:Chern}, which is manifestly U(1) gauge invariant. However, it perturbs the local profile of the wave function away from the target state. The perturbed state is therefore no longer stabilized by the localized number operators $\hat{\cal N}_{\bfr,\nu,\pm}$ of the OW mode $\hat{\chi}_{\bfr,\nu,\pm}$ centered around different $\bfr$ ($\hat{\chi}_{\bfr,\nu,\pm}$ does not always commute with $U(\theta_{\bfr,\mu})$ because the unitary is only U(1) invariant on locally on each site $\mu$), and thus serves as a perturbation to the adaptive protocol to counteract its steering capability toward the target state.

\begin{figure}[ht]
    \centering
    \includegraphics[width=3.4in]{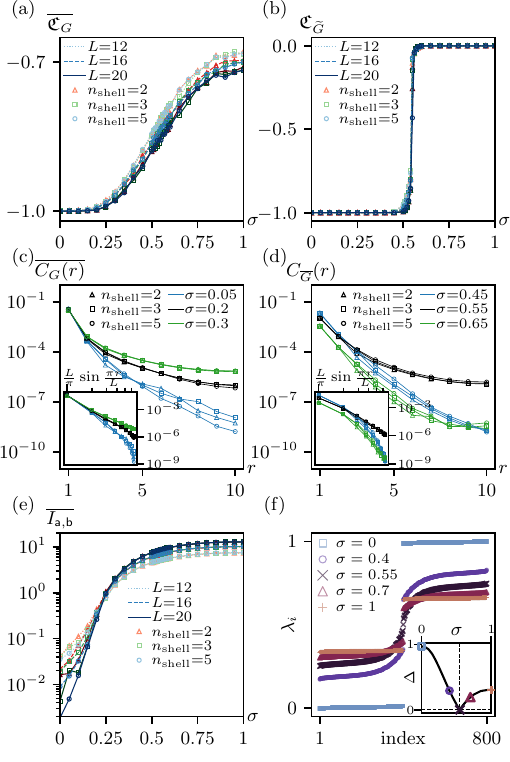}
    \caption{
        (a) Trajectory-resolved Chern number $\overline{\mathfrak{C}_G}$ and (b) regularized trajectory-averaged Chern number $\mathfrak{C}_{\widetilde{G}}$ (see Eq.~\eqref{eq:regularized_correlation_matrix}) as a function of the perturbation strength $\sigma$ at $\alpha=1$ in a square lattice of $L=12$ (dotted lines), $L=16$ (dashed lines), and $L=20$ (solid lines) with localized modes $\hat{\chi}_{\bfr,\nu,\pm}$ truncated to finite ranges corresponding upto the 2nd (red triangles), 3rd (green squares), and 5th (blue circles) shells.
        (c) Trajectory-resolved correlation function $\overline{C_{G}(r)}$ and (d) regularized trajectory-averaged correlation function $C_{\overline{G}}(r)$ as a function of the perturbation strength $\sigma$ at $\alpha=1$.
        (e) Trajectory-resolved bipartite mutual information $\overline{I_{\mathsf{a},\mathsf{b}}}$.
        (f) Eigenvalue spectrum $\lambda_i$ of the trajectory-averaged correlation matrix $\overline{G}$ for the physical system $\mathcal{P}$ under various $\sigma$ = 0 (squares), 0.4 (circles), 0.55 (crosses), 0.7 (triangles), and 1 (plus sign). 
        Inset shows the spectral gap $\Delta$ (Eq.~\eqref{eq:spectral_gap})  as a function of $\sigma $ using the same color and marker, with a dashed line indicating the vanishing of the spectral gap around $\sigma\approx0.55$.}
    \label{fig:robustness}
\end{figure}

We numerically study the stability of the adaptive protocol by starting with the unperturbed (finite-range) topological adaptive circuit at $\alpha=1$ and then turning on $\sigma$. We present the results in Fig.~\ref{fig:robustness}, where we study the dynamics from two aspects: The trajectory-resolved quantities are shown in the left column (Fig.~\ref{fig:robustness}(a,c,e)). The trajectory-averaged quantities are shown in the right column (Fig.~\ref{fig:robustness}(b,d,f)).

We find that the trajectory-resolved Chern number $\overline{\mathfrak{C}_G}$ remains robust against the perturbation up to $\sigma_{c,1}\approx0.2$, beyond which it starts to deviate from the target Chern number $-1$ as shown in Fig.~\ref{fig:robustness}(a). However, for a regularized trajectory-averaged Chern number $\mathfrak{C}_{\widetilde{G}}$, it remains quantized at $-1 $ even up to $\sigma_{c,2}\approx0.55$, and then jumps to $0$, as shown in Fig.~\ref{fig:robustness}(b). 

This regularized Chern number $\mathfrak{C}_{\widetilde{G}}$ is introduced to characterize whether the trajectory-averaged mixed-state density matrix of the system has the targeted topology and chirality of the (noise-free) adaptive circuit. Technically, $\mathfrak{C}_{\widetilde{G}}$ is directly calculated using the trajectory-averaged two-point functions $\overline{G}$ (which only depends on the the trajectory-averaged mixed-state density matrix). Since the real-space Chern number formula Eq.~\eqref{eq:Chern} was originally introduced for pure Gaussian states whose correlation matrices only have 0 and 1 eigenvalues~\cite{kitaev2008anyons}, we regularize $\overline{G}$, which comes from the trajectory-averaged mixed state, before applying Eq.~\eqref{eq:Chern}. We continuously deform the eigenvalue $\lambda_i$ of the trajectory-averaged $\overline{G}$ to be 0 and 1 as per
\begin{equation}\label{eq:regularized_correlation_matrix}
    \widetilde{G} \equiv \frac{\mathbb{1}}{2} + \frac{1}{2}\Sgn\left( \overline{G}- \frac{\mathbb{1}}{2} \right),
\end{equation}
where $\Sgn$ is the sign function applying to the eigenvalues of $\overline{G}$.
With the regularized trajectory-averaged two-point correlation matrix $\widetilde{G}$, we then compute the regularized Chern number $\mathfrak{C}_{\widetilde{G}}$ following Eq.~\eqref{eq:Chern}. We assume that this regularization is a smooth deformation as long as the spectral gap, as defined in Eq.~\eqref{eq:spectral_gap} is finite, which we will show below. A similar regularization procedure was also used to study the Chern numbers of dissipatively prepared mixed states~\cite{barbarino2020preparing}.

The two distinct critical values of $\sigma$ can be also verified from the trajectory-resolved second-moment correlation function $\overline{C_{G}(r)}$, and from the squared trajectory-averaged correlation function $C_{\overline{G}}(r)$, respectively, as shown in Figs.~\ref{fig:robustness}(c) and (d).
In Fig.~\ref{fig:robustness}(c), averaged trajectory-resolved correlation function $\overline{C_{G}(r)}$ remains exponentially decaying until $\sigma\approx0.2$ (black lines in Fig.~\ref{fig:robustness}(c)), and then becomes algebraically decaying, implying a critical phase for $\sigma\gtrsim0.2$. Such a phase was previously found also in 2+1d monitored free-fermion systems and was shown to exhibit $\sim L\log L$ entanglement scaling~\cite{chahine2024entanglement,MirlinAbove1D}. This phase is also referred to as the ``metallic phase". Therefore, our adaptive circuit undergoes a trajectory-resolved topological-metallic transition around $\sigma_{c,1}\approx0.2$.  

In addition, we study the trajectory-resolved bipartite mutual information $\overline{I_{\mathsf{a},\mathsf{b}}}$ in the same geometry as in Fig.~\ref{fig:trajectory_resolved_alpha}(c) with two strip regions $\mathsf{a}$ and $\mathsf{b}$, as shown in Fig.~\ref{fig:robustness}(e). The crossing in the mutual information provides further evidence for the trajectory-resolved topological-metallic transition at around $\sigma_{c,1}\approx0.2$.

In contrast, for the trajectory-averaged correlation function, $C_{\overline{G}}(r)$, we observe a transition from exponential decay to the algebraic decay at $\sigma_{c,2}\approx0.55$ (black lines in Fig.~\ref{fig:robustness}(d)), beyond which it becomes exponentially decaying again. Both the results on the trajectory-averaged correlation function and the regularized Chern number $\mathfrak{C}_{\widetilde{G}}$ together suggest a trajectory-averaged topological transition at $\sigma_{c,2}\approx0.55$. Notably, this trajectory-averaged transition occurs in the regime where the trajectory-resolved behaviors of the circuit already have entered the ``metallic phase". Note that, as we average over trajectories, the effect of the random coherent errors in Eq. \eqref{eq:perturbation} can be identified with that of the incoherent dephasing channels with respect to the occupation $\hat{c}_{\bfr,\mu}^\dagger \hat{c}_{\bfr,\mu}$ of local orbitals. Hence, the result here provides evidence for the stability of our adaptive circuit under incoherent errors as well.

Finally, we present the spectral gap of the $\sigma$-tuned transition in Fig.~\ref{fig:robustness}(f). We find that almost all eigenvalues ${\lambda_i}$ are either 0 or 1 at $\sigma=0$, implying a pure state. This is expected since most of the trajectories are identical in the absence of random unitary perturbations, consistent with Fig.~\ref{fig:trajectory_average_alpha}(b).
The spectral gap $\Delta$ does not immediately collapse after $\sigma$ turns on. A finite gap is sustained until $\sigma_{c,2}\approx0.55$, implying the robustness of our adaptive Algorithm~\ref{algo:chern} against the local perturbations. 
Around $\sigma_{c,2}\approx0.55$, we find that the spectral gap closes and reopens,
consistent with the trajectory-averaged correlation functions $C_{\overline{G}}(r)$ changing from exponential decay to algebraic, and then back to exponential decay again. This $\sigma$-tuned spectral gap in Fig.~\ref{fig:robustness}(f) is much smaller than the one observed in the $\alpha$-tuned transition, implying a much larger trajectory-to-trajectory fluctuation in $G$ due to the random unitaries than the trajectory-average $\overline{G}$ itself. This differs from the $\alpha$-tuned transition with modest trajectory fluctuations. This large trajectory-to-trajectory fluctuation also corroborates the observation that the magnitude $C_{\overline{G}}(r)$ and $\overline{C}_G(r)$ are very different for the same values of $\sigma$. 

In this numerical study, we observed a clear separation between the trajectory-average and trajectory-resolved transitions, similar to earlier works \cite{KhemaniAbsorbing,Iadoecola2023,XiaoChenFeedback,PiroliTriviality,XhekControlling} on non-topological adaptive circuits. However, it is somewhat surprising that, in our adaptive circuit with coherent noise, the regularized trajectory-averaged Chern number remains even after crossing the trajectory-resolved transition. The trajectory-averaged transition occurs in the regime where the trajectory-resolved behaviors belong to the ``metallic phase". In previously studied non-topological settings, the trajectory-averaged transition typically happens before the trajectory-resolved transition as the strength of noise (or other effects that interfere with the adaptive steering in the circuit) increases. Understanding the mechanism that determines the ordering of these transitions is an interesting direction for future work.

\section{Conclusion and Outlook}
\label{sec:outlook}

\subsection{Summary of Results}
In this work, we have developed a novel symmetry classification framework applicable to both interacting and non-interacting fermionic dynamical systems, as well as a general construction of fermionic topological dynamical systems with measurements in arbitrary spacetime dimensions. We discussed two distinct symmetry classification schemes for fermionic dynamics with measurements: the many-body evolution operator (mEO) symmetry class and the single-particle transfer matrix (sTM) symmetry class. The sTM classification is the previously developed symmetry classification of free-fermion dynamical systems with measurements, based on the single-particle description of the dynamics~\cite{jian2022criticality}. It utilizes the general correspondence between the single-particle spacetime propagation in free-fermion dynamics and the single-particle elastic scattering processes in the Anderson localization problems of disordered fermions, which enables the assignment of the sTM class of dynamical systems according to the standard AZ symmetry class for the (static) disordered fermion system. In contrast, the mEO class provides a new symmetry classification scheme directly based on the time-reversal, particle-hole, and chiral symmetries of the dynamical system. The symmetries are examined at the many-body level. Hence, the so-obtained mEO classification naturally extends to interacting fermionic dynamics. We have rigorously shown that the mEO and sTM classifications are in one-to-one correspondence in the non-interacting limit.

The sTM and mEO classes provide complementary perspectives for a topological classification of free-fermion dynamics, emphasizing its different aspects (Table ~\ref{tab: bott periodicity table}). The sTM framework emphasizes the spacetime fermion propagation in the dynamics and their connections to the single-particle elastic scattering in (disordered) topological insulators or superconductors. This perspective has been previously discussed in Ref.~\cite{pan2025topological,KawabataTopoMonitored}. The mEO framework offers a new perspective, emphasizing the topology of the quantum states generated by the fermion dynamics in the long-time limit. Unlike the sTM-based framework, this framework is not fundamentally limited to non-interacting systems. Remarkably, the two complementary perspectives underpin a novel \textit{dynamical} bulk-boundary correspondence: the topology of area-law-entangled Gaussian dynamics in the spacetime bulk manifests in the topology of the steady-state ensemble, effectively living on the temporal boundary of spacetime.

Guided by this correspondence, we constructed Gaussian adaptive circuits with post-selection-free measurements that realize area-law-entangled topological dynamics in mEO classes $\text{A, AI, BDI, and D}$ (which correspond to sTM classes AIII, BDI, D, and DIII, respectively). These adaptive circuits rely on measurement-and-feedforward operations, which steer the systems toward steady states with nontrivial topology and stabilize them. Hence, in addition to realizing the area-law-entangled topological dynamics, our Gaussian adaptive circuit construction provides a general protocol to efficiently prepare and stabilize \textit{any} free-fermion topological state in \textit{any} spacetime dimension using only local operations.

Indeed, our Gaussian adaptive circuits can be generalized to prepare and stabilize free-fermion topological states in the remaining six symmetry classes, $\text{AIII, DIII, AII, CII, C, CI}$, as well. However, such Gaussian circuits involve operations, especially measurements, that break the symmetries in these six mEO classes. Hence, they should not be treated as a realization of the area-law-entangled topological dynamics in mEO classes. Nevertheless, they are valid protocols for topological state preparation and stabilization in any symmetry class.

In fact, we also prove that any Gaussian measurements in the six mEO classes $\text{AIII, DIII, AII, CII, C, CI}$ require post-selection. Therefore, realizing the area-law-entangled topological dynamical phases classified in Table~\ref{tab: bott periodicity table} in the non-interacting limit also needs post-selection. Interacting realizations of these phases, using post-selection-free and symmetry-respecting non-Gaussian measurements, are interesting possibilities for future exploration. 

To demonstrate our construction for the adaptive circuit, we further studied and simulated a concrete 2+1d realization for mEO class A. This adaptive circuit uses measurements and feedforward operations to steer the system toward a 2d Chern insulator, and thereby realizes the mEO-class-A
area-law-entangled topological dynamics in 2+1d according to the dynamical bulk-boundary correspondence. This adaptive circuit applies repeated cycles of stabilizer measurements to detect local deviations from the target state and feedforward operations to correct the deviations. Each operation is exponentially localized, and the number of operations per cycle scales with the system size. Using an effective Lindbladian approach that captures the evolution of the trajectory-averaged density matrix, we have argued that our protocol can prepare topological states within $\mc{O}(1)$ cycles.

The operations in each cycle are not parallelizable due to the (non-commuting) exponential tails of the measurements, which are needed for targeting a single Chern insulator. Interestingly, when the measurement operations are truncated to be finite-range, the adaptive circuit can still steer the system toward a topological state ensemble, with each state sharing the same topology as the original target state. Such a topological steady-state ensemble can be reached in $\mc{O}(1)$ circuit depth (per cycle), which can be confirmed numerically. 

Finally, we conducted extensive numerical simulations of the finite-range version of our adaptive protocol at both the trajectory-resolved and trajectory-averaged levels and confirmed its expected convergence behavior and required circuit depth. By tuning the parameter that controls the Chern number of the target state, we realized \textit{dynamical topological phase transitions}, which we characterized using the real-space Chern number, mutual information, and correlation function scaling of the steady-state ensemble. Furthermore, by spatially varying the control parameter, we realized a 1+1d dynamical topological domain wall between different 2+1d topological dynamical phases in mEO class A. Dynamical domain-wall modes with slow entanglement dynamics, distinct from those in the bulk, were identified. Conceptually, these modes are expected to be the higher-dimensional generalization of the 0+1d dynamical topological domain-wall modes found in Ref.~\cite{pan2025topological}. 

Lastly, we characterized the robustness of our adaptive protocol to local coherent noise and identified finite noise thresholds beyond which trajectory-resolved and trajectory-averaged quantities undergo distinct phase transitions. Similar separation between the trajectory-resolved and trajectory-averaged transitions has been found in non-topological adaptive circuits~\cite{KhemaniAbsorbing,XiaoChenFeedback,PiroliTriviality,Iadoecola2023,XhekControlling}.

\subsection{Outlook and Future Directions}
Our results open several compelling directions for further study. First, we have shown that the area-law-entangled topological dynamical phases in the mEO classes $\text{AIII, DIII, AII, CII, C, CI}$ must involve post-selection in the non-interacting limit. Searching for the interacting and post-selection-free realization of these topological dynamical phases is a natural direction to pursue. More generally, it is well known that the classification of topological insulators and superconductors in equilibrium changes and becomes richer when interactions are included~\cite{FidkowskiZ8,KapustinFermionCobordism,YouXuInteracting,ChongWangInteracting,QRWang2018,Kapustin2017,cheng2018classification}. Now, equipped with the mEO symmetry classification, we can ask if a similar change and enrichment happen to area-law-entangled topological dynamics in each mEO symmetry class. Studying the general classification of interacting area-law-entangled topological dynamics is a promising future direction. 

Second, our numerics provide evidence for 1+1d dynamical topological domain-wall modes in our mEO-class-A topological adaptive circuit, which appear to be dynamical analogues of the chiral edge modes found at interfaces between topologically distinct Chern insulators in equilibrium. However, their precise nature remains unclear. Are they chiral? Is there an analytic description, akin to low-energy edge theory in equilibrium topological phases, that captures their preparation dynamics? Answering these questions could provide valuable examples for a more systematic investigation of topological defects in dynamical quantum systems.

Regarding the topological transitions between different mEO-class-A topological dynamical phases discussed in Sec.~\ref{sec:trajectory_resolved}, due to finite numerical resolution, we cannot conclusively determine whether the system exhibits a single or two distinct dynamical topological phase transitions in our (coherent noise-free) finite-range topological adaptive circuit. Prior works~\cite{KhemaniAbsorbing,PiroliTriviality,Iadoecola2023,XiaoChenFeedback,XhekControlling} on (non-topological) adaptive circuits have found that trajectory-resolved and trajectory-averaged quantities often exhibit distinct transitions. So, further investigation is needed to clarify this structure in our setting. A more careful examination of the trajectory-averaged behavior and the trajectory-to-trajectory fluctuations will be important for this investigation, which we will pursue in the future. Given that our topological adaptive circuit models can be broadly applied to different dimensions and symmetry classes beyond our mEO-class-A example, we can use them as toy models to generalize our studies to a much broader class of dynamical topological phase transitions.

Regarding the noise-induced transitions in our mEO-class-A topological adaptive circuit (see Sec.~\ref{sec:robustness}), the separation between trajectory-averaged and trajectory-resolved transitions is clear. However, their ordering appears unexpected. In many earlier examples~\cite{KhemaniAbsorbing,PiroliTriviality,Iadoecola2023,XiaoChenFeedback,XhekControlling} where the adaptive steering protocols are interfered with by noise (realized, for example, as random unitaries), the trajectory-averaged transition typically happens before the trajectory-resolved transition as the strength of noise increases. The trajectory-averaged transition is also referred to as the {\it absorbing phase transition}. In our setting, the two transitions occur in the opposite order. It is an interesting future direction to understand the potential mechanism that determines the ordering of these two types of transitions.

In the context of quantum state preparation, the ability to prepare free-fermion topological states on-chip is potentially useful for quantum simulation and for engineering topological boundary/defect physics in a controlled setting. Topological free-fermion phases support symmetry-protected edge and defect modes that are robust to local, symmetry-respecting perturbations, making them natural targets for noise-resilient quantum applications. However, topological boundary and defect modes are often difficult to realize and control in conventional solid-state materials. Our adaptive protocol is thus a stepping stone toward preparing controllable topologically-protected edge/defect modes on programmable quantum hardware. More broadly, many topological phases are challenging or currently impossible to realize in conventional solid-state systems. From this perspective, our adaptive circuit construction has the potential to enable controlled studies of topological phases that are otherwise experimentally inaccessible.

Finally, we briefly comment on experimental prospects for our adaptive circuits. Apart from serving as toy models for the study of topology in quantum dynamics, our adaptive circuits also provide general protocols for preparing and stabilizing free-fermion topological states.  Most of the current programmable quantum platforms are based on qubits. Jordan–Wigner or Bravyi–Kitaev mappings~\cite{bravyiFermionicQuantumComputation2002} are necessary for embedding our adaptive circuit into qubit-based or other bosonic hardware. Using such qubit-fermion mappings, 1+1d versions of our Gaussian adaptive circuit could be realized on near-term qubit platforms with mid-circuit readout and ancillas, offering a clear proof-of-principle for 1D free-fermion topological state stabilization. Importantly, because our adaptive protocol is strictly Gaussian, it admits efficient classical simulation. This, in turn, enables systematic benchmarking of gate and measurement fidelities to identify concrete thresholds prior to hardware implementation. On the other hand, there has been notable recent progress on fermionic quantum processors with fermionic degrees of freedom appearing at the hardware level~\cite{gonzalez-cuadraFermionicQuantumProcessing2023,schuckertFermionqubitFaulttolerantQuantum2024,ottErrorcorrectedFermionicQuantum2024}. Our adaptive circuit may find native realization in the future generations of fermionic quantum processors.

\emph{Acknowledgements.}--- C.-M.J. thanks Andreas W. W. Ludwig for collaboration on related topics. A.B. thanks Zhou Yang and Yichen Xu for helpful discussions.  A.B. is supported by the Natural Sciences and Engineering Research Council of Canada through the Postgraduate Scholarships–Doctoral program. H.P. is supported by US-ONR grant No.~N00014-23-1-2357. C.-M.J. is supported by the Alfred P. Sloan Foundation through a Sloan Research Fellowship.

\newpage 
\onecolumngrid
\appendix
\newpage

\section{Correspondence between Single-Particle Transfer Matrix (sTM) Class and Many-Body Evolution Operator (mEO) Class}
\label{app: sTM-mEO_relation}

For free-fermion dynamics, we have introduced two types of symmetry classification: the sTM class and the mEO class. The former is defined using the group structure of the single-particle transfer matrix (sTM), while the latter is defined using the symmetry properties of the many-body evolution operator (mEO). The symmetry constraints on mEOs are formulated in Eqs.~\eqref{eq:trs_constraint},~\eqref{eq:phs_constraint}, and~\eqref{eq:cs_constraint}. In this appendix, we clarify the relation between the two classifications.

For a given free-fermion dynamical system, the sTM class and the mEO class have a one-to-one correspondence, summarized in Table~\ref{tab:mEO_sTM_correspondence}. This table is obtained through the following chain of reasoning: upon specifying an sTM class $\mc{K}_{\mr{sTM}}$, one fixes an sTM Lie group $\mc{G}$ and thereby a Lie algebra $\mf{g}$, which parametrizes an associated mEO via
\begin{equation}
    \hat{V} = \exp\left(\sum_{ij} M_{ij} \hat{\psi}^\dagger_i \hat{\psi}_j \right), \quad M \in \mf{g}.
\end{equation}
The mEO class $\mc{K}_{\mr{mEO}}$ is then determined by the symmetries preserved by the sTM Lie algebra $\mf{g}$---that is, whether $M \in \mf{g}$ satisfies some combination of Eqs.~\eqref{eq:trs_constraint},~\eqref{eq:phs_constraint}, and~\eqref{eq:cs_constraint}.

It is worth noting that $\mc{K}_{\mr{sTM}} \neq \mc{K}_{\mr{mEO}}$. In fact, within the eight real symmetry classes (AI, BDI, D, DIII, AII, CII, C, and CI), the sTM and mEO classes are related by a cyclic permutation. A similar permutation occurs within the two complex classes (A and AIII) as well. We provide a case-by-case proof of this correspondence, with our results summarized in the first two columns of Table~\ref{tab:mEO_sTM_correspondence}.

To prove the correspondence summarized in Table~\ref{tab:mEO_sTM_correspondence} case by case, we first set up our notation. We denote the $2 \times 2$ Pauli matrices as $\sigma^{x,y,z}$ and refer to TRS or PHS of $\pm1$ type as even $(+)$ and odd $(-)$. The $N \times N$ identity matrix is denoted by $\Id_N$. We also define $\Omega \equiv \ii \sigma^y$, which is commonly referred to as the \textit{symplectic form} in mathematics and quantum optics literature~\cite{weedbrook2012gaussian}. Elements of sTM groups $\mc{G}$ will be denoted by $\mf{t}$, while $M = \log(\mf{t})$ will denote elements of the corresponding Lie algebras $\mf{g}$. Definitions for each of the ten classical groups can be found in, for example, Ref.~\cite{ludwig2013lyapunov}.
\begin{table}[ht]
    \centering
    \caption{One-to-one correspondence between the many-body evolution operator (mEO) class $\mathcal{K}_{\rm mEO}$ and the single-particle transfer matrix (sTM) class $\mathcal{K}_{\rm sTM}$ for free-fermion dynamics. Within both the complex and real AZ symmetry classes, the two classifications are related by a cyclic permutation. The associated sTM groups $\mc{G}$ and maximal compact subgroups $\mc{U}\subset\mc{G}$ are also listed. Definitions for both $\mc{G}$ and $\mc{U}$ can be found in Ref.~\cite{ludwig2013lyapunov}.}
    \label{tab:mEO_sTM_correspondence}
    \begin{tabular}{c@{\hspace{12pt}}|@{\hspace{12pt}}c@{\hspace{12pt}}|@{\hspace{12pt}}c@{\hspace{12pt}}|c}
        \hline\hline
        \textbf{mEO Class $\mathcal{K}_{\rm mEO}$} & \textbf{sTM Class $\mathcal{K}_{\rm sTM}$} & $\mathcal{G}$ & $\mathcal{U} \cong$ \\
        \hline
        A    & AIII & $\mathrm{GL}(N, \mathbb{C})$ & $\mathrm{U}(N)$ \\
        AIII & A    & $\mathrm{U}(N, N)$ & $\mathrm{U}(N) \times \mathrm{U}(N)$ \\
        \hline
        AI   & BDI  & $\mathrm{GL}(N, \mathbb{R})$ & $\mathrm{O}(N)$ \\
        BDI  & D    & $\mathrm{O}(N, N)$ & $\mathrm{O}(N) \times \mathrm{O}(N)$ \\
        D    & DIII & $\mathrm{O}(N, \mathbb{C})$ & $\mathrm{O}(N)$ \\
        DIII & AII  & $\mathrm{SO}^*(2N)$ & $\mathrm{U}(N)$ \\
        AII  & CII  & $\mathrm{U}^*(2N)$ & $\mathrm{Sp}(2N)$ \\
        CII  & C    & $\mathrm{Sp}(2N, 2N)$ & $\mathrm{Sp}(2N) \times \mathrm{Sp}(2N)$ \\
        C    & CI   & $\mathrm{Sp}(2N, \mathbb{C})$ & $\mathrm{Sp}(2N)$ \\
        CI   & AI   & $\mathrm{Sp}(2N, \mathbb{R})$ & $\mathrm{U}(N)$ \\
        \hline\hline
    \end{tabular}
\end{table}

\noindent \textbf{\textbf{Correspondence between sTM Class A and mEO Class AIII}}: 
In a lattice system of $2N$ complex fermion modes, the sTM group associated with sTM class A is given by

\begin{equation}
    \mathrm{U}(N,N) = \{\mathfrak{t}\in\mathrm{GL}(2N,\mathbb{C})\ | \ \mathfrak{t}(\sigma^z\otimes\Id_N)\mathfrak{t}^\dagger =(\sigma^z\otimes\Id_N) \},
\end{equation}
where $\sigma^z$ acts on the two sublattice indices. The matrix $M$, associated with the sTM via $\mathfrak{t}= e^M$, belongs to the Lie algebra of the sTM group $\mathrm{U}(N,N)$:
\begin{equation}
    \mathfrak{u}(N,N) = \{M\in \mathbb{C}^{2N\cross 2N} \ | \ M(\sigma^z\otimes\Id_N) + (\sigma^z\otimes\Id_N)M^\dagger = 0\}.
\end{equation}
Now, we define the CS with its 1st-quantized action given by $\mc{U}_S = (\sigma^z\otimes\Id_N)$. One can check that the Lie algebra implies that $M\in  \mathfrak{u}(N,N)$ satisfies
\begin{align}
     M = - (\cU_S^\dagger M^*\cU_S)^\intercal.
\end{align}
According to Eq. \eqref{eq:cs_constraint}, the mEO $\hat{V} = e^{\sum_{ij} M_{ij}\hat{\psi}^\dag_i \hat{\psi}_j}$ respects CS. TRS and PHS are generically broken in mEOs that correspond to class-A sTMs. Therefore, sTM class A corresponds to mEO class AIII.\\

\medskip

\noindent \textbf{Correspondence between sTM class AIII and mEO class A:}
In a lattice of $N$ complex fermion modes, the sTM group associated with sTM class AIII is the general linear group over the complex field $\mathrm{GL}(N,\mathbb{C})$. The associated Lie algebra is $\mathfrak{gl}(N,\mathbb{C}) =\mathbb{C}^{N\cross N}$, meaning $M\in \mathbb{C}^{N\cross N}$ is a square complex matrix. It is easy to show that the corresponding mEO $\hat{V}$ does not preserve TRS, PHS, or CS. Therefore, sTM class AIII corresponds to mEO class A.\\

\medskip 

\noindent\textbf{Correspondence between sTM Class AI and mEO Class CI:} In a lattice of $2N$ complex fermion modes, the sTM group associated with sTM class AI is given by
\begin{equation}
    \mathrm{Sp}(2N,\mathbb{R}) = \{\mf{t}\in\mathrm{GL}(2N,\mathbb{R})\ | \ \mf{t}(\Omega\otimes\Id_N)\mf{t}^\intercal =\Omega\otimes\Id_N \}.
\end{equation}
where $\Omega$ acts on two sublattice degrees of freedom. The associated Lie algebra is 
\begin{equation}
    \mathfrak{sp}(2N,\mathbb{R}) = \{M\in \mathbb{R}^{2N\cross 2N} \ | \ M(\Omega\otimes\Id_N) + (\Omega\otimes\Id_N)M^\intercal = 0\}.
\end{equation}
Because $M\in\mathfrak{sp}(2N,\mathbb{R})$ is real, even TRS is trivially satisfied with $\mc{U}_T=\Id_{2N}$ since $M=M^*$. Odd PHS is satisfied by defining $\mc{U}_C=\Omega\otimes\Id_N$ which implies $\mc{U}_C\mc{U}_C^*=-\Id_{2N}$. One can check that the Lie algebra implies that $M\in\mathfrak{sp}(2N,\mathbb{R})$ satisfies
\begin{equation}
    M =  -(\mc{U}_C^\dagger M \mc{U}_{C})^\intercal
\end{equation}
Since both TRS and PHS are preserved, CS is also preserved --- these are symmetries of class CI. Therefore, sTM class AI corresponds to mEO class CI.\\

\medskip

\noindent\textbf{Correspondence between sTM Class BDI and mEO Class AI:} In a lattice of $N$ complex fermion modes, the sTM group associated with sTM class BDI is the general linear group over the real numbers $\mathrm{GL}(N,\mathbb{R})$. The associated Lie Algebra is $\mathfrak{gl}(N,\mathbb{R}) =\mathbb{R}^{N\cross N}$, meaning $M\in \mathbb{R}^{N\cross N}$ is a square real matrix. It is easy to show that $M$ only preserves even TRS with $\mc{U}_T=\Id_N$, since $M=M^*$ and $\mc{U}_T^2=\Id_N$---these are the symmetries of class AI. Therefore, sTM class BDI corresponds to mEO class AI.\\

\medskip

\noindent\textbf{Correspondence between sTM Class D and mEO Class BDI:} In a lattice of $2N$ Majorana fermion modes, the sTM group associated with sTM class D is given by:
\begin{equation}
    \mathrm{O}(N,N)=\{\mf{t}\in\mathrm{GL}(2N,\mathbb{R}) \ | \ \mf{t}^{\intercal}(\sigma^z\otimes\Id_N) \mf{t}=(\sigma^z\otimes\Id_N)  \}.
\end{equation}
The associated Lie algebra is
\begin{equation}
    \mathfrak{o}(N,N) = \{M \in \mathbb{R}^{2N\times 2N} \ | \ M^\intercal(\sigma^z\otimes\Id_N) + (\sigma^z\otimes\Id_N)M = 0 \}.\label{eq: o(N,N) def}
\end{equation}
Since $M$ is real, even TRS is trivially preserved with $\mc{U}_T=\Id_{2N}$. Even PHS is satisfied by defining $\mc{U}_C=\sigma^z\otimes\Id_N$. Then, one can check for $M\in \mathfrak{o}(N,N)$ that $M$ satisfies
\begin{equation}
    M = - (\mc{U}_C^\dagger M\mc{U}_C)^\intercal.
\end{equation}
Even TRS and even PHS (and thus, CS) correspond to the symmetries of class BDI. Therefore, sTM class D corresponds to mEO class BDI.\\

\medskip

\noindent\textbf{Correspondence between sTM Class DIII and mEO Class D:} In a lattice of $N$ Majorana fermion modes, the sTM group associated with sTM class DIII is given by:
\begin{equation}
    \mathrm{O}(N,\mathbb{C})=\{\mf{t}\in\mathrm{GL}(N,\mathbb{C}) \ | \ \mf{t}^{\intercal}\mf{t}=\Id_N \}.
\end{equation}
The associated Lie algebra is
\begin{equation}
    \mathfrak{o}(N,\mathbb{C}) = \{M \in \mathbb{C}^{N\cross N} \ | \ M=-M^\intercal\}
\end{equation}
Only even PHS is preserved by $M\in\mathfrak{o}(N,\mathbb{C})$, with $\mc{U}_C=\Id_N$, while TRS and CS are not preserved --- these are the symmetries of class D. Therefore, sTM class DIII corresponds to mEO class D.\\

\medskip

\noindent\textbf{Correspondence between sTM Class AII and mEO Class DIII:} In a lattice of $4N$ spin-1/2 Majorana fermion modes, the sTM group associated with sTM class AII is given by:
\begin{equation}
    \mathrm{SO}^*(2N)=\{\mf{t}\in\mathrm{SO}(4N,\mathbb{C}) \ | \ \mf{t}^{\dagger}(\Omega\otimes\Id_{2N})\mf{t}=(\Omega\otimes\Id_{2N}) \}.
\end{equation}
The associated Lie algebra is
\begin{equation}
    \mathfrak{so}^*(4N) = \{M \in \mathbb{C}^{4N\cross 4N} \ | \ M=-M^\intercal, \  M^\dagger(\Omega\otimes\Id_{2N})+(\Omega\otimes\Id_{2N})M =0\}
\end{equation}
Odd TRS is satisfied by defining $\mc{U}_T=\Omega\otimes\Id_{2N}$, which obeys $\mc{U}_T\mc{U}^*_T=-\Id_{4N}$. Even PHS is satisfied by defining $\mc{U}_C=\Id_{4N}$. Then one can check that for $M\in\mathfrak{so}^*(4N)$, $M$ satisfies 
\begin{align}
    M &= \mc{U}_T^\dagger M^*\mc{U}_T\\  
    M &= - (\mc{U}_C^\dagger M\mc{U}_C)^\intercal
\end{align}
Odd TRS and even PHS (and thus, CS as well) are the symmetries of class DIII. Therefore, sTM class AII corresponds to mEO class DIII.\\

\medskip

\noindent\textbf{Correspondence between sTM Class CII and mEO Class AII:} In a lattice of $2N$ spinful complex fermion modes, the sTM group associated with sTM class CII is given by:
\begin{equation}
    \mathrm{U}^*(2N)=\{\mf{t}\in\mathrm{GL}(2N,\mathbb{C}) \ | \ (\Omega \otimes \Id_{N})^\intercal \mf{t}^{*}(\Omega \otimes \Id_{N})= \mf{t} \}
\end{equation}
where $\Omega$ acts on the spin-1/2 degrees of freedom.
The associated Lie algebra is
\begin{equation}
    \mathfrak{u}^*(2N) = \{M \in \mathbb{C}^{2N\cross 2N} \ | \ (\Omega \otimes \Id_{N})^\intercal M^{*}(\Omega \otimes \Id_{N})= M\}
\end{equation}
where $\Omega=\ii\sigma^y$. Odd TRS can be satisfied by defining $\mc{U}_T=\Omega \otimes \Id_{N}$, which obeys $\mc{U}_T\mc{U}_T^*=-\Id_{2N}$. One can check that for $M\in\mathfrak{u}^*(2N)$, $M$ satisfies 
\begin{equation}
    M = \mc{U}_T^\dagger M^* \mc{U}_T
\end{equation}
and no other symmetries --- these are the symmetries of class AII. Therefore, sTM class CII corresponds to mEO class AII.\\

\medskip

\noindent\textbf{Correspondence between sTM Class C and mEO Class CII:} In a lattice of $4N$ spinful complex fermion modes, the sTM group associated with sTM class C is given by:
\begin{equation}
    \mathrm{Sp}(2N,2N)=\{\mf{t}\in\mathrm{GL}(4N,\mathbb{C}) \ | \ \mf{t}^\dagger (\sigma^z\otimes\Id_{2N})\mf{t} =\sigma^z\otimes\Id_{2N}, \ (\sigma^z\otimes\Omega\otimes\Id_{N})^\intercal \mf{t}^* (\sigma^z\otimes\Omega\otimes\Id_{N}) = \mf{t}\}.
\end{equation}
The associated Lie algebra is
\begin{equation}
    \mathfrak{sp}(2N,2N) = \{M \in \mathbb{C}^{4N\times 4N} \ | \ M^\dagger (\sigma^z\otimes\Id_{2N}) + (\sigma^z\otimes\Id_{2N})M = 0, \ (\sigma^z\otimes\Omega\otimes\Id_{N})^\intercal M^* (\sigma^z\otimes\Omega\otimes\Id_{N}) = M\}.
\end{equation}
Here, matrices of the form $X\otimes \Id_{2N}$ act nontrivially only on the two sublattice degrees of freedom, while $\Id_2\otimes X\otimes\Id_{N}$ act nontrivially only on the two spin-1/2 degrees of freedom. The two conditions on $M\in\mathfrak{sp}(2N,2N)$ imply that 
\begin{equation}
    M = -(\Id_2\otimes \Omega \otimes \Id_N)^\intercal M^\intercal(\Id_2\otimes \Omega \otimes \Id_N).
\end{equation}
Thus, $M\in\mathfrak{sp}(2N,2N)$ is invariant under odd TRS with $\mc{U}_T= \sigma^z\otimes\Omega\otimes\Id_{N}$, and invariant under odd PHS with $\mc{U}_C = \Id_2\otimes \Omega \otimes \Id_N $ --- these are the symmetries of class CII. Therefore, sTM class C corresponds to mEO class CII.\\

\medskip

\noindent\textbf{Correspondence between sTM Class CI and mEO Class C:} In a lattice of $2N$ complex fermion modes, the sTM group associated with sTM class CII is given by:
\begin{equation}
    \mathrm{Sp}(2N,\mathbb{C})=\{\mf{t}\in\mathrm{GL}(2N,\mathbb{C}) \ | \ \mf{t}^\intercal (\Omega\otimes\Id_N)\mf{t} =\Omega\otimes\Id_N\}.
\end{equation}
The associated Lie algebra is
\begin{equation}
    \mathfrak{sp}(2N,\mathbb{C}) = \{M \in \mathbb{C}^{2N\times 2N} \ | \ M^\intercal(\Omega\otimes\Id_N)+ (\Omega\otimes\Id_N)M=0\}.
\end{equation}
The Lie algebra implies that 
\begin{equation}
    M = -(\Omega\otimes\Id_N)^\intercal M^\intercal (\Omega\otimes\Id_N).
\end{equation}
One can check that $M\in\mathfrak{sp}(2N,\mathbb{C})$ preserves odd PHS with $\mc{U}_C=\Omega\otimes\Id_N$ and does not preserve any other symmetries --- these are the symmetries of class C. Therefore, sTM class CI corresponds to mEO class C.

\section{Gaussian POVM Admissibility in different Symmetry Classes}
\label{app: Gaussian POVM admissibility}
\subsection{Problem Statement}
In this appendix, we determine which symmetry classes admit ensembles of symmetry-invariant many-body evolution operators (mEOs) that support nontrivial, post-selection-free Gaussian measurements---that is, Gaussian POVMs. Here, ``nontrivial'' excludes the case where the mEO ensemble consists purely of Gaussian unitary mEOs, which trivially satisfy the POVM condition. We henceforth assume nontriviality implicitly. Moreover, for symmetry classes that do not admit Gaussian POVMs, we do not rule out the possibility that \textit{interacting}, symmetry-compatible mEOs could still give rise to a valid POVM---this is why we specify ``Gaussian'' POVM throughout. An example where interactions enable POVM admissibility despite symmetry constraints is discussed in Sec.~\ref{Sec: POVMadmissibility}. Note that much of the Lie group/algebra notation used in this appendix will follow App.~\ref{app: sTM-mEO_relation}.

Our starting point is the continuum generalization of the discrete POVM condition introduced in Eq.~\eqref{eq:discrete POVM main text}:
\begin{equation}
    \int dM\omega(M) \hat{K}^\dagger(M) \hat{K}(M) \overset{!}{=} \hat{\Id}. \label{eq:POVM cond cont}
\end{equation}
In the above, $\hat{K}(M)$ is a Gaussian mEO parameterized by $M\in\mf{g}$, where $\mf{g}$ is the Lie algebra associated with the corresponding sTM group. {We use the notation \(\overset{!}{=}\) to denote a required equality, i.e., a constraint that must be satisfied for a valid Gaussian POVM to exist.} Furthermore, $dM\omega(M)$ is an unspecified measure over $\mf{g}$. Equivalently, we define $\omega(M)$ to be an unspecified nonnegative scalar weight function over $\mf{g}$---a continuum generalization of the discrete weight function introduced in Eq.~\eqref{eq:discrete POVM main text}. In the above, the mEO ensemble $\{\hat{K}(M)\}_{M\in\mf{g}}$ is element-wise invariant with respect to the discrete symmetries associated with mEO class $\mc{K}_{\mr{mEO}}$. We will refer to a weight function as \textit{trivial} if it constrains $\hat{K}(M)$ to be unitary for all $M$, and as \textit{nontrivial} if not. The problem now is to determine which mEO classes can admit Gaussian POVMs via analysis of Eq.~\eqref{eq:POVM cond cont} for each symmetry class.

\medskip

\noindent \textbf{Criterion for Gaussian POVM Admissibility:}  
A given mEO class $\mc{K}_{\mr{mEO}}$ admits a Gaussian POVM if and only if there exists an ensemble of Gaussian Kraus operators $\{\hat{K}(M)\}_{M\in\mf{g}}$ that is invariant under the symmetries of $\mc{K}_{\mr{mEO}}$ and a \textit{nontrivial} weight function $\omega(M)$ such that Eq.~\eqref{eq:POVM cond cont} is satisfied. Recall that we define a nontrivial weight function to be a weight function that does not constrain the entire ensemble $\{\hat{K}(M)\}_{M\in\mf{g}}$ to be purely unitary. Conversely, we say that mEO class $\mc{K}_{\mr{mEO}}$ does not admit Gaussian POVMs if Eq.~\eqref{eq:POVM cond cont} is not satisfied for any nontrivial choice of $\omega(M)$.

\medskip

We now turn to our analysis of Gaussian POVM admissibility for each mEO class. As discussed in Sec.~\ref{Sec: POVMadmissibility}, we find that mEO classes $\mc{K}_{\mr{mEO}}\in\{\mr{A},\mr{AI},\mr{BDI},\mr{D}\}$ admit Gaussian POVMs, while symmetry classes $\mc{K}_{\mr{mEO}} \in \{\mathrm{AIII}, \mathrm{C}, \mathrm{CI}, \mathrm{CII}, \mathrm{DIII}, \mathrm{AII}\}$ do not. For Gaussian POVM admissible classes, it suffices to establish the existence of a symmetry-compatible Gaussian Kraus operator $\hat{K}(M)$ and a nontrivial weight function $\omega(M)$ such that the POVM condition Eq.~(\ref{eq:POVM cond cont}) is satisfied. We begin with the symmetry classes that admit Gaussian POVMs.

For the classes that do not admit Gaussian POVMs, we employ two distinct strategies to establish Gaussian POVM inadmissibility. The first applies to mEO classes $\mc{K}_{\mr{mEO}}\in\{\mr{AIII}, \ \mr{C}, \ \mr{CI}, \ \mr{CII}\}$, while the second applies to classes $\mc{K}_{\mr{mEO}}\in\{\mr{DIII}, \mr{AII}\}$.

\newpage

\begin{table}[ht]
    \centering
    \caption{Listed here are the many-body evolution operator (mEO) classes, their associated discrete symmetries, their corresponding single-particle transfer matrix (sTM) classes, and whether the mEO class admits nontrivial Gaussian POVMs (post-selection-free Gaussian measurements).}
    \label{mEO_class_POVM_table}
    \begin{tabular}{c@{\hspace{10pt}}c@{\hspace{10pt}}c@{\hspace{10pt}}c@{\hspace{10pt}}c@{\hspace{10pt}}c}
        \hline\hline
        \textbf{mEO Class} $\mc{K}_{\mr{mEO}}$ & \textbf{TRS} & \textbf{PHS} & \textbf{CS} & \textbf{sTM Class} $\mc{K}_{\mr{sTM}}$ & \makecell{ \textbf{Post-Selection-Free} \\  \textbf{Gaussian Measurements}} \\
        \hline
        A    & 0     & 0     & 0 & AIII & \cmark \\
        AI   & $+1$  & 0     & 0 & BDI  & \cmark \\
        BDI  & $+1$  & $+1$  & 1 & D    & \cmark \\
        D    & 0     & $+1$  & 0 & DIII & \cmark \\
        \hline
        AIII & 0     & 0     & 1 & A    & \xmark \\
        C    & 0     & $-1 $ & 0 & CI   & \xmark \\
        CI   & $+1$  & $-1 $ & 1 & AI   & \xmark \\
        CII  & $-1 $ & $-1 $ & 1 & C    & \xmark \\
        DIII & $-1 $ & $+1$  & 1 & AII  & \xmark \\
        AII  & $-1 $ & 0     & 0 & CII  & \xmark \\
        \hline\hline
    \end{tabular}
\end{table}

\subsection{Establishing Existence of Gaussian POVMs for mEO Classes A, AI, BDI, D}

\medskip

\noindent \textbf{mEO Class A Admits Gaussian POVMs:} Consider a system of $N$ complex fermion modes with mEO class A symmetries. For this class, we have $\mc{G}=\mr{GL}(N,\mathbb{C})$ and thus $M \in\mf{gl}(N,\mathbb{C})$. We show that there exists a symmetry-compatible $\hat{K}(M)$ and weight function $\omega(M)$ that admits a Gaussian POVM for mEO class A. Consider the following choices for $N=1$, with the $1\times1$ `matrix' $M$ being parameterized by a real number $\lambda\in\mathbb{R}$ and a phase $\phi\in\mr{S}^1$
\begin{equation}
    \hat{K}(\lambda,\phi)=e^{(\lambda+\ii\phi)\hat{n}}, \quad \omega(\lambda,\phi) = \frac{1}{2\pi}\frac{1}{2\cosh(\alpha)}\left[\delta(\lambda+\alpha) + \delta(\lambda-\alpha)\right]e^{-\lambda}, \ \alpha\in\mathbb{R}\label{eq:mEO class A examples}
\end{equation}
where $\hat{n}$ is the fermion number operator. Physically, such a choice of Kraus operator and weight function corresponds to weak measurements of single-mode fermion occupation, with measurement strength $\lambda$, followed by a random unitary rotation parameterized by a phase $\theta$. 

Clearly, our choice of $\hat{K}$ is compatible with the symmetries of the mEO class A. Substitution of Eq.~(\ref{eq:mEO class A examples}) into Eq.~(\ref{eq:POVM cond cont}) leads to
\begin{equation}
\begin{split}
    \int_0^{2\pi}d\phi\int_\mathbb{R}d\lambda\omega(\lambda,\phi) \ \hat{K}^\dagger(\lambda,\phi)\hat{K}(\lambda,\phi) &=
    \int_0^{2\pi}\frac{d\phi}{2\pi}\int_{\mathbb{R}} d\lambda \frac{1}{2\cosh(\alpha)}\left[\delta(\lambda+\alpha) + \delta(\lambda-\alpha)\right]e^{\lambda(2\hat{n}-1)}\\ 
    &= \frac{1}{2\cosh(\alpha)}\left[e^{-\alpha(2\hat{n}-1)}+e^{\alpha(2\hat{n}-1)}\right]\\ 
    \Rightarrow \int_0^{2\pi}d\phi\int_\mathbb{R}d\lambda\omega(\lambda,\phi) \ \hat{K}^\dagger(\lambda,\phi)\hat{K}(\lambda,\phi) &= \hat{\Id}.
\end{split}
\end{equation}
Therefore, \textit{Gaussian Kraus operators with mEO class A symmetries can admit a Gaussian POVM.}\\

\medskip

\noindent \textbf{mEO Class AI Admits Gaussian POVMs:} Consider a system of $N$ complex fermion modes with mEO class AI symmetries. For this class, we have $\mc{G}=\mr{GL}(N,\mathbb{R})$ and thus $M \in\mf{gl}(N,\mathbb{R})$. We show that there exists a symmetry-compatible $\hat{K}(M)$ and weight function $\omega(M)$ that admits a Gaussian POVM for mEO class AI. Consider the following choices for $N=1$, with the $1\times1$ `matrix' $M$ being parameterized by a real number $\lambda\in\mathbb{R}$
\begin{equation}
    \hat{K}(\lambda)=e^{\lambda\hat{n}}, \quad \omega(\lambda) = \frac{1}{2\cosh(\alpha)}\left[\delta(\lambda+\alpha) + \delta(\lambda-\alpha)\right]e^{-\lambda}, \ \lambda,\alpha\in\mathbb{R}.\label{eq:mEO class AI examples}
\end{equation}
While similar to mEO class A, mEO class AI symmetries enforce a realness condition of $\hat{K}$, leading to a purely nonunitary weak measurement of single-mode fermion occupation with measurement strength $\lambda$. 

Our choice of $\hat{K}$ is compatible with the symmetries of the mEO class AI. One can verify that substitution of Eq.~(\ref{eq:mEO class AI examples}) into Eq.~(\ref{eq:POVM cond cont}) satisfies the POVM condition, following the same logic as mEO class A.\\
Therefore, \textit{Gaussian Kraus operators with mEO class AI symmetries can admit a Gaussian POVM}.\\

\medskip

\noindent \textbf{mEO Class BDI Admits Gaussian POVMs:}
Consider a system of $2N$ Majorana fermion modes with mEO class BDI symmetries. For this class, we have $\mc{G}=\mr{O}(N,N)$ and thus $M\in\mf{o}(N,N)$. We show that there exists a symmetry-compatible $\hat{K}(M)$ and weight function $\omega(M)$ that admits a Gaussian POVM for mEO class BDI. However, to do so, we will have to work in a rotated representation of the Lie algebra $\mf{o}(N,N)$, in contrast to the definition provided in Eq.~\eqref{eq: o(N,N) def}, but more on this later. 

Consider the following choices of Kraus operator and weight function for $N=1$, with $2\times 2$ generator $M=\ii\lambda\Omega$ being parameterized by a single real number $\lambda\in\mathbb{R}$
\begin{equation}
    \hat{K}(\lambda) = \exp\left[-\frac{1}{2}\ii\lambda(\hat{\gamma}_1 \ \hat{\gamma}_2)\Omega\begin{pmatrix} \hat{\gamma}_1\\ \hat{\gamma}_2\end{pmatrix}\right], \quad \omega(\lambda)=\frac{1}{2\cosh(2\alpha)}\left[\delta(\lambda-\alpha)+\delta(\lambda+\alpha)\right], \  \alpha\in\mathbb{R}\label{eq:mEO class BDI examples}
\end{equation}
where the Majorana fermions $\hat{\gamma}_i$ satisfy $\{\hat{\gamma}_i,\hat{\gamma}_j\}=2\delta_{ij}$. Such a choice of Kraus operator and weight function physically corresponds to a weak measurement of single-mode fermion parity, with measurement strength $\lambda$.

Recall that the discrete symmetries of class BDI are even TRS and even PHS (and thus, CS). One can check that the Hermitian single-particle generator $M=\ii\lambda\Omega$ satisfies class BDI symmetries with $\mc{U}_T=\sigma^z$ for even TRS and $\mc{U}_{C}=\Id_2$ for even PHS. Now, consider the product $\hat{K}^\dagger(\lambda)\hat{K}(\lambda)$ involved in the POVM condition Eq.~(\ref{eq:POVM cond cont}). One can show
\begin{equation}
    \hat{K}^\dagger(\lambda)\hat{K}(\lambda) = \cosh(2\lambda)-\sinh(2\lambda)\ii\hat{\gamma}_1\hat{\gamma}_2\label{eq:mEO class BDI POVM simplification step}
\end{equation}
where we have use $\hat{\gamma}_i^2=1$. Substitution of Eq.~(\ref{eq:mEO class BDI POVM simplification step}) and our choice of weight function defined in Eq.~(\ref{eq:mEO class BDI examples}) into Eq.~(\ref{eq:POVM cond cont}) leads to  
\begin{equation}
\begin{split}
    \int_{\mathbb{R}}d\lambda \omega(\lambda) \ \hat{K}^\dagger(\lambda)\hat{K}(\lambda)=    \int_{\mathbb{R}} d\lambda \frac{1}{2\cosh(2\alpha)}\left[\delta(\lambda-\alpha)+\delta(\lambda+\alpha)\right]\left(\cosh(2\lambda)-\sinh(2\lambda)\ii\hat{\gamma}_1\hat{\gamma}_2\right)= \hat{\Id}
\end{split}
\end{equation}
Therefore, \textit{Gaussian Kraus operators with mEO class BDI symmetries can admit a Gaussian POVM.}

We now show that $M$ lives in a rotated representation of the Lie algebra $\mf{o}(N,N)$ associated with mEO class BDI. For a system of $2N$ Majorana fermion modes, we have constrained $M$ to live in the Hermitian subspace of
\begin{equation}
    \left\{M\in\mathbb{C}^{2N\times 2N} \ | \ (\sigma_z\otimes\Id_N)M^*(\sigma_z\otimes\Id_N)=M, \ M=-M^\intercal, \right\}\label{eq:mEO class BDI weird algebra}
\end{equation}
with $\sigma_z$ acting on two sublattice degrees of freedom. This particular rotation switches from a Majorana fermion basis to a complex fermion basis.

We can show that $M$ lives in a different representation of the Lie algebra $\mathfrak{o}(N,N)$ associated with the mEO class BDI. Indeed, consider the rotation 
\begin{equation}
    M = e^{-\ii\frac{\pi}{4}(\sigma^z\otimes\Id_N)}M'e^{\ii\frac{\pi}{4}(\sigma^z\otimes\Id_N)} 
\end{equation}
Then the algebra given in Eq.~(\ref{eq:mEO class BDI weird algebra}) implies that $M'$ obeys the algebra 
\begin{equation}
    \{M'\in \mathbb{R}^{2N\times 2N} \ | \ (\sigma^z\otimes\Id_N)M'(\sigma^z\otimes\Id_N)=-M'^\intercal\}
\end{equation}
which is precisely the definition of Lie algebra $\mathfrak{o}(N,N)$ associated with mEO class BDI provided in Eq.~\eqref{eq: o(N,N) def}. Now, even TRS is implemented via single-particle action $\mc{U}_T=\Id_{2N}$, and even PHS is implemented via $\mc{U}_{C}=\sigma^z\otimes\Id_{N}$. Note that if we constrain $M'$ to live in the Hermitian subspace of $\mathfrak{o}(N,N)$, the only choice that can satisfy such constraints is the trivial zero matrix $M'=0$. This is the reason we chose to work in a rotated representation of $\mathfrak{o}(N,N)$, i.e., Eq.~\eqref{eq:mEO class BDI weird algebra}.\\

\medskip

\noindent \textbf{mEO Class D Admits Gaussian POVM:}
Consider a system of $N$ Majorana fermion modes with mEO class D symmetries. For this class, we have $\mc{G}=\mr{O}(N,\mathbb{C})$ and thus $M\in\mf{o}(N,\mathbb{C})$. We show that there exists a symmetry-compatible $\hat{K}(M)$ and weight function $\omega(M)$ that admits a Gaussian POVM for mEO class D. 

Consider the following choices of Kraus operator and weight function for $N=2$, with the $2\times 2$ single-particle generator $M=\ii(\lambda+\ii\phi)\Omega$ being parameterized by a real number $\lambda\in\mathbb{R}$ and a phase $\phi\in\mr{S}^1$
\begin{equation}
    \hat{K}(\lambda) = \exp\left[-\frac{1}{2}\ii(\lambda+\ii\phi)(\hat{\gamma}_1 \ \hat{\gamma}_2)\Omega\begin{pmatrix} \hat{\gamma}_1\\ \hat{\gamma}_2\end{pmatrix}\right], \quad \omega(\lambda,\phi)=\frac{1}{2\pi}\frac{1}{2\cosh(2\alpha)}\left[\delta(\lambda-\alpha)+\delta(\lambda+\alpha)\right], \  \alpha\in\mathbb{R}\label{eq:mEO class D examples}
\end{equation}
where the Majorana fermions $\hat{\gamma}_i$ satisfy $\{\hat{\gamma}_i,\hat{\gamma}_j\}=2\delta_{ij}$. Such a choice of Kraus operator and weight function physically corresponds to a weak measurement of single-mode fermion parity, with measurement strength $\lambda$, followed by a random unitary rotation parameterized by a phase $\theta$. Unlike mEO class BDI, mEO class D is not invariant under TRS, allowing for operations that combine measurement and unitary evolution.

To verify that our choices of Kraus operator and weight function defined above satisfy the POVM condition Eq.~(\ref{eq:POVM cond cont}), one can show that 
\begin{equation}
    \int_{0}^{2\pi}d\phi\int_{\mathbb{R}}d\lambda \omega(\lambda,\phi)\hat{K}^\dagger(\lambda,\phi)\hat{K}(\lambda,\phi)=\hat{\Id}.
\end{equation}
The algebra that follows is identical to the algebra used to verify the POVM condition for mEO class BDI. It follows that \textit{Gaussian Kraus operators with mEO class D symmetries can admit a Gaussian POVM}.

\subsection{Proofs of Gaussian POVM Inadmissibility for mEO Classes AIII, C, CI, CII, DIII, and AII}
\subsubsection{Cartan Decomposition of Classical Lie Groups}

In the following proofs of Gaussian POVM inadmissibility, we will repeatedly make use of the degeneracy structure of the spectrum of Hermitian Gaussian operators of the form $\hat{K}^\dagger(M)\hat{K}(M)$, where $\hat{K}(M)$ is a symmetry-invariant Kraus operator in mEO class $\mathcal{K}_{\mathrm{mEO}}$ parameterized by a single-particle generator $M$. These operators arise naturally in the integrand of the POVM condition, and can always be diagonalized in the number basis as
\begin{equation}
    \hat{K}^\dagger(M)\hat{K}(M) \sim \exp\left[\sum_{\alpha}\lambda_{\alpha}(M)\hat{n}_{\alpha}\right],
\end{equation}
where $\{\lambda_{\alpha}\}_\alpha$ constitutes the spectrum of the single-particle generator
\begin{equation}
    H \equiv \log\bigl(e^M e^{M^\dagger}\bigr)\label{eq: hermitian single-particle generator}
\end{equation}
of $\hat{K}^\dagger(M)\hat{K}(M)$~\footnote{Here, Eq.~\eqref{eq: hermitian single-particle generator} is defined assuming the fermion operators are appropriately normalized i.e., satisfying $\{\hat{\psi}^\dagger_i,\hat{\psi}_j\}=\delta_{ij}$}. Equivalently, $\{\lambda_{\alpha}\}_\alpha$ represents the \textit{log-singular value} spectrum of the sTM $\mf{t}=e^M$ associated with the mEO class of interest. To determine whether $\hat{K}^\dagger(M)\hat{K}(M)$ is compatible with POVM normalization for a given mEO class, we must understand the allowed degeneracy patterns in the spectrum $\{\lambda_\alpha\}_\alpha$---specifically, whether they must appear in symmetry-related multiplets. To this end, we make use of the \textit{Cartan decomposition} of Hermitian elements in Lie algebras associated with each symmetry class. Each of the ten Altland–Zirnbauer classes defines a non-compact Lie group $\mathcal{G}$ that contains the single-particle transfer matrices of Gaussian Kraus operators. The single-particle generators $H$ defined above live in the Hermitian subspace of the corresponding Lie algebra $\mathfrak{g}$. Every such group admits a Cartan (or polar) decomposition of the form
\[
  \mathcal G = \mathcal U \mathcal A \mathcal U,
  \quad
  \mathcal A = e^{\mathfrak a},
\]
where $\mathcal{U} \subset \mathcal{G}$ is a maximal compact subgroup and $\mathfrak{a} \subset \mathfrak{g}$ is a maximal abelian subspace. Consequently, every Hermitian $H \in \mathfrak{g}$ can be brought into the form
\begin{equation}
  H = U^\dagger A(D) U,
  \label{eq:cartan-decomp}
\end{equation}
where $U \in \mathcal{U}$ and $A(D) \in \mathfrak{a}$ is a linear function of a real diagonal matrix $D = \mathrm{diag}(\lambda_1, \lambda_2, \dots)$ constrained by symmetry. The existence of this decomposition is guaranteed by standard Cartan theory~\cite{helgason1979differential,knapp1996lie}, though explicit construction can be subtle. For each symmetry class, we have derived a canonical form for $A(D)$ and found that it generically takes the form
\begin{equation}
  A(D) = V_1 \otimes D \otimes V_2,
  \label{eq:A-decomp}
\end{equation}
where $V_1$ and $V_2$ are fixed matrices determined by symmetry. Physically, the entries $\lambda_i$ of $D$ encode measurement strength: $D = 0$ if and only if $\hat{K}$ is unitary. The matrices $V_{1,2}$ control the degeneracy structure of the sTM log-singular value spectrum. In practice, they are typically simple: $1 \times 1$ or $2 \times 2$ identity matrices, Pauli matrices (up to a factor of $\ii$), or weight-2 Pauli strings. To illustrate how this decomposition works in detail, we provided derivations of Cartan decompositions for mEO classes DIII and AIII next.

\medskip

\noindent\textbf{Example I: Cartan Decomposition of mEO class DIII.}  
Consider a mEO class DIII dynamical system acting on a lattice of \(4N\) spin-1/2 Majorana modes. The dynamics are constrained by even PHS and odd TRS. From Table~\ref{tab:mEO_sTM_correspondence}, we identify the corresponding sTM class as \(\mathcal{K}_{\mathrm{sTM}} = \mathrm{AII}\). The associated sTM group is \(\mathcal{G}^{\mathrm{AII}} = \mathrm{SO}^*(4N)\), with Lie algebra \(\mathfrak{g}^{\mathrm{AII}} = \mathfrak{so}^*(4N)\). The maximal compact subgroup is
\( \mathcal{U}^{\mathrm{AII}} = \mathrm{SO}(4N) \cap \mathrm{Sp}(4N, \mathbb{R}) \cong \mathrm{U}(2N).\)
For convenience, we recall the definitions of \(\mathcal{G}^{\mathrm{AII}}\), \(\mathcal{U}^{\mathrm{AII}}\), and \(\mathfrak{g}^{\mathrm{AII}}\):
\begin{align}
    \mathcal{G}^{\mathrm{AII}} &= \left\{ \mathfrak{t} \in \mathrm{SO}(4N, \mathbb{C}) \ \middle| \ \mathfrak{t}^\dagger(\Omega \otimes \Id_{2N}) \mathfrak{t} = \Omega \otimes \Id_{2N} \right\}, \\
    \mathcal{U}^{\mathrm{AII}} &= \left\{ \mathfrak{t} \in \mathrm{SO}(4N) \ \middle| \ [\mathfrak{t}, \Omega \otimes \Id_{2N}] = 0 \right\}, \label{eq: U^AII} \\
    \mathfrak{g}^{\mathrm{AII}} &= \left\{ M \in \mathbb{C}^{4N \times 4N} \ \middle| \ M = -M^\intercal, \ M^\dagger(\Omega \otimes \Id_{2N}) + (\Omega \otimes \Id_{2N}) M = 0 \right\}.
\end{align}
Here, \(\Omega \otimes \Id_{2N}\) acts nontrivially only on the spin degrees of freedom of the Majorana modes and represents the first-quantized action of TRS on the system.

Given a Hermitian element \(H \in \mathfrak{g}^{\mathrm{AII}}\), one can show that \(H\) admits the following Cartan decomposition:
\begin{equation}
    H = \ii O^\intercal \left[ \sigma^z \otimes D \otimes \Omega \right] O,
    \label{eq: cartan decomp DIII claim}
\end{equation}
where \(D \geq 0\) is a real \(N \times N\) diagonal matrix and \(O \in \mathcal{U}^{\mathrm{AII}}\). Here, we identify \(V_1 = \sigma^z\) and \(V_2 = \ii\Omega\).

\textit{(Proof).} Let \(H = H^\dagger \in \mathfrak{g}^{\mathrm{AII}}\). Since \(H\) is Hermitian and antisymmetric, it must be purely imaginary and can be written as \(H = \ii \tilde{H}\), where \(\tilde{H}\) is real and antisymmetric. Any such matrix admits the canonical form \(\tilde{H} = O^\intercal [D \otimes \Omega] O\), with \(D \geq 0\) real diagonal and \(O \in \mathrm{SO}(4N)\)~\cite{bravyi2005lagrangian}. The condition \(H \in \mathfrak{g}^{\mathrm{AII}}\) further requires \(\{H, \Omega \otimes \Id_{2N}\} = 0\), which in turn implies that the decomposition \(O^\intercal (D \otimes \Omega) O\) must also satisfy this symmetry. To ensure this, we impose \([O, \Omega \otimes \Id_{2N}] = 0\), which restricts \(O\) to lie in the maximal compact subgroup \(\mathcal{U}^{\mathrm{AII}}\) (see Eq.~\eqref{eq: U^AII}), as required by the Cartan decomposition Eq.~\eqref{eq:cartan-decomp}. For consistency, the matrix \(D \otimes \Omega\) must then anticommute with \(\Omega \otimes \Id_{2N}\). Since $D$ is diagonal, the constraint $\{D \otimes \Omega,\Omega\otimes\Id_{2N}\}=0$ can only be satisfied if \(D = \sigma^z \otimes \tilde{D}\), with \(\tilde{D} \geq 0\) a real diagonal \(N \times N\) matrix. Substituting back into $H$ yields the claimed decomposition in Eq.~\eqref{eq: cartan decomp DIII claim}. \hfill\(\square\)

Since we are working in a Majorana fermion basis, we can pair each mode in the subspace on which \(\Id_{2N} \otimes \Omega\) acts nontrivially. Switching to the number basis then leads to a log-singular value spectrum that comes in \(\pm\) pairs.

\medskip

\noindent\textbf{Example II: Cartan Decomposition of mEO class AIII.}
Consider a mEO class AIII dynamical system acting on a lattice of \(2N\) complex fermion modes with two sublattice degrees of freedom. The dynamics are constrained by CS only. From Table~\ref{tab:mEO_sTM_correspondence}, we identify the corresponding sTM class as \(\mathcal{K}_{\mathrm{sTM}} = \mathrm{A}\). The associated sTM group is $\mathcal{G}^{\mathrm{A}} = \mathrm{U}(N,N)$, with Lie algebra $\mathfrak{g}^{\mathrm{A}} = \mathfrak{u}(N,N)$. The maximal compact subgroup is
\( \mathcal{U}^{\mathrm{A}} = \mathrm{U}(N)\times\mathrm{U}(N)\).
Recall the definitions of \(\mathcal{G}^{\mathrm{A}}\), \(\mathcal{U}^{\mathrm{A}}\), and \(\mathfrak{g}^{\mathrm{A}}\):
\begin{align}
    \mathcal{G}^{\mathrm{A}} &= \left\{ \mathfrak{t} \in \mathrm{GL}(2N, \mathbb{C}) \ \middle| \ \mathfrak{t}(\sigma^z \otimes \Id_{N}) \mathfrak{t}^\dagger = \sigma^z \otimes \Id_{N} \right\}, \\
    \mathcal{U}^{\mathrm{A}} &= \left\{ \mathfrak{t} \in \mathrm{GL}(2N, \mathbb{C}) \ \middle| \ \mf{t}=\mf{t}_A\oplus\mf{t}_B, \ \mf{t}_A,\mf{t}_B\in \mr{U}(N) \right\}, \label{eq: U^A} \\
    \mathfrak{g}^{\mathrm{A}} &= \left\{ M \in \mathbb{C}^{2N \times 2N} \ \middle| \ M(\sigma^z \otimes \Id_{N}) + (\sigma^z \otimes \Id_{N}) M^\dagger = 0 \right\}.
\end{align}
Here, \(\sigma^z \otimes \Id_{N}\) acts nontrivially only on the sublattice degrees of freedom of the complex fermion modes and represents the first-quantized action of CS on the system. Furthermore, $\oplus$ denotes the direct sum of matrices, i.e., block-diagonal concatenation.

Given a Hermitian element \(H \in \mathfrak{g}^{\mathrm{A}}\), one can show that \(H\) admits the following Cartan decomposition:
\begin{equation}
    H = \left(U_A\oplus U_B\right)^\dagger \left[ \sigma^x \otimes D \right] \left(U_A\oplus U_B\right),
    \label{eq: cartan decomp AIII claim}
\end{equation}
where \(D \geq 0\) is a real \(N \times N\) diagonal matrix and \(U_A\oplus U_B \in \mathcal{U}^{\mathrm{A}}\). Here, we identify \(V_1 = \sigma^x\) and \(V_2 = 1\).

\textit{(Proof).} Let \(H = H^\dagger \in \mathfrak{g}^{\mathrm{A}}\). Since \(H\) is Hermitian, the constraints placed on $H$ by $\mf{g}^{\mr{A}}$ imply that $H$ anticommutes with $\sigma^z\otimes \Id_N$. It follows that $H$ can be decomposed as
\begin{equation}
    H = \begin{pmatrix}
        \ & \tilde{H}\\
        \tilde{H}^\dagger & \
        \end{pmatrix}
\end{equation}
where $\tilde{H}\in\mathbb{C}^{N\times N}$. Now apply a singular value transformation to $\tilde{H}$ such that $\tilde{H}=U_{A}^\dagger D U_B$, where $U_A$, $U_B$ are $N\times N$ unitary matrices that are not necessarily identical, and $D\geq0$ is a real diagonal matrix. It follows that we can decompose $H$ into the form of Eq.~\eqref{eq: cartan decomp AIII claim}.\hfill\(\square\)

Switching to the number basis (i.e., diagonalizing $\sigma^x$ in Eq.~\eqref{eq: cartan decomp AIII claim}) then leads to a log-singular value spectrum that comes in \(\pm\) pairs.

\medskip

It is a straightforward (though tedious) exercise to derive the Cartan decompositions in Eqs.~\eqref{eq:cartan-decomp} and \eqref{eq:A-decomp} for each symmetry class. Once obtained, one can determine the degeneracy structure of the spectrum of \(H\) (equivalently, the log-singular value spectrum of \(\mathfrak{t} \in \mathcal{G}\)). This spectral structure will be essential in the proofs below.

\subsubsection{Gaussian POVM Inadmissibility for mEO Classes AIII, C, CI, CII}

\medskip

\noindent\textbf{Proof of Gaussian POVM Inadmissibility for mEO Classes AIII, C, CI:}
Here, we present \textit{proofs by contradiction} for mEO classes AIII, C, and CI. Specifically, we assume for each class that a Gaussian POVM exists, and then derive two mutually incompatible conditions that arise from the imposed symmetry constraints and the POVM condition. The resulting contradiction invalidates the initial assumption, thereby proving that these mEO classes do not admit Gaussian POVMs.

Suppose, on the contrary, that mEO classes AIII, C, and CI do admit Gaussian POVMs, i.e., the POVM condition
\begin{equation}
    \int dM\omega(M) \hat{K}^\dagger(M) \hat{K}(M) = \hat{\Id}, \label{eq:POVM cond cont 2}
\end{equation}
holds for a choice of nontrivial weight function. Here, the Kraus operators respect the symmetries of the mEO class of interest, and $\omega(M)$ is an unspecified weight function. We will refer to a weight function as \textit{trivial} if it constrains $\hat{K}(M)$ to be unitary for all $M$, and as \textit{nontrivial} if not.

Gaussian Kraus operators invariant under the symmetries associated with mEO classes AIII, C, and CI require $\mr{U}(1)$ symmetry and have the following generic form:
\begin{equation}
    \hat{K}(M)=\exp\left(-\sum_{ij}M_{ij}\hat{c}^\dagger_i\hat{c}_j\right), \ M\in\mf{g},
\end{equation}
where $\hat{c}_{i=1,2,\dots}$ are a set of complex fermion operators satisfying $\{\hat{c}^\dagger_i,\hat{c}_j\}=\delta_{ij}$ and $\mf{g}$ is the Lie algebra associated with the mEO class of interest. Let us define the Fock vacuum as $\vac$, where the double ket notation denotes a many-body Fock state. Since $\hat{K}^\dagger(M)\hat{K}(M)\vac=\vac$, it follows that if Eq.~\eqref{eq:POVM cond cont 2} holds, the weight function must be normalized:
\begin{equation}
    \int dM \omega(M) = 1. \label{eq:POVM constraint 1 U(1)}
\end{equation}
We will refer to this as the first constraint. Next, we compute the Fock space trace of $\hat{K}^\dagger(M)\hat{K}(M)$. If Eq.~\eqref{eq:POVM cond cont 2} holds, then it follows that
\begin{equation}
    \int dM\omega(M) \ \mr{Tr}\left(\hat{K}^\dagger(M)\hat{K}(M)\right) = \mr{Tr} \left(\hat{\Id}\right). \label{eq:POVM constraint 2 U(1)}
\end{equation}
This will be the second constraint. Because the argument of the trace is Hermitian, we can diagonalize $\hat{K}^\dagger(M)\hat{K}(M)$ and rewrite Eq.~\eqref{eq:POVM constraint 2 U(1)} as 
\begin{equation}
    \int dM\omega(M) \ \mr{Tr}\exp\left[\sum_{\alpha}\lambda_{\alpha}(M)\hat{n}_\alpha\right] = \mr{Tr} \left(\hat{\Id}\right), \label{eq:POVM constraint 2 U(1) rewritten}
\end{equation}
where $\{\lambda_{\alpha}\}_{\alpha}$ are the log-singular values of the sTM $\mf{t}=e^M$ associated with the Kraus operator $\hat{K}(M)$. For simplicity, we do not transform the measure on $M$ in the above and treat each $\lambda_{\alpha}=\lambda_{\alpha}(M)$ as an implicit function of $M$.

We show that Eq.~\eqref{eq:POVM constraint 1 U(1)} and Eq.~\eqref{eq:POVM constraint 2 U(1) rewritten} cannot simultaneously hold for classes AIII, C, and CI, except in the case of a trivial weight function, contradicting the original assumption that the mEO class of interest admits a Gaussian POVM and proving the desired statements. To proceed, we specialize to each mEO class.

One can show that for the three symmetry classes of interest $\{$AIII, C, CI$\}$, the spectrum $\{\lambda_{\alpha}\}_{\alpha}$ has a doublet structure and comes in $\pm$ pairs via Cartan decomposition. Suppose our system of interest has $2N$ complex fermion modes, with two internal degrees of freedom that we label as $A$ and $B$. Splitting the index $\alpha \rightarrow (A/B, i=1,2,\dots,N)$ and relabeling $\lambda_{\alpha}$ as $\lambda_i \equiv \lambda_{A,i} = -\lambda_{B,i}$, we can rewrite Eq.~\eqref{eq:POVM constraint 2 U(1) rewritten} as
\begin{equation}
    \int dM\omega(M) \ \mr{Tr}\prod_{i=1}^N\exp\left[\lambda_{i}(M)(\hat{n}_{A,i}-\hat{n}_{B,i})\right] = \mr{Tr} \left(\hat{\Id}\right) = 2^{2N}.
\end{equation}
The integrand is lower bounded as
\begin{equation}
     \mr{Tr}\prod_{i=1}^N\exp\left[\lambda_{i}(M)(\hat{n}_{A,i}-\hat{n}_{B,i})\right] = 2^N\prod_{i=1}^N \left(1+\cosh(\lambda_i)\right)\geq 2^{2N},
\end{equation}
with equality only when $\lambda_i = 0$ for all $i=1,2,\dots,N$. Integrating both sides of the second equality above with respect to the weight function $\omega(M)$ and applying the weight function normalization condition Eq.~\eqref{eq:POVM constraint 1 U(1)} gives
\begin{equation}
     \int dM\omega(M) \ 2^N\prod_{i=1}^N \left(1+\cosh(\lambda_i)\right)\geq\mr{Tr}(\hat{\Id}),
\end{equation}
with equality holding only for the choice of trivial weight function $\omega(M)=\prod_{i=1}^N\delta(\lambda_i(M))$. 

Thus, the normalization condition Eq.~\eqref{eq:POVM constraint 1 U(1)} and the trace constraint Eq.~\eqref{eq:POVM constraint 2 U(1) rewritten} cannot simultaneously be satisfied unless the weight function is trivial, contradicting the original assumption of Gaussian POVM admissibility for mEO classes AIII, C, and CI. It follows that \textit{mEO classes AIII, C, and CI cannot admit Gaussian POVMs.} \hfill\(\square\)\\

\medskip

\noindent\textbf{Proof of Gaussian POVM Inadmissibility for mEO Class CII:}
The proof of Gaussian POVM inadmissibility for mEO class CII is quite similar to the proof for mEO classes AIII, C, and CI, since class CII requires $\mr{U}(1)$ symmetry as well. However, our system of fermions will need additional degrees of freedom to satisfy class CII symmetries. Consider a system of $4N$ complex fermion modes with two sublattice degrees of freedom labeled as $A/B$ and two spin-1/2 degrees of freedom labeled as $\uparrow/\downarrow$. Because of $\mr{U}(1)$ symmetry, the weight function normalization constraint Eq.~\eqref{eq:POVM constraint 1 U(1)} still applies. However, the evaluation of Eq.~\eqref{eq:POVM constraint 2 U(1) rewritten} changes slightly—the Cartan decomposition of class CII leads to a spectrum $\{\lambda_{\alpha}\}_{\alpha}$ that consists of $\pm$-paired eigenvalues, each with double degeneracy. Indeed, Eq.~\eqref{eq:POVM constraint 2 U(1) rewritten} becomes:
\begin{equation}
    \int dM\omega(M) \ \mr{Tr}\left(\prod_{i=1}^N\exp\left(-\lambda_i(M)\left(\hat{n}_{A,\uparrow,i}-\hat{n}_{B,\uparrow,i}\right)\right)\exp\left(\lambda_i(M)\left(\hat{n}_{A,\downarrow,i}-\hat{n}_{B,\downarrow,i}\right)\right)\right) = \mr{Tr}\left(\hat{\Id}\right) = 2^{4N}\label{eq:class CII POVM constraint rewritten}
\end{equation}
The integrand is lower bounded as
\begin{equation}
    \mr{Tr}\left(\prod_{i=1}^N\exp\left(-\lambda_i\left(\hat{n}_{A,\uparrow,i}-\hat{n}_{B,\uparrow,i}\right)\right)\exp\left(\lambda_i\left(\hat{n}_{A,\downarrow,i}-\hat{n}_{B,\downarrow,i}\right)\right)\right)=2^{2N}\prod_{i=1}^N\left(1+\cosh(\lambda_i)\right)^2\geq 2^{4N}
\end{equation}
with equality only when $\lambda_i=0$ for all $i=1, 2, \dots, N$. Integrating both sides of the second equality above with respect to the weight function $\omega(M)$ and applying the weight function normalization condition Eq.~\eqref{eq:POVM constraint 1 U(1)} gives us
\begin{equation}
     \int dM\omega(M) \ 2^{2N}\prod_{i=1}^N \left(1+\cosh(\lambda_i)\right)^2\geq\mr{Tr}(\hat{\Id})
\end{equation}
with equality only holding for the choice of trivial weight function $\omega(M)=\prod_{i=1}^N\delta(\lambda_i(M))$. Thus, Eq.~\eqref{eq:POVM constraint 1 U(1)} and Eq.~\eqref{eq:class CII POVM constraint rewritten} cannot simultaneously hold except for the choice of trivial weight function, contradicting our original assumption of Gaussian POVM admissibility for mEO class CII—it follows that \textit{mEO class CII cannot admit Gaussian POVMs}. \hfill\(\square\)\\

\subsubsection{Gaussian POVM Inadmissibility for mEO Classes DIII and AII}

\noindent\textbf{Proof of Gaussian POVM Inadmissibility for mEO Class DIII:}
Unlike mEO classes AIII, C, CI, CII, and AII, mEO class DIII does not respect $\mr{U}(1)$ symmetry, as imposing $\mr{U}(1)$ on a class DIII system renders it effectively described by a class AII system. For this reason, the weight function normalization constraint is not implied if we suppose that class DIII admits Gaussian POVMs. Instead, we develop an alternative proof by contradiction, deriving two mutually incompatible constraints implied by the POVM condition. 

A Kraus operator that respects class DIII symmetries takes the form
\begin{equation}
    \hat{K}(M)=\exp\left(-\frac{1}{2}\sum_{\alpha\beta}^{}M_{\alpha\beta}\hat{\gamma}_{\alpha}\hat{\gamma}_{\beta}\right).\label{eq:class DIII Kraus operator}
\end{equation}
Here $\{\hat{\gamma}_{\alpha}\}_{\alpha}$ denotes a collection of Majorana fermions satisfying $\hat{\gamma}_{\alpha}^2=\hat{\Id}$ for all $\alpha$, such that 
\begin{equation}
    \{\hat{\gamma}_\alpha,\hat{\gamma}_\beta\}=2\delta_{\alpha\beta}.
\end{equation}
Suppose, for contradiction, that mEO class DIII admits Gaussian POVMs, i.e., the POVM condition
\begin{equation}
    \int dM\omega(M) \hat{K}^\dagger(M) \hat{K}(M) = \hat{\Id}, \label{eq:POVM cond cont 3 class DIII}
\end{equation}
holds for a choice of nontrivial weight function $\omega(M)$. Here, the Kraus operators respect mEO class DIII symmetries.

Consider a system of $4N$ spin-1/2 Majorana fermion modes with mEO class DIII symmetries. Splitting the index $\alpha\rightarrow(\uparrow/\downarrow, \ i=1,2,\dots,2N)$, one can show that the integrand of the POVM condition can be decomposed as 
\begin{equation}
    \hat{K}^\dagger(M) \hat{K}(M) = \hat{U}_O\left[\prod_{i=1}^N \exp\left({-\lambda_i(M)\ii\hat{\gamma}_{\uparrow,2i-1}\hat{\gamma}_{\uparrow,2i}}\right)\exp\left({\lambda_i(M)\ii\hat{\gamma}_{\downarrow,2i-1}\hat{\gamma}_{\downarrow,2i}}\right)\right]\hat{U}_O^\dagger\label{eq:class DIII integrand decomp}
\end{equation}
via Cartan decomposition of the Hermitian single-particle generator $H=\frac{1}{2}\log(e^{2M}e^{2M^\dagger})\in\mf{so}^*(4N)$, which the Lie algebra associated with mEO class DIII. Here, $\hat{U}_O$ is a Gaussian unitary in mEO class DIII, with first-quantized action $\hat{U}_O\hat{\gamma}_{\alpha}\hat{U}_{O}^{-1}=\sum_{\beta}O_{\alpha\beta}\hat{\gamma}_{\beta}$. Here, $O\in \mr{SpO}(4N)\equiv \mr{Sp}(4N,\mathbb{R}) \cap \mr{SO}(4N,\mathbb{R})\cong\mr{U}(2N)$ i.e., $O$ lives in the maximal compact subgroup associated with mEO class DIII (sTM class AII). The log-singular values of the associated transfer matrix $\mf{t}=e^M$ satisfy $\lambda_i \equiv \lambda_{\uparrow,i} = -\lambda_{\downarrow,i}$.

Tracing over the Fock space then yields the constraint 
\begin{equation}
    \int dM\omega(M) \ \prod_{i=1}^N\cosh^2\left(\lambda_i(M)\right) = 1\label{eq:class DIII constraint 1}.
\end{equation}
This will be our first constraint implied by Eq.~\eqref{eq:POVM cond cont 3 class DIII}.

Next, we construct a many-body operator that (up to a factor of $\ii$) generates the odd TRS of class DIII and commutes with $\hat{U}_O$. Define
\begin{equation}
    \hat{\Gamma} = \frac{1}{2}\sum_{i=1}^{2N}\sum_{a,b=\uparrow,\downarrow}\Omega_{ab}\ii\hat{\gamma}_{a,i}\hat{\gamma}_{b,i} = \sum_{i=1}^{2N}\ii\hat{\gamma}_{\uparrow,i}\hat{\gamma}_{\downarrow,i} = \sum_{i=1}^N\left[\ii\hat{\gamma}_{\uparrow,2i-1}\hat{\gamma}_{\downarrow,2i-1} + \ii\hat{\gamma}_{\uparrow,2i}\hat{\gamma}_{\downarrow,2i}\right].\label{eq:class DIII Gamma operator}
\end{equation}
One can verify that 
\begin{equation}
    [\hat{U}_O,\hat{\Gamma}]=0
\end{equation}
for any $O\in\mr{SpO}(4N)$. This follows from the fact that $\mr{SpO}(4N)$ is the subgroup of orthogonal matrices that commute with the symplectic form $\Omega$, i.e., $[O,(\Omega\otimes\Id_{2N})]=0$, where $\Omega$ acts on the spin-1/2 subspace.

Now multiply both sides of Eq.~\eqref{eq:POVM cond cont 3 class DIII} by $\hat{\Gamma}^2$ and take the trace over Fock space. Using $[\hat{U}_O, \hat{\Gamma}]=0$, we obtain:
\begin{equation}
    \int dM\omega(M) \ \mr{Tr}\left(\left[\prod_{i=1}^N \exp\left({-\lambda_i(M)\ii\hat{\gamma}_{\uparrow,2i-1}\hat{\gamma}_{\uparrow,2i}}\right)\exp\left({\lambda_i(M)\ii\hat{\gamma}_{\downarrow,2i-1}\hat{\gamma}_{\downarrow,2i}}\right)\right]\hat{\Gamma}^2\right) = \mr{Tr}\left(\hat{\Gamma}^2\right).\label{eq:class DIII proof trace both sides with gamma squared}
\end{equation}
The RHS is straightforward: $\mr{Tr}\left(\hat{\Gamma}^2\right) = 2N\mr{Tr}(\hat{\Id})$.

To evaluate the LHS, we expand
\begin{equation}
\begin{split}
\hat{\Gamma}^2 &= \sum_{j,k=1}^N \Big[
\left( \ii \hat{\gamma}_{\uparrow,2j-1} \hat{\gamma}_{\downarrow,2j-1} \right)
\left( \ii \hat{\gamma}_{\uparrow,2k-1} \hat{\gamma}_{\downarrow,2k-1} \right)
+ \left( \ii \hat{\gamma}_{\uparrow,2j} \hat{\gamma}_{\downarrow,2j} \right)
\left( \ii \hat{\gamma}_{\uparrow,2k-1} \hat{\gamma}_{\downarrow,2k-1} \right) \\
&\quad + \left( \ii \hat{\gamma}_{\uparrow,2j-1} \hat{\gamma}_{\downarrow,2j-1} \right)
\left( \ii \hat{\gamma}_{\uparrow,2k} \hat{\gamma}_{\downarrow,2k} \right)
+ \left( \ii \hat{\gamma}_{\uparrow,2j} \hat{\gamma}_{\downarrow,2j} \right)
\left( \ii \hat{\gamma}_{\uparrow,2k} \hat{\gamma}_{\downarrow,2k} \right)
\Big].\label{eq: Gamma expansion}
\end{split}
\end{equation}
After substituting the above into Eq.~\eqref{eq:class DIII proof trace both sides with gamma squared}, we evaluate it term-by-term. The key idea is that only terms involving bilinears acting on the same set of Majorana modes contribute a nonzero trace. We will make use of the following expansion of the integrand in Eq.~\eqref{eq:class DIII proof trace both sides with gamma squared}
\begin{equation}
\begin{split}
    \exp\left({-\lambda_i(M)\ii\hat{\gamma}_{\uparrow,2i-1}\hat{\gamma}_{\uparrow,2i}}\right)\exp\left({\lambda_i(M)\ii\hat{\gamma}_{\downarrow,2i-1}\hat{\gamma}_{\downarrow,2i}}\right) &= 
    \cosh^2 \lambda_i 
    - \sinh \lambda_i \cosh \lambda_i \left( \ii \hat{\gamma}_{\uparrow,2i-1} \hat{\gamma}_{\uparrow,2i} 
    - \ii \hat{\gamma}_{\downarrow,2i-1} \hat{\gamma}_{\downarrow,2i} \right)\\
    &- \sinh^2 \lambda_i 
    \left( \ii \hat{\gamma}_{\uparrow,2i-1} \hat{\gamma}_{\uparrow,2i} \right)
    \left( \ii \hat{\gamma}_{\downarrow,2i-1} \hat{\gamma}_{\downarrow,2i} \right).
\end{split}
\end{equation}
The above expansion follows by identifying each Majorana pair $\hat{\gamma}_{a,2i-1},\hat{\gamma}_{a,2i}$ with a single fermion mode, such that $\ii\hat{\gamma}_{a,2i-1}\hat{\gamma}_{a,2i} = 2\hat{n}_{a,i}-\hat{\Id}$. It follows that
\begin{equation}
\begin{split}
\mr{\textbf{Term 1:}}\quad  &\operatorname{Tr} \Bigg( \prod_{i=1}^N \Big[ 
\cosh^2 \lambda_i 
- \sinh \lambda_i \cosh \lambda_i \left( \ii \hat{\gamma}_{\uparrow,2i-1} \hat{\gamma}_{\uparrow,2i} 
- \ii \hat{\gamma}_{\downarrow,2i-1} \hat{\gamma}_{\downarrow,2i} \right) \\
&\qquad - \sinh^2 \lambda_i 
\left( \ii \hat{\gamma}_{\uparrow,2i-1} \hat{\gamma}_{\uparrow,2i} \right)
\left( \ii \hat{\gamma}_{\downarrow,2i-1} \hat{\gamma}_{\downarrow,2i} \right) 
\Big]
\left( \ii \hat{\gamma}_{\uparrow,2j-1} \hat{\gamma}_{\downarrow,2j-1} \right) 
\left( \ii \hat{\gamma}_{\uparrow,2k-1} \hat{\gamma}_{\downarrow,2k-1} \right) 
\Bigg) \\
&= \delta_{jk}\prod_{i=1}^N\cosh^2(\lambda_i)\mr{Tr}(\hat{\Id}).
\end{split}
\end{equation}

\begin{equation}
\begin{split}
\mr{\textbf{Term 2:}}\quad&\operatorname{Tr} \Bigg( \prod_{i=1}^N \Big[ 
\cosh^2 \lambda_i 
- \sinh \lambda_i \cosh \lambda_i \left( \ii \hat{\gamma}_{\uparrow,2i-1} \hat{\gamma}_{\uparrow,2i} 
- \ii \hat{\gamma}_{\downarrow,2i-1} \hat{\gamma}_{\downarrow,2i} \right) \\
&\qquad - \sinh^2 \lambda_i 
\left( \ii \hat{\gamma}_{\uparrow,2i-1} \hat{\gamma}_{\uparrow,2i} \right)
\left( \ii \hat{\gamma}_{\downarrow,2i-1} \hat{\gamma}_{\downarrow,2i} \right) 
\Big]
\left( \ii \hat{\gamma}_{\uparrow,2j} \hat{\gamma}_{\downarrow,2j} \right) 
\left( \ii \hat{\gamma}_{\uparrow,2k-1} \hat{\gamma}_{\downarrow,2k-1} \right) 
\Bigg) \\
&= \delta_{jk}\sinh^2(\lambda_j)\left[\prod_{i\neq j}^N\cosh^2(\lambda_i(M))\right]\mr{Tr}(\hat{\Id}).
\end{split}
\end{equation}
Terms 3 and 4 follow similarly since Eq.~\eqref{eq: Gamma expansion} is invariant under swapping indices $j\leftrightarrow k$. 

\noindent Combining all terms and simplifying, Eq.~\eqref{eq:class DIII proof trace both sides with gamma squared} becomes:
\begin{equation}
\int dM\omega(M)  \ \left(\prod_{i=1}^N\cosh^2(\lambda_i(M))\right)\left(1+\frac{1}{N}\sum_{j=1}^N\tanh^2\left(\lambda_j(M)\right)\right) = 1\label{eq:class DIII constraint 2},
\end{equation}
which is our second constraint.

We now show that Eq.~\eqref{eq:class DIII constraint 1} and Eq.~\eqref{eq:class DIII constraint 2} cannot simultaneously hold unless $\omega(M)$ is trivial. Subtracting Eq.~\eqref{eq:class DIII constraint 1} from Eq.~\eqref{eq:class DIII constraint 2}, we obtain:
\begin{equation}
     \int dM\omega(M) \ \left(\prod_{i=1}^N\cosh^2(\lambda_i)\right)\left(\sum_{j=1}^N\tanh^2(\lambda_j)\right) = 0.
\end{equation}
However, the integrand satisfies
\begin{equation}
\left(\prod_{i=1}^N\cosh^2(\lambda_i)\right)\left(\sum_{j=1}^N\tanh^2(\lambda_j)\right)\geq 0,
\end{equation}
with equality only when $\lambda_i=0$ for all $i$. Hence,
\begin{equation}
     \int dM\omega(M) \ \left(\prod_{i=1}^N\cosh^2(\lambda_i)\right)\left(\sum_{j=1}^N\tanh^2(\lambda_j)\right) \geq 0,
\end{equation}
with equality only for the trivial weight function $\omega(M)=\prod_{i=1}^N\delta(\lambda_i)$. Therefore, Eq.~\eqref{eq:class DIII constraint 1} and Eq.~\eqref{eq:class DIII constraint 2} cannot both hold unless $\omega(M)$ is trivial, contradicting our original assumption. We conclude that \textit{mEO class DIII does not admit Gaussian POVMs.} \hfill\(\square\)

\medskip

\noindent\textbf{Proof of Gaussian POVM Inadmissibility for mEO Class AII:}
Finally, we show that mEO class AII does not admit Gaussian POVMs. Although class AII requires $\mathrm{U}(1)$ symmetry, the argument used for classes AIII, C, CI, and CII does not directly apply due to the doubly degenerate log-singular value spectrum of the sTM group associated with mEO class AII. Instead, we demonstrate that mEO class AII can be viewed as a constrained subclass of DIII when expressed in a Majorana basis. Since we have already shown that class DIII mEOs do not admit Gaussian POVMs, the same conclusion follows for class AII.

Consider a system of $2N$ spinful complex fermions with mEO class AII symmetries. A Gaussian Kraus operator that respects mEO class AII symmetries will have the form
\begin{equation}
    \hat{K}(M)=\exp\left(-\sum_{ij=1}^N\sum_{ab=\uparrow,\downarrow} M_{ab,ij}\hat{c}^\dagger_{a,i}\hat{c}_{b,j} \right)
\end{equation}
where $a,b$ denote spin-1/2 degrees of freedom. Suppose, for contradiction, that mEO class AII admits Gaussian POVMs, i.e., the POVM condition
\begin{equation}
    \int dM\omega(M) \hat{K}^\dagger(M) \hat{K}(M) = \hat{\Id}, \label{eq:POVM cond cont 3 class AII}
\end{equation}
holds for a choice of nontrivial weight function $\omega(M)$. Here, the Kraus operators respect mEO class AII symmetries.

The integrand of the POVM condition can be written as 
\begin{equation}
    \hat{K}^\dagger(M) \hat{K}(M) = \exp\left(-\sum_{ij=1}^N \sum_{ab=\uparrow,\downarrow} H_{ab,ij} \, \hat{c}^\dagger_{a,i}\hat{c}_{b,j} \right),
\end{equation}
with $H = \log(e^M e^{M^\dagger})$ being a Hermitian single-particle generator. Now, introduce spinful Majorana fermions via 
$\hat{c}_{a,i}=\hat{\gamma}_{a,2i-1}+\ii \hat{\gamma}_{a,2i}$ and rewrite the above as
\begin{equation}
    \hat{K}^\dagger(M) \hat{K}(M) 
    = \exp\left(-\sum_{ij=1}^{2N}\sum_{ab=\uparrow,\downarrow} 
    \tilde{H}_{ab,ij}\, \hat{\gamma}_{a,i}\hat{\gamma}_{b,j} \right),
\end{equation}
where the augmented single-particle generator $\tilde{H}\in\mathbb{C}^{4N\times 4N}$ is given by
\begin{equation}
    \tilde{H} = \ii \big[ \mathbb{1}_2 \otimes \mathrm{Im}H + \Omega \otimes \mathrm{Re}H \big].
\end{equation}
Since $H = H^\dagger$, we have $\mathrm{Im}H$ antisymmetric, while $\mathrm{Re}H$ is symmetric. Recall that the sTM Lie algebra associated with mEO class DIII (sTM class AII) is 
\begin{equation}
    \mf{so}^*(4N) 
    = \big\{M \in \mathbb{C}^{4N \times 4N} \,\big|\, 
    M = -M^{\top}, \;
    M^{\dagger}(\Omega \otimes \mathbb{1}_{2N}) 
    + (\Omega \otimes \mathbb{1}_{2N}) M = 0 
    \big\}.
\end{equation}
It is clear that $\tilde{H}$ lies in a constrained Hermitian subspace of $\mf{so}^*(4N)$. The extra constraint imposed by $\mathrm{U}(1)$ symmetry forces $\tilde{H}$ to be parametrized entirely by $H$. Nevertheless, $\tilde{H}$ is a valid Hermitian element of $\mf{so}^*(4N)$, which implies that the proof used for general mEO class DIII POVM inadmissibility applies here as well. Therefore, we conclude that \emph{mEO class AII does not admit Gaussian POVMs.} \hfill $\square$

\section{Topological Obstructions in Chern Bands and Overcomplete Wannier Functions} \label{app: topological obstructions}
In this appendix, we briefly review the topological obstruction to constructing exponentially localized Wannier functions in Chern bands and how it can be circumvented by relaxing the requirement of completeness. While Chern bands forbid a \textit{complete} set of exponentially localized Wannier functions~\cite{thoulessWannierFunctionsMagnetic1984}, one can instead construct an \textit{overcomplete} set of non-orthogonal but exponentially localized coherent states~\cite{rashba1997orthogonal, qi2011generic, jian2013crystalsymmetry, li2024statistical}. We refer to such states as overcomplete Wannier (OW) functions in this paper.

To make this construction explicit, it is convenient to use the projection method for Wannier function construction~\cite{marzari2012maximally,li2024constraints}. \textbf{We now introduce notation that is independent from the main text.} Let $\ket{\psi_n(\mbf{k})}$ denote the single-particle Bloch wavefunction at momentum $\mbf{k}$ in the $n$th band of the parent Hamiltonian $\h{\mc{H}}_{\rm CI}$, with $n = \pm$. The corresponding single-particle momentum-space projectors to the upper and lower bands are given by
\begin{equation}
    P_{\pm}(\mbf{k}) = \ket{\psi_{\pm}(\mbf{k})}\bra{\psi_{\pm}(\mbf{k})},
\end{equation}
consistent with the definition in Eq.~\eqref{eq:CI_ParentHamiltonian_Projectors} of the main text.

Using these projectors, we define the (normalized) single-particle OW wavefunctions, which form an overcomplete basis for the upper ($+$) and lower ($-$) bands:
\begin{equation}
    \ket{W_{\mbf{r},\nu,\pm}} = \int\frac{d^2\mbf{k}}{(2\pi)^2}e^{\ii\mbf{k}\cdot\mbf{r}}  f_{\nu,\pm}(\mbf{k})\ket{\psi_{\pm}(\mbf{k})},
\end{equation}
where $f_{\nu,n}(\mbf{k})$ is a form factor. The form factor is defined as the overlap between the Bloch state and a choice of ``trial'' local orbital $\ket{\tau_\nu(\mbf{k})}$
\begin{equation}
    f_{A/B,\pm}(\mbf{k})\propto\bra{\psi_{\pm}(\mbf{k})}\tau_{A/B}(\mbf{k})\rangle,
\end{equation}
where $\{\ket{\tau_\nu}\}_{\nu=A,B}$ form an orthonormal basis and $\nu=A,B$ denotes distinct choices of local orbitals. In the main text, we choose $\ket{\tau_A} = (1, \ 1)^\intercal/\sqrt{2}$ and $\ket{\tau_B} = (1, \ -1)^\intercal/\sqrt{2}$ for our adaptive protocol. The proportionality reflects that each form factor is $L^2$-normalized over the Brillouin zone: $\int\frac{d^2\mbf{k}}{(2\pi)^2} \left|f_{\nu,n}(\mbf{k})\right|^2 = 1$.

In Chern bands, a topological obstruction manifests directly in the form factors $f_{\nu,n}(\mbf{k})$ used to define OW functions~\cite{li2024statistical}. For bands with Chern number 1, the form factor $f_{\nu,n}(\mbf{k})$ vanishes at a unique momentum $\mbf{k}^0_{\nu,n}$ for each choice of local orbital $\nu$ and for each band $n$:
\begin{equation}
    f_{\nu,n}(\mbf{k}^0_{\nu,n}) = 0.
\end{equation}
As a result, each OW function becomes orthogonal to exactly one Bloch state per band in the thermodynamic limit 
\begin{equation}
    \langle W_{\nu,n}(\mbf{k}^0_{\nu,n})|\psi_n(\mbf{k}^0_{\nu,n}) \rangle = 0,
\end{equation}
where $\ket{W_{\nu,n}(\mbf{k})}$ is the Fourier transform of the real space OW function. Hence, an OW function basis constructed using only a single orbital choice (e.g., $\nu = A$) is \textit{incomplete}, missing one Bloch state per band. That is, the set
\begin{equation}
    \left\{ \ket{W_{\mbf{r},A,n}} \right\}_{\mbf{r}\in\mathbb{Z}^2, n=\pm}
\end{equation}
fails to span the full single-particle Hilbert space by two states. However, the zeroes of $f_{\nu,n}(\mbf{k})$ differ for $\nu=A$ and $\nu = B$. Thus, by including OW functions from both orbitals, we obtain the combined set
\begin{equation}
        \left\{ \ket{W_{\mbf{r},A,n}} \right\}_{\mbf{r}\in\mathbb{Z}^2, n=\pm} \cup     \left\{ \ket{W_{\mbf{r},B,n}} \right\}_{\mbf{r}\in\mathbb{Z}^2, n=\pm},
\end{equation}
which \textit{does} span the full single-particle Hilbert space. Intuitively, the $\nu = A$ basis ``patches'' the ``hole'' in the $\nu = B$ basis, and vice-versa. While this resolves the topological obstruction, it also introduces redundancy. To quantify the resulting overcompleteness, consider a system with $N$ unit cells and two orbitals per unit cell. Thus, the total number of OW functions is $4N$,
\begin{equation}
    \left| \left\{ \ket{W_{\mbf{r},\nu,n}} \right\}_{\mbf{r}\in\mathbb{Z}^2, \nu = A, B, n=\pm } \right| = 4N,
\end{equation}
whereas the Hilbert space dimension is $2N$. The OW function basis is thus overcomplete by $2N$ elements, and this redundancy persists in the thermodynamic limit $N \to \infty$. Nevertheless, the resulting basis consists of exponentially localized states.

For a system of many Chern bands with total Chern number $|\mf{C}| > 1$, each form factor $f_{\nu,n}(\mbf{k})$ may vanish at multiple momenta, with the total degree of the zeroes summing to $|\mf{C}|$~\cite{li2024constraints}. As a result, OW functions from a single orbital $\nu$ will be orthogonal to up to $|\mf{C}|$ Bloch states per band and fail to span the full Hilbert space. To patch these ``holes'' in the basis set, one can combine OW functions from multiple orbitals whose form factor zeroes occur at distinct locations. The minimal number of such orbital types needed depends on the number and degree of zeroes in $f_{\nu,n}(\mbf{k})$.

\section{Details on Lindbladian Analysis for Adaptive Chern Insulator Preparation}\label{app:lindblad}
In this appendix, we provide further details on the effective Lindbladian analysis of our adaptive protocol to prepare Chern insulators. Note that this analysis applies to the \textit{untruncated} version of our protocol, and so we retain the exponential tails of overcomplete Wannier functions in our analysis. We will begin by constructing the quantum channel that describes each trajectory-averaged cycle of our protocol. Following this, we approximate the ancillary system as a Markovian bath with fixed local particle density and trace it out, yielding an effective channel that only acts on the physical layer. The effective channel is then embedded into a continuous-time description, yielding an effective Lindbladian that drives the physical system toward the Chern insulator state. We then estimate the time-scale of convergence for our adaptive protocol by analyzing the two-point operator dynamics generated by this effective Lindbladian. 

\subsection{Effective Quantum Channel on Physical Layer via Markovian Approximation}

The complete quantum channel, which describes the trajectory-averaged dynamics for a single cycle of our adaptive protocol, consists of four elementary measurement‐feedforward operations (depletion and filling) per unit cell, followed by a sequence of random charge-conserving Gaussian unitary gates that redistribute charge in the ancillary system, and then a measurement of all local ancillary occupations. Formally, we write the quantum channel $\mc{E}^{\mr{cycle}}$ representing a single cycle of our circuit as 

\begin{equation}
    \mc{E}^{\mr{cycle}} = \mc{E}_{\mr{a}}^{\mr{meas.}}\mc{E}_{\mr{a}}^{\mr{rand.}}\left[\prod_{\mathbf{r}\in\mathbb{Z}^2}\prod_{\nu=A,B}\prod_{n=\pm}\mc{E}_{\mbf{r},\nu,n}\right] \label{circuit time step channel}
\end{equation}
where the channels $\mc{E}_{\mbf{r},\nu,n}$ are defined in Eq.~\eqref{eq:filling channel} and Eq.~\eqref{eq:depletion channel}). In this section, the ordered `multiplication' of channels denotes repeated application---one should think of it as a sequence of compositions of channels. The ancillary-only channels $\mc{E}_{\mr{a}}^{\mr{meas.}}, \ \mc{E}_{\mr{a}}^{\mr{rand.}}$ are given by
\begin{align}
    \mc{E}_\mr{a}^{\mr{meas.}} &= \prod_{\mbf{r\in\mathbb{Z}^2}}\prod_{\nu=A,B}\mc{E}_{\mr{a},\mbf{r},\nu} \ , \quad \mc{E}_{\mr{a},\mbf{r},\nu}(\h \rho)=\h n_{\mr{a},\mbf{r},\nu}\h\rho \h n_{\mr{a},\mbf{r},\nu} + (1-\h n_{\mr{a},\mbf{r},\nu})\hat{\rho}(1-\h n_{\mr{a},\mbf{r},\nu})\\ 
    \mc{E}^{\mr{rand.}}_\mr{a}(\h \rho) &= \int d\mu(\bs{\theta}) \ \hat{U}_\mr{a}(\bs{\theta})\hat{\rho}\hat{U}^\dagger_{\mr{a}}(\bs{\theta}).
\end{align}
The channel $\mc{E}_\mr{a}^{\mr{meas.}}(\h \rho)$ represents the trajectory-averaged dynamics of the local ancillary occupation measurements $\hat{n}_{\mr{a},\mbf{r},\nu}\equiv d^\dagger_{\mbf{r},\nu}d_{\mbf{r},\nu}$, while the channel $\mc{E}^{\mr{rand.}}_\mr{a}(\h \rho)$ represents the trajectory-averaged random local charge redistribution in the ancillary system. In the above, $\hat{U}_\mr{a}(\bs{\theta})$ is product of $\mr{U}(1)$-symmetric local Gaussian unitaries that is parameterized by a set of independently random phases $\bs{\theta}=(\theta_1,\theta_2,\dots)$ distributed according to a normalized measure $d\mu(\bs{\theta})$---see Eq.~\eqref{eq:random unitary on bottom layer} for an explicit choice of $\hat{U}_\mr{a}(\bs{\theta})$. Both $\mc{E}_\mr{a}^{\mr{meas.}}(\h \rho)$ and $\mc{E}^{\mr{rand.}}_\mr{a}(\h \rho)$ act only on the ancillary system.

A direct calculation shows that the unique fixed point of $\mathcal{E}^{\mathrm{cycle}}$ is
\begin{equation}\label{eq:fixed_point_full}
    \hat\rho_{\mathrm{ss}}
    = \ket{\mathrm{CI}}\!\rangle\langle\!\bra{\mathrm{CI}}\otimes\hat\sigma_{\mr{a}}, 
    \quad 
    \hat\sigma_{\mathrm{a}} \propto \hat{\Id}_{\{\h Q_{\mr{a}}=\Delta{Q}\}}
\end{equation}
where $\ket{\mathrm{CI}}\!\rangle$ is the Chern insulator many‐body state on the physical layer, and the steady-state ancillary-system density matrix $\hat\sigma_{\mathrm{a}}$ is a maximally-mixed state over a definite global charge sector with charge $\Delta{Q}$. Here, $\Delta Q=Q-N_{\mr{uc}}$ denotes the total charge of the ancillary-system steady-state, with $Q$ being the total initial charge of the whole system, and $N_{\mr{uc}}$ denotes the number of unit cells in each layer. Furthermore, $\hat Q_{\mathrm{a}}=\sum_{\mathbf{r}\in\mathbb{Z}^2}\sum_{\nu=A,B} \h n_{\mr{a},\mbf{r},\nu}$ denotes the global charge operator on the ancillary system.

To derive an effective channel acting only on the physical layer, we adopt a \emph{Markovian approximation} for each elementary measurement‐feedforward step. At every elementary step within a single cycle, we assume the ancillary system is instantaneously refreshed as a product Gaussian state with fixed local charge density, i.e., the total density matrix can always be written as 
\begin{equation}\label{eq:bottom_markov}
    \hat\rho_{\mr{tot}} = \hat\rho_{\mathrm{ph}}\otimes\hat\rho_{\mathrm{a}}, 
    \quad \mathrm{Tr}\bigl[\hat\rho_{\mathrm{a}}\h n_{\mr{a},\mbf{r},\nu}\bigr]
    =\bar n_{\mathrm{a}} \ \forall \ \nu,\mathbf{r},
\end{equation}
during its evolution within a cycle, where $\bar n_{\mathrm{a}}\in[0,1]$ is the uniform occupation probability for each local ancilla. 

Since both the filling and depletion channels decompose into two non-commuting channels (corresponding to different orbital combinations $\nu = A,B$), we minimally enlarge the ancillary Hilbert space by adding two additional orbitals per unit cell and modify the protocol such that each fSWAP acts on an independent ancillary mode, removing correlations introduced by reusing the same ancillary-layer orbitals for both $n$ and fixed $\mbf{r},\nu$. With this setup, the ancillary system can be traced out channel-by-channel, yielding an effective channel that acts solely on the physical layer. The result is
\begin{equation}
    \tilde{\mc{E}}^{\mr{cycle}} = \prod_{\mbf{r}\in\mathbb{Z}^2}\prod_{\nu=A,B}\prod_{n=\pm}\tilde{\mc{E}}_{\mbf{r},\nu,n}
\end{equation}
where the effective channels $\tilde{\mc{E}}_{\mbf{r},\nu,n}$ are defined in Eqs.~(\ref{eq:effective filling channel}, \ref{eq:effective depletion channel}).

\subsection{Continuous‐Time Embedding and Effective Lindbladian Theory}
Inspired by the framework of \textit{continuous-time error correction}~\cite{paz1998continuous,ahn2003quantum,oreshkov2013continuoustime,ippoliti2015perturbative}, we embed the discrete-time channel dynamics into a continuous-time setting by promoting the quantum channel to the generator of a continuous-time semigroup, which necessarily takes the form of a Lindbladian generator~\cite{wolf2008dividing}. To embed the trajectory-averaged dynamics into an effective Lindbladian, we assign each local elementary measurement-feedforward channel an infinitesimal success probability $dt$~\footnote{This infinitesimal update resembles the ensemble-averaged dynamics of a Poisson process, where each local channel $\tilde{\mc{E}}_{\mbf{r},\nu,n}$ acts stochastically with rate $dt$. In this sense, the effective Lindbladian emerges from averaging over an underlying stochastic process.}, resulting in the following update rule for the density matrix:
\begin{equation}
    \h\rho_{\mr{ph}}(t + dt) = (1 - dt) \h\rho_{\mr{ph}}(t) + dt \tilde{\mc{E}}_{\mbf{r},\nu,n}(\h\rho_{\mr{ph}}(t)).\label{eq:infinitesimal_update}
\end{equation}
This corresponds to the embedding,
\begin{equation}
    \tilde{\mc{E}}_{\mbf{r},\nu,n} \longrightarrow e^{ dt \mc{L}_{\mbf{r},\nu,n}}, \quad \mc{L}_{\mbf{r},\nu,n} \equiv \tilde{\mc{E}}_{\mbf{r},\nu,n} - \mr{Id}
\end{equation}
where $\mc{L}_{\mbf{r},\nu,n}$ defines the local Lindbladian~\footnote{This construction closely parallels the standard continuous-time embedding of discrete-time Markov chains, where transition matrices give rise to continuous-time semigroup generators~\cite{norris1998markov}} and $\mr{Id}$ denotes the identity channel, i.e., $\mr{Id}(\h \rho) = \h\rho$. Consequently, the complete channel over one infinitesimal circuit time-step is identified with
\begin{equation}
    \tilde{\mc{E}}^{\text{cycle}} \longrightarrow \exp\left({dt  \mc{L}^{\text{cycle}}}\right), \quad \mc{L}^{\text{cycle}}=\sum_{\mbf{r}\in\mathbb{Z}^2}\sum_{\nu = A,B}\sum_{n=\pm}\left(\tilde{\mc{E}}_{\mbf{r},\nu,n} - \mr{Id}\right).
\end{equation}
We can reorganize $\mc{L}^{\text{cycle}}$ into three separate terms
\begin{equation}
    \mathcal{L}^{\mathrm{cycle}} = \mathcal{L}^{\mathrm{gain}} + \mathcal{L}^{\mathrm{loss}} + \mathcal{L}^{\mathrm{decoh.}},\label{eq:app lindblad cycle}
\end{equation}
with
\begin{align}
&\mathcal{L}^{\mathrm{gain}} = \bar{n}_\mathrm{a}\sum_{\mbf{r}\in\mathbb{Z}^2}\sum_{\nu=A,B}\mathcal{D}[\hat{\chi}_{\bfr,\nu,-}^\dagger]\label{eq:app lindblad gain}, \\ 
&\mathcal{L}^{\mathrm{loss}} = (1-\bar{n}_\mathrm{a})\sum_{\mbf{r}\in\mathbb{Z}^2}\sum_{\nu=A,B}\mathcal{D}[\hat{\chi}_{\bfr,\nu,+}]\label{eq:app lindblad deplete}, \\ 
&\mathcal{L}^{\mathrm{decoh.}} = (2-\bar{n}_\mathrm{a})\sum_{\mbf{r}\in\mathbb{Z}^2}\sum_{\nu=A,B}\mathcal{D}[\hat{\mathcal{N}}_{\bfr,\nu,-}] + (1+\bar{n}_\mathrm{a})\sum_{\mbf{r}\in\mathbb{Z}^2}\sum_{\nu=A,B}\mathcal{D}[\hat{\mathcal{N}}_{\bfr,\nu,+}]\label{eq:app lindblad decoh},
\end{align}
and where
\begin{equation}
    \mathcal{D}[\hat{L}](\h\rho)=\hat{L}\rho\hat{L}^\dagger - \frac{1}{2}\{\hat{L}^\dagger\hat{L},\h\rho\}
\end{equation}
is the standard dissipator. The dissipator superoperator can be seen to arise naturally once the identity channel is subtracted from the operator-sum description of the local channel. Indeed, the Kraus operators of a channel $\mc{E}$ become the jump operators for the corresponding \sout{local} Lindbladian $\mc{L}_{\mc{E}}=\mc{E}-\mr{Id}$. 

The time-local Lindbladian in Eq.~(\ref{eq:app lindblad cycle}) governs the coarse-grained evolution of the physical density matrix and provides an analytically tractable description of the adaptive circuit dynamics. Here, $\mathcal{L}^{\mathrm{gain}}$ represents the filling of lower band orbitals through particle swaps from the ancillary system, occurring with a rate proportional to the fixed local ancillary orbital occupation probability $\bar{n}_\mathrm{a}$. Similarly, $\mathcal{L}^{\mathrm{loss}}$ describes the depletion of upper band orbitals, proportional to the complementary occupation probability $(1 - \bar{n}_\mathrm{a})$. The decoherence term $\mathcal{L}_{\mathrm{decoh.}}$ encodes additional measurement-induced decoherence arising from randomness in the overcomplete Wannier function occupation measurement outcomes. It is easy to verify that the unique fixed point of $\mc{L}^{\text{cycle}}$ is $\ket{\mr{CI}}\!\rangle\langle\!\bra{\mr{CI}}$, as guaranteed by the stabilizer conditions given by Eq.~\eqref{eq:CIStabilizerCondition1} and Eq.~\eqref{eq:CIStabilizerCondition2}.

Having constructed an effective Lindbladian for the trajectory-averaged dynamics, we now estimate the convergence timescale of the discrete-time adaptive protocol by determining the convergence timescale of its effective Lindbladian description, which we find to be $\mc{O}(1)$. This is in agreement with the numerical evidence presented in Fig.~\ref{fig:dynamics}, which finds that the discrete-time adaptive protocol exponentially converges to the fixed point in $\mc{O}(1)$ cycles. Moreover, the effective Lindbladian description offers useful intuition: it reveals that including both orbital combinations $\nu = A,B$---i.e., the overcompleteness of the Wannier function basis---is essential for a gapped relaxation time-scale. 

\subsection{Two-Point Operator Dynamics and Convergence Time-Scale}

To extract the convergence time-scale, we analyze the Heisenberg evolution of the two-point operators under the dynamics induced by the effective Lindbladian Eq.~\eqref{eq:app lindblad cycle}. While the gain and loss components of $\mc{L}^{\mr{cycle}}$ preserve Gaussianity---since their jump operators are linear in fermion operators---the decoherence term $\mc{L}^{\mr{decoh.}}$ does not. Nevertheless, because the decoherence jump operators are Hermitian, the two-point function still evolves under a closed equation of motion~\cite{dolgirev2020nongaussian,barthel2022solving}. Although the many-body state may become non-Gaussian during its evolution, the Heisenberg equations of motion of higher-point correlators form a closed hierarchy: the evolution of $2k$-point functions depends only on correlators of order $\leq 2k$. In particular, the two-point dynamics are independent of higher moments and admit a closed-form expression. Since the steady state is Gaussian, it is fully determined by its two-point function. Hence, we expect that analyzing the convergence of the two-point operators will suffice to estimate the timescale at which the adaptive protocol prepares the Chern insulator state. We will use the notation established in App.~\ref{app: topological obstructions} in this subsection.

Exploiting translation invariance, we work in momentum space. We define momentum space operators for the $\sigma$th fermion mode in the physical layer as 
\begin{equation}
    \hat{c}_{\mbf{r},\sigma} = \int \frac{d^2\mbf{k}}{(2\pi)^2} e^{\ii\mbf{k}\cdot\mbf{r}} \ \hat{c}_\sigma(\mbf{k})
\end{equation}
and similarly for the momentum OW operators $\hat{\chi}_{\nu,-}^\dagger(\mbf{k})$. The $\mbf{k}$-space two-point operator is then defined as 
\begin{equation}
    \h G_{\sigma\eta}(\mbf{k},\mbf{k'}) = \h c_{\sigma}^\dagger(\mbf{k}) \h c_{\eta}(\mbf{k}')
\end{equation}
where $\sigma,\eta = 1,2$ label orbitals. The two-point function is then just the expected value of the two-point operator $G\equiv\langle \hat{G}\rangle$, where the expectation value is taken with respect to the physical density matrix. If the density matrix evolves in time according to $\mc{L}^{\text{cycle}}$ i.e.,
\begin{equation}
    d \h\rho_{\mr{ph}} = dt\mc{L}^\text{cycle}(\h \rho_{\mr{ph}}(t)),
\end{equation}
it follows that the two-point function time-evolves in the Heisenberg picture according to the \textit{dual} Lindbladian
\begin{equation}
   \dot{G}_{\sigma\eta}(\mbf{k},\mbf{k'},t) = \left\langle\mc{L}^{\text{cycle}\dagger}\left(\h G_{\sigma\eta}(\mbf{k},\mbf{k'},t)\right)\right\rangle \label{eq: heisenberg two point evolution}
\end{equation}
where the dual dissipator is defined as $\mc{D}^\dagger[\hat{L}](\h\rho)=\hat{L}^\dagger\h\rho\h L - \frac{1}{2}\{\hat{L^\dagger}\hat{L}, \h\rho\}$~\footnote{The dual or adjoint dissipator is defined with respect to the Hilbert-Schmidt inner product---see~\cite{guoDesigningOpenQuantum2025} for more technical details}. 
Using the fact that the dual dissipator can be expressed in terms of commutators
\begin{equation}
    \mc{D}^\dagger[\hat{L}](\h\rho) = \frac{1}{2}\left(\hat{L}^\dagger[\h \rho, \h L] + [\h L^\dagger,\h \rho]\h L\right)
\end{equation}
we can write the components of the dual effective Lindbladian $\mc{L}^{\mr{cycle}\dagger}$ in momentum space as 
\begin{align}
\mathcal{L}^{\mathrm{gain}\dagger}(\h \rho) & = \bar{n}_\mathrm{a}\int\frac{d^2\mbf{k}}{(2\pi)^2}\sum_{\nu=A,B}\frac{1}{2}\left(\hat{\chi}_{\nu,-}(\mbf{k})\left[\h\rho \ ,\hat{\chi}^\dagger_{\nu,-}(\mbf{k}) \right] + \left[\hat{\chi}_{\nu,-}(\mbf{k}) \ , \ \h\rho\right]\hat{\chi}^\dagger_{\nu,-}(\mbf{k})\right) ,\label{eq:app lindblad gain k-space}\\ 
\mathcal{L}^{\mathrm{loss}\dagger}(\h \rho) &= (1-\bar{n}_\mathrm{a})\int\frac{d^2\mbf{k}}{(2\pi)^2}\sum_{\nu=A,B}\frac{1}{2}\left(\hat{\chi}^\dagger_{\nu,+}(\mbf{k})\left[\h\rho \ ,\hat{\chi}_{\nu,+}(\mbf{k}) \right] + \left[\hat{\chi}^\dagger_{\nu,+}(\mbf{k}) \ , \ \h\rho\right]\hat{\chi}_{\nu,+}(\mbf{k})\right) ,\label{eq:app lindblad deplete k-space}
\end{align}
\begin{equation}
\begin{split}
    \mathcal{L}^{\mathrm{decoh.}}(\hat \rho) = -(2-\bar{n}_\mathrm{a})\int\prod_{i=1}^4\frac{d^2\mbf{k}_i}{(2\pi)^2}\sum_{\nu=A,B}\frac{1}{2}\left[\hat{\chi}^\dagger_{\nu,-}(\mbf{k}_1)\hat{\chi}_{\nu,-}(\mbf{k}_2) \ , \ \left[\hat{\chi}^\dagger_{\nu,-}(\mbf{k}_3)\hat{\chi}_{\nu,-}(\mbf{k}_4) \ , \ \h\rho\right]\right](2\pi)^2\delta(\mbf{k}_1-\mbf{k}_2+\mbf{k}_3-\mbf{k}_4)\\
    - (1+\bar{n}_\mathrm{a})\int\prod_{i=1}^4\frac{d^2\mbf{k}_i}{(2\pi)^2}\sum_{\nu=A,B}\frac{1}{2}\left[\hat{\chi}^\dagger_{\nu,+}(\mbf{k}_1)\hat{\chi}_{\nu,+}(\mbf{k}_2) \ , \ \left[\hat{\chi}^\dagger_{\nu,+}(\mbf{k}_3)\hat{\chi}_{\nu,+}(\mbf{k}_4) \ , \ \h\rho\right]\right](2\pi)^2\delta(\mbf{k}_1-\mbf{k}_2+\mbf{k}_3-\mbf{k}_4)\label{eq:app lindblad decoh k-space}
\end{split}
\end{equation}
where we note that $\mc{L}^\mr{decoh.}=\mc{L}^\mr{decoh.\dagger}$ is self-dual. 

Assuming a translation-invariant initial state~\footnote{This assumption suffices to characterize convergence for more general initial states, since Cauchy-Schwarz implies $|G_{\sigma\eta}(\mbf{k},\mbf{k}')|^2 \leq G_{\sigma\sigma}(\mbf{k}) G_{\eta\eta}(\mbf{k}')$, so decay of the diagonal two-point functions bounds the off-diagonals.} we restrict attention to the diagonal $\mbf{k}$-space components $G_{\sigma\eta}(\mbf{k}) \equiv G_{\sigma\eta}(\mbf{k},\mbf{k})$ and analyze their Heisenberg evolution under the effective Lindbladian $\mc{L}^{\mr{cycle}}$.
Because the dual Lindbladian $\mc{L}^{\mr{cycle}\dagger}$ just consists of commutators with single fermion operators and Hermitian fermion bilinears, the action of $\mc{L}^{\text{cycle}\dagger}$ on the two-point operator is equivalent to a c-number valued superoperation on the two-point function~\cite{barthel2022solving}.
Expanding the overcomplete Wannier field operators in terms of fermion orbital modes via 
\begin{equation}
    \hat{\chi}^{\dagger}_{\nu,\pm}(\mbf{k})=\sum_{\sigma} f_{\nu,\pm}(\mbf{k})\ket{\psi_{\pm}(\mbf{k})}_{\sigma}\hat{c}^{\dagger}_{\sigma}(\mbf{k)}
\end{equation}
where $\ket{\psi_{\pm}(\mbf{k})}_{\sigma}$ denotes the $\sigma$th component of the two-component Bloch state $\ket{\psi_{\pm}(\mbf{k})}$, it follows from Eqs.~\eqref{eq: heisenberg two point evolution}, \eqref{eq:app lindblad gain k-space}, \eqref{eq:app lindblad deplete k-space}, \eqref{eq:app lindblad decoh k-space} that the Heisenberg evolution of the two-point function is given by
\begin{equation}
    \begin{split}
        \dot{G}_{\sigma\eta}(\mbf{k},t) &= \bar{n}_{\mr{a}}\sum_{\nu=A,B}|f_{\nu,-}(\mbf{k})|^2\left[P_{-}^*(\mbf{k})-\frac{1}{2}\{P_{-}^*(\mbf{k}),G(\mbf{k})\}\right]_{\sigma\eta}\\
        &-(1-\bar{n}_{\mr{a}})\sum_{\nu=A,B}|f_{\nu,+}(\mbf{k})|^2\left[\frac{1}{2}\{P^*_{+}(\mbf{k}),G(\mbf{k},t)\}\right]_{\sigma\eta}\\
        &+(2-\bar{n}_\mr{a})\sum_{\nu=A,B}|f_{\nu,-}(\mbf{k})|^2\left[\left(\int\frac{d^2\mbf{q}}{(2\pi)^2}|f_{\nu,-}(\mbf{q})|^2\bra{\psi_{-}(\mbf{q})}G^*(\mbf{q},t)\ket{\psi_{-}(\mbf{q})}\right)^*P^*_{-}(\mbf{k})-\frac{1}{2}\{P_{-}^*(\mbf{k}),G(\mbf{k},t)\}\right]_{\sigma\eta}\\
        &+(1+\bar{n}_\mr{a})\sum_{\nu=A,B}|f_{\nu,+}(\mbf{k})|^2\left[\left(\int\frac{d^2\mbf{q}}{(2\pi)^2}|f_{\nu,+}(\mbf{q})|^2\bra{\psi_{+}(\mbf{q})}G^*(\mbf{q},t)\ket{\psi_{+}(\mbf{q})}\right)^*P^*_{+}(\mbf{k})-\frac{1}{2}\{P_{+}^*(\mbf{k}),G(\mbf{k},t)\}\right]_{\sigma\eta}
        \label{eq:translation-invariant L_cycle heisenber evo}
    \end{split}
\end{equation}
where we follow the notation established in App.~\ref{app: topological obstructions}. In particular, $f_{\nu,n}(\mbf{k})$ are $L^2$-normalized form factors of the corresponding overcomplete Wannier functions, $P_{\pm}(\mbf{k})\equiv\ket{\psi_{\pm}(\mbf{k}}\bra{\psi_{\pm}(\mbf{k})}$ denote projections on to the upper $(+)$ and lower $(-)$ band Bloch states at a momentum $\mbf{k}$, and $(...)^*$ denotes complex conjugation. It is easy to verify that Chern insulator two-point function $G^{\mr{CI}}(\mbf{k})=P^*_{-}(\mbf{k})$ is the unique steady-state of the Lindbladian evolution. This follows from the fact that at the many-body level, the Chern insulator state $\ket{\mr{CI}}\!\rangle\langle\!\bra{\mr{CI}}$ is the simultaneous steady state of all local terms in the Lindbladian~\eqref{eq:app lindblad cycle}. 

We now extract the relaxation timescale of the two-point function dynamics. First, complex conjugate both sides of Eq.~\eqref{eq:translation-invariant L_cycle heisenber evo} (for simplicity of notation) and expand $G^*(\mbf{k},t)$ in terms of band basis matrix elements
\begin{equation}
    G^*(\mbf{k},t) = g_{+}(\mbf{k},t)P_{+}(\mbf{k}) + g_{-}(\mbf{k},t)P_{-}(\mbf{k}) + \alpha(\mbf{k},t)\ket{\psi_+(\mbf{k})}\bra{\psi_{-}(\mbf{k})} + \alpha^*(\mbf{k},t)\ket{\psi_{-}(\mbf{k})}\bra{\psi_{+}(\mbf{k})}
\end{equation}
where $g_{\pm}(\mbf{k},t)\equiv\bra{\psi_{\pm}(\mbf{k})}G^*(\mbf{k},t)\ket{\psi_{\pm}(\mbf{k})}$ are scalar functions representing the diagonal components and $\alpha(\mbf{k},t)\equiv\bra{\psi_{+}(\mbf{k})}G^*(\mbf{k},t)\ket{\psi_{-}(\mbf{k})}$ represent the off-diagonal components. Note that the unique steady-states for each scalar function are given by
\begin{equation}
    g_{+}(\mbf{k},t\to\infty) = 0, \quad g_{-}(\mbf{k},t\to\infty) = 1, \quad \alpha(\mbf{k},t\to\infty) = 0 \label{eq: two point component steady-states}
\end{equation}
which gives us $G^*(\mbf{k},t\to\infty)=P_{-}(\mbf{k})$, as desired. We can derive equations of motion for $g_{\pm}(\mbf{k},t)$ and $\alpha(\mbf{k},t)$ by projecting both sides of Eq.~\eqref{eq:translation-invariant L_cycle heisenber evo} onto band basis matrix elements. This leads to
\begin{align}
    \dot{g}_+(\mbf{k},t) &= -2\sum_{\nu=A,B}|f_{\nu,+}(\mbf{k})|^2 g_{+}(\mbf{k},t) +(1+\bar{n}_\mr{a})\sum_{\nu=A,B}|f_{\nu,+}(\mbf{k})|^2\left[\int\frac{d^2\mbf{q}}{(2\pi)^2} \ |f_{\nu,+}(\mbf{q})|^2g_{+}(\mbf{q},t)\right]\label{eq: upper band two point EOM}\\
    \dot{g}_-(\mbf{k},t) &= \sum_{\nu=A,B}|f_{\nu,-}(\mbf{k})|^2 \left[\bar{n}_{\mr{a}}-2g_{-}(\mbf{k},t)\right] +(2-\bar{n}_\mr{a})\sum_{\nu=A,B}|f_{\nu,-}(\mbf{k})|^2\left[\int\frac{d^2\mbf{q}}{(2\pi)^2} \ |f_{\nu,-}(\mbf{q})|^2g_{-}(\mbf{q},t)\right]\label{eq: lower band two point EOM}\\
    \dot{\alpha}(\mbf{k},t)&=-\sum_{n=\pm}\sum_{\nu=A,B}|f_{\nu,n}(\mbf{k})|^2\alpha(\mbf{k},t)\label{eq: off-diagonal two point EOM}
\end{align}
The off-diagonal equation of motion Eq.~\eqref{eq: off-diagonal two point EOM} can be solved exactly, leading to
\begin{equation}
    \alpha(\mbf{k},t) = \alpha(\mbf{k},0)e^{-\left[\sum_{n=\pm}\gamma_{n}(\mbf{k})\right]t}\label{eq: off-diagonal two point solution}
\end{equation}
where we have defined the dissipation rates
\begin{equation}
    \gamma_{\pm}(\mbf{k})\equiv\sum_{\nu=A,B}|f_{\nu,\pm}(\mbf{k})|^2.\label{eq: sum of square form factors}
\end{equation}
These dissipation rates will show up again, and we pause to comment on their relationship with topological obstructions. In our setting, the dissipation rates remain nonzero for all $\mathbf{k}$. While the individual form factors $f_{\nu,\pm}(\mathbf{k})$ do possess zeros in the topological phase, these zeros occur at distinct momenta for different orbital combinations $\nu$---see App.~\ref{app: topological obstructions} for a detailed discussion. Through Eq.~\eqref{eq: sum of square form factors}, we can see that it is crucial that we use both local orbitals $\nu=A,B$ such that $\gamma_{\pm}(\mbf{k})\neq0$ for any $\mbf{k}$. The relaxation time will be shown to be the reciprocal of the minimal value of  $\gamma_{n}(\mbf{k})$, so using both $\nu=A,B$ ensures that the relaxation time is \textit{finite}.

We can show that the dissipation rates are $\mc{O}(1)$ quantities. First, because the form factors have distinct zeroes for different $\nu$, we have that $\gamma_{\pm}(\mbf{k})>0$ for all $\mbf{k}$. Second, since the form factors are normalized i.e., satisfy $\int\frac{d^2\mbf{k}}{(2\pi)^2} \ |f_{\nu,\pm}(\mbf{k})|^2=1$, we have that $|f_{\nu,\pm}(\mbf{k})|^2$ is independent of system size i.e., they are $\mc{O}(1)$ quantities. As a result, the dissipation rates $\gamma_{\pm}(\mathbf{k})$ are strictly positive and $\mc{O}(1)$ (away from the gap closing points).

Since $\gamma_{\pm}(\mathbf{k})>0$, we see that from Eq.~\eqref{eq: off-diagonal two point solution} that $\alpha(\mbf{k},t\to\infty)\to0$ for all $\mbf{k}$, as expected from Eq.~\eqref{eq: two point component steady-states}.
In contrast, we cannot solve the equations of motion for the diagonal components $g_\pm(\mbf{k},t)$ explicitly. However, we can still determine the relaxation time of the occupation \textit{densities} in each band. Consider the upper band component equation of motion Eq.~\eqref{eq: upper band two point EOM} and integrate both sides of Eq.~\eqref{eq: upper band two point EOM} with respect to $\mbf{k}$ over the entire two-dimensional Brillouin zone. Making use of the fact that the form factors $f_{\nu,n}(\mbf{k})$ are $L^2$-normalized with respect to $\mbf{k}$, we can simplify the result to
\begin{equation}
    \int\frac{d^2\mbf{k}}{(2\pi)^2} \ \dot{g}_{+}(\mbf{k},t) =-(1-\bar{n}_{\mr{a}})\int\frac{d^2\mbf{k}}{(2\pi)^2} \ \gamma_{+}(\mbf{k})g_{+}(\mbf{k},t)\label{eq: upper band component two-point analysis}
\end{equation}
Now define the $\mbf{k}$-space averaged diagonal components
\begin{equation}
    \bar{g}_{\pm}(t)=\int \frac{d^2\mbf{k}}{(2\pi)^2} \ g_{\pm}(\mbf{k},t)
\end{equation}
which physically correspond to the total occupation densities of the $n=\pm$ bands. Furthermore, we define the dissipation gaps
\begin{equation}
    \delta_{\pm} \equiv \underset{\mbf{k}}{\mr{min}} \ \gamma_{\pm}(\mbf{k}).
\end{equation}
Then it follows from Eq.~\eqref{eq: upper band component two-point analysis} that $\dot{\bar{g}}_{+}(t)$ is upper bounded as
\begin{equation}
    \dot{\bar{g}}_{+}(t) \leq -(1-\bar{n}_{\mr{a}})\delta_{+}\bar{g}_+(t)
\end{equation}
which yields the solution
\begin{equation}
    \bar{g}_+(t)\leq\bar{g}_+(0)e^{-(1-\bar{n}_{\mr{a}})\delta_{+}t}\label{eq: upper band density upper bound}.
\end{equation}
We see that the upper band occupation density is upper bounded by exponential decay, with a decay rate set by the product of the ancillary ``hole'' occupancy $(1-\bar{n}_{\mr{a}})$ and the dissipation gap $\delta_{+}>0$.

A similar analysis can be performed for the lower band component equation of motion. Integrating both sides of Eq.~\eqref{eq: lower band two point EOM} with respect to $\mbf{k}$ and applying normalization constraints on the form factors, we find 
\begin{equation}
    \int\frac{d^2\mbf{k}}{(2\pi)^2} \ \dot{g}_{-}(\mbf{k},t) =2n_{\mr{a}}-\bar{n}_{\mr{a}}\int\frac{d^2\mbf{k}}{(2\pi)^2} \ \gamma_{-}(\mbf{k})g_{-}(\mbf{k},t).\label{eq: lower band component two-point analysis}
\end{equation}
Separating out the steady state solution $g_{-}(\mbf{k},t) = 1+\delta g_{-}(\mbf{k},t)$ and simplifying leads to
\begin{equation}
    \int\frac{d^2\mbf{k}}{(2\pi)^2} \ \delta\dot{g}_{-}(\mbf{k},t) =-\bar{n}_{\mr{a}}\int\frac{d^2\mbf{k}}{(2\pi)^2} \ \gamma_{-}(\mbf{k})\delta g_{-}(\mbf{k},t).\label{eq: lower band component two-point analysis 2}
\end{equation}
Now, define the $\mbf{k}$-space averages of the steady state deviation $\overline{\delta g}(t)$. It follows that one can upper-bound the lower band occupancy density deviation as
\begin{equation}
    \overline{\delta g}_{-}(t)\leq \overline{\delta g}_-(0)e^{-\bar{n}_{\mr{a}}\delta_{-}t}.\label{eq: lower band density upper bound}
\end{equation}
We see that the lower band occupation density is upper bounded by exponential decay, with a decay rate set by the product of the ancillary particle occupancy $\bar{n}_{\mr{a}}$ and the dissipation gap $\delta_{-}>0$.

Taking into account Eqs.~\eqref{eq: off-diagonal two point solution}, \eqref{eq: upper band density upper bound}, \eqref{eq: lower band density upper bound}, we infer that the longest relaxation time-scale $T_{\mr{conv.}}$ of the effective Lindbladian is given by
\begin{equation}
    1/T_{\mr{conv.}} = \mr{min}\left[(1-n_{\mr{a}})\delta_{+} \ , \ n_{\mr{a}}\delta_{-}\right] \propto \underset{\mbf{k},n}{\mr{min}}\left[\gamma_{n}(\mbf{k})\right]
\end{equation}
which is equivalent to the result in the main text Eq.~\eqref{eq: main text convg timescale}.

We have already argued that for any finite-size system, the dissipation rates are $\mc{O}(1)$ quantities. Since $\bar{n}_\mr{a}\in[0,1]$, we have that $T_{\mr{conv.}}\sim\mc{O}(1)$ and is thus independent of system size. We therefore estimate that the discrete-time adaptive protocol reaches the steady state Chern insulator in $\mathcal{O}(1)$ measurement-feedforward cycles, in agreement with numerics---see Fig.~\ref{fig:dynamics}.

\twocolumngrid
\bibliography{References_PRR}
\end{document}